\documentclass{aa}  
\bibpunct{(}{)}{;}{a}{}{,} 

\usepackage{graphicx}
\usepackage{txfonts}
\usepackage{color}

\newcommand{\cmap}{chromosome map}

\newcommand{\teff}{T_\mathrm{eff}}
\newcommand{\logg}{\log g}
\newcommand{\feh}{[\mathrm{Fe}/\mathrm{H}]}
\newcommand{\vrad}{v_\mathrm{rad}}
\newcommand{\sumew}{\Sigma \mathrm{EW}}

\newcommand{\vhb}{V_\mathrm{HB}}
\newcommand{\dex}{\,\mathrm{dex}}

\begin{document}

\title{A stellar census in globular clusters with MUSE}
\subtitle{Extending the CaT-metallicity relation below the horizontal branch and applying it to multiple populations}

\author{Tim-Oliver Husser\inst{1}\thanks{E-mail: thusser@uni-goettingen.de}
        \and Marilyn Latour\inst{1} 
        \and Jarle Brinchmann\inst{2,3}
        \and Stefan Dreizler\inst{1}
        \and Benjamin Giesers\inst{1}
        \and Fabian G\"ottgens\inst{1}
        \and Sebastian Kamann\inst{4}
        \and Martin M.\ Roth\inst{5}
        \and Peter M.\ Weilbacher\inst{5}
        \and Martin Wendt\inst{6}
    }

\institute{
Institut f\"ur Astrophysik, Georg-August-Universit\"at G\"ottingen, Friedrich-Hund-Platz 1, 37077 G\"ottingen, Germany
\and
Leiden Observatory, Leiden University, PO Box 9513, NL-2300 RA Leiden, the Netherlands
\and
Instituto de Astrof\'isica e Ci\^{e}ncias do Espa\c{c}o, Universidade do Porto, CAUP, Rua das Estrelas, PT4150-762 Porto, Portugal
\and
Astrophysics Research Institute, Liverpool John Moores University, 146 Brownlow Hill, Liverpool L3 5RF, UK
\and
Leibniz-Institut f\"ur Astrophysik Potsdam (AIP), An der Sternwarte 16, 14482 Potsdam, Germany
\and
Institut f\"ur Physik und Astronomie, Universit\"at Potsdam, Karl-Liebknecht-Str. 24/25, 14476 Golm, Germany
}

\date{Received September 15, 1996; accepted March 16, 1997}

\abstract{}
{
We use the spectra of more than 30,000 red giant branch (RGB) stars in 25 globular clusters (GC), obtained within the MUSE survey of Galactic globular clusters, to calibrate the \ion{Ca}{ii} triplet (CaT) metallicity relation and derive metallicities for all individual stars. We investigate the overall metallicity distributions as well as those of the different populations within each cluster.
}
{
The \ion{Ca}{ii} triplet in the near-infrared at 8498, 8542, and 8662 \AA\ is visible in stars with spectral types between F and M and can be used to determine their metallicities. In this work, we calibrate the relation using average cluster metallicities from literature and MUSE spectra, and extend it below the horizontal branch -- a cutoff that has traditionally been made to avoid a non-linear relation --  using a quadratic function. In addition to the classic relation based on $V-\vhb$ we also present calibrations based on absolute magnitude and luminosity. The obtained relations were used to calculate metallicities for all the stars in the sample and to derive metallicity distributions for different populations within a cluster, which have been separated using so-called ``chromosome maps'' based on HST photometry.
}{
We show that despite the relatively low spectral resolution of MUSE ($R=1900\mbox{--}3700$) we can derive single star metallicities with a mean statistical intra-cluster uncertainty of $\sim 0.12\dex$. We present metallicity distributions for the RGB stars in 25 GCs and investigate the different metallicities of the populations P3 (and higher) in so-called metal-complex or Type~II clusters, finding metallicity variations in all of them. We also detected unexpected metallicity variations in the Type~I cluster \object{NGC\,2808} and confirm the Type~II status of \object{NGC\,7078}.
}{}


\maketitle

\section{Introduction}
Over the last two decades, the Hubble Space Telescope (HST) has proven to be a game-changer in the research of globular clusters (GCs). Not only did it open the window to an unprecedented view into the crowded centres of the clusters, which today allows us to derive detailed proper motions for single stars \citep{2014ApJ...797..115B}, but it also provided stellar magnitudes \citep{2007AJ....133.1658S,2018MNRAS.tmp.2405N} with enough precision to distinguish complex structures in the CMDs of globular clusters.

Following on the early findings of a bimodal main sequence (MS) in the color-magnitude diagram (CMD) of \object{$\omega$~Centauri} \citep{1997PhDT.........8A,2004ApJ...605L.125B}, splits on the MSs, the subgiant (SGB) and red giant branches (RGBs), and even on the asymptotic giant branches and horizontal branches (HBs) have been found for several clusters \citep[see, e.g.,][]{2012A&ARv..20...50G,2013ApJ...775...15P,2014MNRAS.437.1609M,2015MNRAS.447..927M}. According to the most recent studies, it appears that nearly all GCs (older than about 2 Gyrs) show structures in their CMD suggesting the presence of multiple groups of stars that are usually referred to as different generations or populations \citep{2017MNRAS.464.3636M}. The presence of these multiple populations has been further supported by spectroscopic results, especially by the discovery of light element variations that have been observed in all investigated clusters, emerging in the form of anti-correlations of elemental abundances, such as Na-O and Mg-Al \citep{2010A&A...516A..55C}.

Most of the scenarios that have been proposed to explain these observations are built on the assumption of a process of ``self-enrichment'' of the interstellar medium and multiple star formation events, thus explaining the use of the term ``generations''. Possible candidates for the polluters range from massive asymptotic giant branch stars \citep{2010MNRAS.407..854D} to fast-rotating massive stars \citep{2007A&A...464.1029D} to interacting massive binary stars \citep{2009A&A...507L...1D}. However, \citet{2015MNRAS.449.3333B} showed that none of these scenarios alone can reproduce the observed abundance trends in all GCs.

\citet{2015ApJ...808...51M} showed that a pseudo-CMD, constructed from two peusdo-colors calculated by combining four filters covering wavelength ranges from the near-UV to the optical, allows the different populations to be easily separated, at least on the RGB and along the lower MS. These pseudo-CMDs are commonly referred to as ``chromosome maps''. For the majority of the globular clusters investigated by \citet{2017MNRAS.464.3636M}, the authors divided the clusters' RGB stars into two populations. They called the bulk of stars near the origin of the chromosome map the ``first generation'' stars, and all the others, usually extending above, the ``second generation'' stars. We will refer to them as populations 1 (P1) and 2 (P2). The stars belonging to the P1 populations show a normal abundance pattern while the other stars have a chemistry showing signs of processing such as enhanced Na abundances \citep[see, e.g.,][]{2019MNRAS.487.3815M}.

Additionally, \citet{2017MNRAS.464.3636M} identified more than two populations in some of the clusters in their sample. 
These clusters were referred to as Type~II or metal-complex (as compared to the Type~I clusters containing only 2 populations).
The Type~II clusters show a split in their subgiant branch in both optical and UV CMDs and the faint SGB connects with a red-RGB. The stars belonging to the red-RGB form one, or sometimes more, additional population(s). The stars belonging to this additional population (that we will refer to as P3) have been investigated in a few clusters and some of them appear to be enriched in iron, s-process elements, and some also in their C+N+O abundances \citep[e.g.,][]{2018ApJ...859...81M,2015MNRAS.450..815M,2014MNRAS.441.3396Y}.
A few other clusters, although they are not identified as Type~II, also have additional populations that were investigated with the help of their chromosome maps. For example, previous studies have identified five populations in \object{NGC\,2808} and \object{NGC\,7078} \citep{2015ApJ...808...51M,2018MNRAS.tmp.2405N}.
Although variations in metallicity have not been reported so far in Type~I clusters, the presence of an iron-spread has been suggested to explain the extension of the P1 stars in the chromosome map of some GCs \citep{2015ApJ...808...51M,2019MNRAS.487.3815M}. However, whether iron or helium variations are responsible for the color spread of the P1 stars is still a matter of debate \citep[see e.g.,][]{2018A&A...616A.168L,2018MNRAS.481.5098M}. 
A more detailed investigation of metallicities in globular clusters would certainly help to constrain the possible formation scenarios of their multiple populations.

The common way of deriving metallicities from observed medium resolution spectra is to compare them with models \citep[see, e.g.,][]{2016A&A...588A.148H}. However, it is useful to have an alternative method available that is independent of model assumptions and only relies on observations. One of these alternatives is the infrared \ion{Ca}{ii} triplet (CaT) lines at 8498, 8542, and 8662~\AA, which is often used as a proxy for metallicity measurements.
These three lines are among the most prominent features in the spectra of G, K, and M stars \citep{2005A&A...430..669A} and are easily visible even on low resolution or noisy stellar spectra. 

\citet{1988AJ.....96...92A} analysed the integrated-light spectra of GCs and found that the measured EWs of the CaT lines strongly correlate with the cluster metallicity $\feh$. Building on this result, \citet{1991AJ....101.1329A} focused on individual RGB stars and revealed an additional dependence between EWs and brightness. Plotting their EWs as function of the magnitude difference to the HB, $V-\vhb$, they found that the intercepts of the linear fit with the ordinate, which they called the reduced equivalent width ($W'$), nicely correlate with the metallicity.
Studies using the CaT to infer metallicities usually only include RGB stars brighter than the HB \citep{2009ApJ...705.1481D} and use a linear relation between the measured EWs and $W'$. Since this excludes a large amount of RGB stars, \citet{2007AJ....134.1298C} suggested the use of a quadratic relation and including all stars on the RGB.

In this paper, we combine stellar metallicities, derived from the CaT-metallicity relation, with chromosome maps to investigate the metallicity distributions of the populations within GCs. To achieve this, we use a homogeneous sample of RGB spectra obtained as part of the MUSE survey of Galactic globular clusters. We first calibrated the CaT-metallicity relation using the spectra of RGB stars in 19 GCs and provided a calibration that extends below the HB as well as calibrations based on absolute magnitude and luminosity. We then used these relations to derive metallicities for more than 30 000 RGB stars in our total sample of 25 GCs and investigate the metallicity distribution of these clusters. Only 21 clusters in our sample have the necessary photometric data to create chromosome maps. For these, we also obtained the metallicity distributions of their individual populations. Our approach is valid as long as the Ca abundances [Ca/Fe] do not vary from star to star. This is not expected for Type~I clusters and, indeed, \citet{2019MNRAS.487.3815M} found no significant Ca variation between the P1 and P2 stars. However, for at least two Type~II clusters, namely \object{NGC\,5139} (\object{$\omega$~Centauri}) and \object{NGC\,6715} (M 54, not in our sample), they found an increase in Ca from the blue- to the red-RGB stars. We note that \object{$\omega$~Centauri} is not used for the calibration of the CaT-metallicity relation and our approach should not be affected by changes in Ca abundances. Another Type~II cluster with reported variations is \object{NGC\,6656} \citep[\object{M\,22}, see][]{2009Natur.462..480L, 2011A&A...532A...8M}.

The paper is organized as follows. We first describe the observations and the data reduction in Sect.~\ref{sec:observations}. The process of creating chromosome maps from HST photometry is discussed in Sect.~\ref{sec:cmaps}. The measurements of EWs, the CaT calibration itself, and its extension below the HB are presented in Sect.~\ref{sec:calib}. Section~\ref{sec:feh_dists} gives a general overview of the metallicity distributions for all clusters, while in Sect.~\ref{sec:1g_variations}, we investigate on the possibility of a metallicity trend within the primordial populations of the clusters. Finally, Sect.~\ref{sec:individual} includes short discussions on the results for all 25 individual clusters in our sample and we briefly present our conclusion in Sect.~8.

All the results from this paper are available as tables in VizieR and on our project homepage\footnote{\url{https://musegc.uni-goettingen.de/}}, containing columns for cluster names, star IDs, RA/Dec coordinates, the measured EWs of the CaT lines (both from Voigt profiles and from simple integration), the derived reduced equivalent widths, and the final metallicities, relative to their respective cluster means.

\section{Observations and data reduction}
\label{sec:observations}
\begin{table}
 \caption{Overview of observed RGB stars in the MUSE survey for the 25 GCs investigated in this paper.}
 \begin{center}
 \begin{tabular}{rlrrr}
 \hline \hline
    NGC &   Name &    RGB & Valid & $V-\vhb<0.2$ \\
   (1) &    (2) &    (3) &  (4)  &   (5) \\ \hline
   104 & 47 Tuc &   2587 &     2538 &    354 (13.9\%) \\
   362 &        &   1236 &     1144 &    237 (20.7\%) \\
  1851 &        &   1454 &     1358 &    273 (20.1\%) \\
  1904 &        &    454 &      430 & --- \\
  2808 &        &   2788 &     2512 &    713 (28.4\%) \\
  3201 &        &    139 &      137 &     41 (29.9\%) \\
  5139 & $\omega$ Cen &   1485 &     1421 & --- \\
  5286 &        &   1376 &     1153 &    212 (18.4\%) \\
  5904 &    M 5 &    937 &      870 &    198 (22.8\%) \\
  6093 &        &   1315 &     1071 &    248 (23.2\%) \\
  6218 &   M 12 &    245 &      236 & --- \\
  6254 &   M 10 &    439 &      399 &     90 (22.6\%) \\
  6266 &   M 62 &   2314 &     2191 & --- \\
  6293 &        &    230 &      168 & --- \\
  6388 &        &   4668 &     4098 &    741 (18.1\%) \\
  6441 &        &   4978 &     4408 &   1047 (23.8\%) \\
  6522 &        &    536 &      481 & --- \\
  6541 &        &    910 &      820 &    135 (16.5\%) \\
  6624 &        &    581 &      539 &     72 (13.4\%) \\
  6656 &   M 22 &    423 &      397 &     83 (20.9\%) \\
  6681 &   M 70 &    344 &      327 &     71 (21.7\%) \\
  6752 &        &    578 &      539 &     82 (15.2\%) \\
  7078 &   M 15 &   1685 &     1318 &    337 (25.6\%) \\
  7089 &    M 2 &   1908 &     1727 &    377 (21.8\%) \\
  7099 &   M 30 &    341 &      290 &     71 (24.5\%) \\
\hline \textbf{Total} & &  33951 &  30572 &   5382 (17.6\%) \\ \hline
 \end{tabular}
 \end{center} 
 \label{table:numstars}
 Notes. (1) NGC number. (2) Alternative identifier (if any). (3) Total number of observed RGB stars, (4) of which have valid EW measurements, (5) of which are brighter than the HB (percentage relative to column 4).
\end{table}
Within the guaranteed time observations for MUSE, we are currently carrying out a massive spectroscopic survey (PI: S.\ Dreizler, S.\ Kamann) of 29 GCs in the Milky Way and beyond. The survey itself, the obtained data, and the following data reduction are discussed in detail in \citet{2018MNRAS.473.5591K}. However, more observations have been carried out since that publication, so the current study includes all data gathered until September 2018.

The data analysis was performed using a procedure similar to the one described in \citet{2016A&A...588A.148H}. After a basic reduction using the standard MUSE pipeline \citep{2012SPIE.8451E..0BW,2014ASPC..485..451W} we extracted the spectra from the MUSE data cubes using \textsc{PampelMuse}\footnote{\url{https://gitlab.gwdg.de/skamann/pampelmuse}} \citep{2013A&A...549A..71K}. For this step, we need catalogs of high-resolution photometry for the positions and magnitudes of the stars and, where possible, we used data from the ACS survey of Galactic globular clusters \citep{2007AJ....133.1658S,2008AJ....135.2055A}. For some of our clusters, these were not available; the list of additional photometry that we used is listed in \citet{2018MNRAS.473.5591K}. The extraction yields spectra with the wavelength ranging from 4750 to 9350\,\AA, a spectral sampling of 1.25\,\AA, and a resolution of 2.5\,\AA, which is equivalent to $R \approx 1900 \mbox{--} 3700$. 

For each cluster in our sample we found an isochrone from \citet{2017ApJ...835...77M} matching the HST photometry by \citet{2007AJ....133.1658S}. This photometry has already been used before for the extraction process, so it is readily available. Values for effective temperatures ($\teff$) and surface gravities ($\logg$) were obtained by finding the nearest neighbour for each star on the isochrone in the CMD. Using these values, template spectra were taken from the \textsc{G\"ottingen Spectral Library}\footnote{\url{http://phoenix.astro.physik.uni-goettingen.de/}} of PHOENIX spectra and then used for performing a cross-correlation on each spectrum, yielding a radial velocity ($\vrad$). These results were used as initial guesses for a full-spectrum fit with \textsc{spexxy}\footnote{\url{https://github.com/thusser/spexxy}} using the grid of PHOENIX spectra, yielding final values for $\teff$, $\feh$, and $\vrad$. The surface gravity was taken from the comparison with the isochrone due to problems with fitting $\logg$ from low-resolution spectra.

Two of our clusters, namely \object{NGC\,6388} and \object{NGC\,6441}, are almost twins in many regards \citep[see, e.g.,][]{2013ApJ...765...32B,2017MNRAS.465.1046T}: they are both old, massive, and metal-rich bulge clusters. \citet{2008AJ....135.2055A} comment on difficulties when creating the catalogs due to blending in the crowded centers, especially at absolute magnitudes about $-12.5\,\mathrm{mag}$ in F606W and F814W. Probably as a result, we see extremely broadened main sequences and giant branches in the CMDs of both clusters, which makes further analyses challenging.

In order to get high signal-to-noise spectra for each star, we combined all the spectra that we obtained for a single star. During the full-spectrum fit, \textsc{spexxy} also fits a model for the telluric absorption lines and a polynomial that, multiplied with the model spectrum, best reproduce the observed spectrum. This polynomial eliminates the effects of reddening and ensures that we fit only spectral lines and not the continuum, which is therefore completely ignored during the fit. Before combining the individual spectra, we first removed the tellurics and divided the spectra by the polynomial in order to get rid of a wavy structure that we sometimes observe in MUSE spectra. Finally, we co-add the individual raw spectra with their respective signal-to-noise ratios (S/N) as weights.

\begin{figure}
 \includegraphics{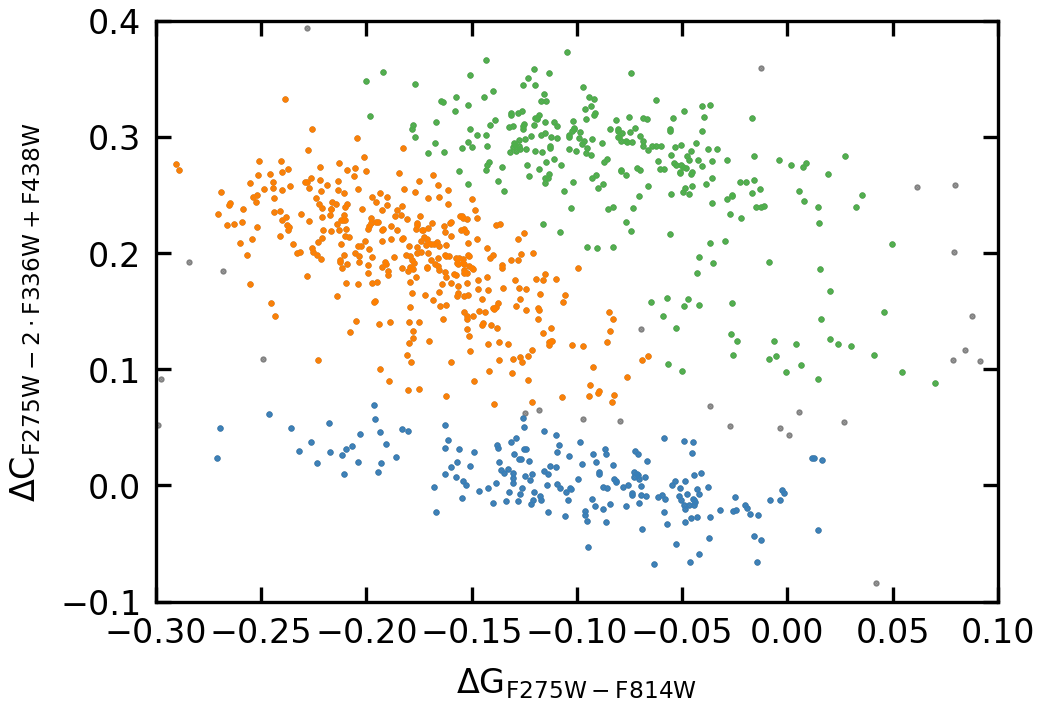} \\
 \includegraphics{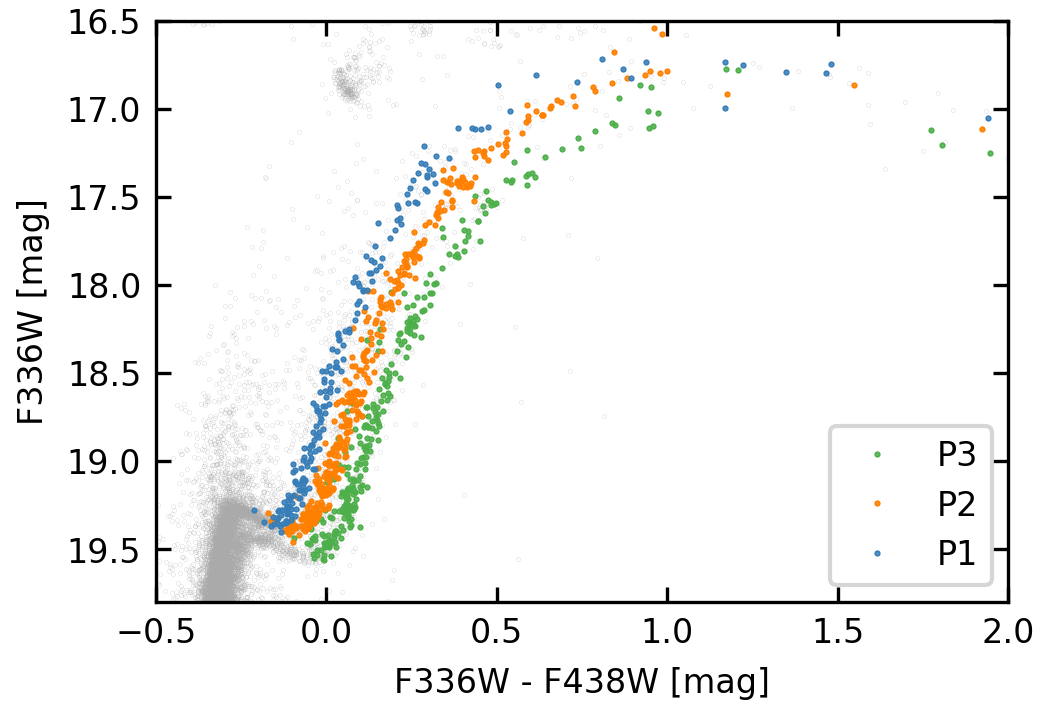}
 \caption{In the upper panel the chromosome map of \object{NGC\,1851} is shown with the three identified populations marked in different colors (see text in Sect.~\ref{sec:cmaps} for explanation). The same color-coding is used for the color-magnitude diagram in the lower panel, where the populations can also easily be distinguished.}
 \label{fig:ngc1851_cmap}
\end{figure}

We selected the RGB stars for this study by manually drawing a corresponding region into the CMD of each cluster. Furthermore, we determined membership of the stars to the cluster using $\feh$ and $\vrad$ from the full-spectrum fits \citep[see][]{2018MNRAS.473.5591K}, and removed non-members from the sample. The total number of observed RGB stars per cluster is given in column (3) of Table~\ref{table:numstars}, adding up to a full sample of almost 34,000 stars.

\citet{2001MNRAS.326..959C} suggested that the best targets for a CaT analysis are stars with spectral types between F5 ($\teff \approx 6500$K) and M2 ($\teff \approx 3700$K). We used only stars within this given temperature range, according to the effective temperatures derived from our full-spectrum fits as described in \citet{2016A&A...588A.148H}. We also excluded the very brightest stars with $V-\vhb<-3$ or $\log L/L_\odot>\sim 3$ (depending on the cluster). Column (4) of Table~\ref{table:numstars} gives the numbers of remaining stars, for which we obtained a valid EW measurement (see Sect.~\ref{sec:ews}).

\section{Chromosome maps}
\label{sec:cmaps}
The pseudo-two-color diagrams introduced by \citet{2015MNRAS.447..927M,2015ApJ...808...51M} and then termed as chromosome map \citep{2016IAUS..317..170M} proved to be an optimal way to distinguish the various populations of a given RGB of a GC. These maps are built using a combination of HST filters ($F275W$, $F336W$, $F438W$, and $F814W$) that are sensitive to spectral features affected by the chemical variations that characterize the different populations (see, e.g., \citealt{2018MNRAS.481.5098M}). 

The details for creating the chromosome maps used in this study are described in \citet{2019A&A...631A..14L}, but we will summarize them here. We used the astrophotometric catalogs presented by \citet{2018MNRAS.tmp.2405N} that are part of the HST UV Globular cluster Survey \citep[HUGS, see][]{2015AJ....149...91P}. First, we cleaned the data, then we constructed the chromosome maps following the approach described in \citet{2017MNRAS.464.3636M}. We created two CMDs using the 
F814W magnitude and the two pseudo-colors $\Delta \mathrm{G}_\mathrm{F275W-F814W}$ and $\Delta$C$_\mathrm{F275W - 2 \cdot F336W + F438W}$. Then both CMDs are verticalized using red and blue fiducial lines (i.e., they are stretched and shifted so that these fiducial lines become straight vertical lines) and the results are combined to become the \cmap.
Figure~\ref{fig:ngc1851_cmap} shows the \cmap\ and the corresponding CMD for the Type~II cluster \object{NGC\,1851}, using the same colors for the three populations in both panels. The populations are well defined in the \cmap\ and also separate nicely in the CMD.

In order to use the chromosome maps with our data, we had to match the ACS catalog \citep{2007AJ....133.1658S} used for identifying our stars, with the HUGS catalog. Some stars could not be unambiguously identified in both catalogs and were not used for the multiple populations study.

\section{Calibrating the CaT-metallicity relation}
\label{sec:calib}

\subsection{Measuring equivalent widths}
\label{sec:ews}
\begin{figure}
 \includegraphics{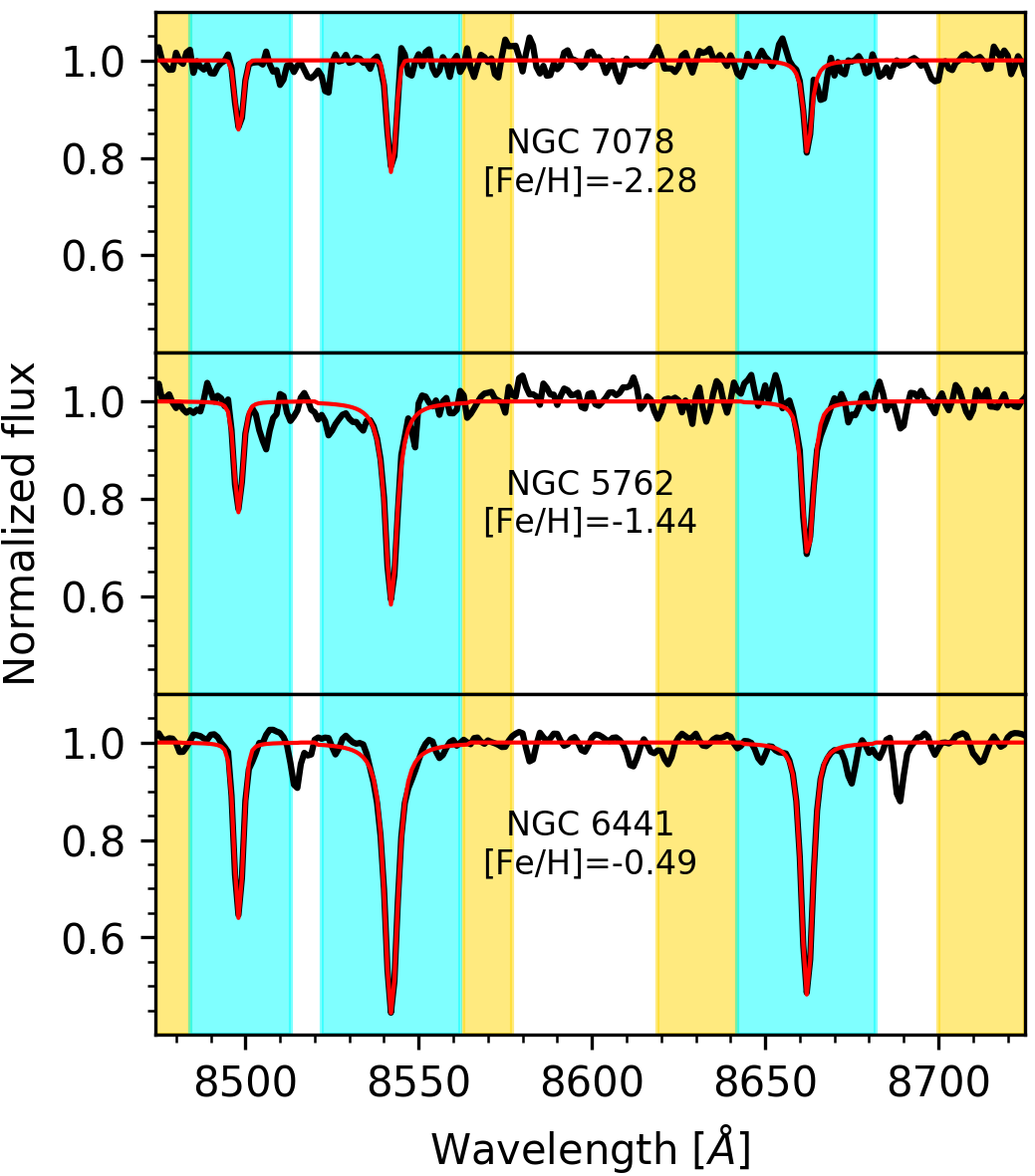}
 \caption{Three example spectra with S/N$\approx$50 from different clusters covering the whole range of metallicities in our sample. The observed spectra are shown in black, overplotted with the best fitting Voigt profiles. The areas marked in yellow were used for the continuum correction, while those in blue define the line bandpasses that were used for fitting the Voigt profiles and calculating the equivalent widths. The given metallicities are mean cluster metallicities from \protect\citet{2016A&A...590A...9D}.
 }
 \label{fig:example-fits}
\end{figure}
\begin{table}
 \centering
 \caption{Line and continuum bandpasses from \protect\citet{2007AJ....134.1298C}.}
 \begin{tabular}{cc}
  \hline \hline
  Line bandpasses & Continuum bandpasses \\ 
  \hline
  8484--8513 \AA & 8474--8484 \AA\\
  8522--8562 \AA & 8563--8577 \AA\\
  8642--8682 \AA & 8619--8642 \AA\\
                 & 8700--8725 \AA\\
                 & 8776--8792 \AA\\ \hline
 \end{tabular}
 \label{table:bandpasses}
\end{table}

In the past, different functions have been used for fitting the Ca lines. While, for instance, \citet{1991AJ....101.1329A} used Gaussians, \citet{2004MNRAS.347..367C} found that for high metallicities these deviate strongly from the real line shapes due to strong damping in the wings. As an alternative, they suggest to use the sum of a Gaussian and a Lorentzian, which was adopted by many later studies \citep[see, e.g.,][]{2007AJ....134.1298C,2009A&A...500..735G}. \citet{2012A&A...540A..27S} distinguished between low and high metallicity clusters and respectively fitted Gaussians and Gaussians+Lorentzians. We found a problem with this approach for spectra with relatively low S/N in which case a broad Lorentzian often just fitted the noise. \citet{1997PASP..109..883R} and others used a Moffat function. We decided to adopt the method from \citet{2016MNRAS.460.1846Y} and used Voigt profiles, representing the convolution of the thermal and pressure broadening.

In order to fit profiles to the lines, we needed to define the bandpasses for both the lines and the pseudo-continuum, which were used for normalizing the spectra. \citet{2007AJ....134.1298C} compared the bandpasses given by \citet{1988AJ.....96...92A}, \citet{1997PASP..109..883R}, and \citet{2001MNRAS.326..959C}. Following their argument that only the line bandpasses of \citet{2001MNRAS.326..959C} cover the wings of the lines completely, we adopted those for our analysis (see Table~\ref{table:bandpasses}).

For determining the equivalent widths of the three Ca lines, we first fit a low-degree polynomial to the continuum bandpasses to remove the continuum. Then we fit a Voigt profile to each line individually within its given bandpass using a Levenberg-Marquardt optimisation. This is done with the VoigtModel profile within LMFIT \citep{newville_2014_11813}. The integration of the fitted Voigt profiles (also in the given bandpasses) yields the equivalent widths of the lines.

In Fig.~\ref{fig:example-fits}, some spectra from three globular clusters covering the whole range of metallicities in our sample are shown, together with the bandpasses for continuum and lines, and the best-fitting Voigt profiles.

\begin{figure}
 \includegraphics{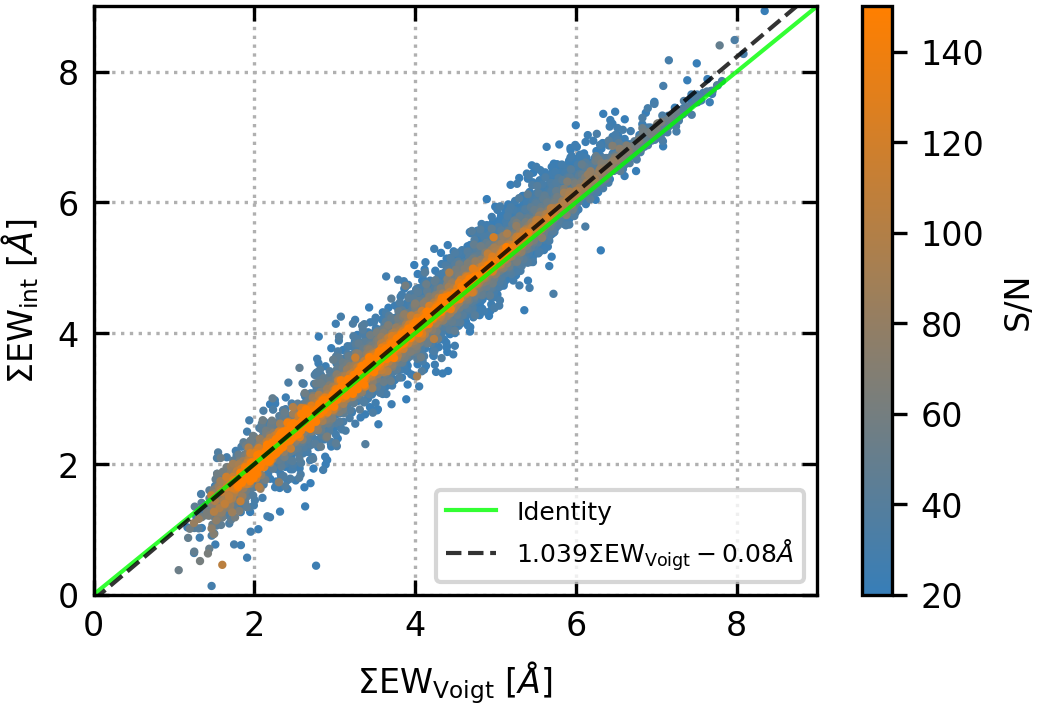}
 \caption{Comparison between equivalent widths derived from simply integrating the lines in the given bandpasses ($\sumew_\mathrm{int}$) and from fitting Voigt profiles ($\sumew_\mathrm{Voigt}$) for all spectra with S/N>20 and the S/N as color-coding. Note that we clipped the color range to a maximum value of 150, although we reach S/N of up to 400 for single spectra. The dashed black line provides a linear fit to the data, while the green one indicates the identity.}
 \label{fig:integr_vs_model}
\end{figure}
There has been some discussion in the literature on whether to use the (weighted) sum of the equivalent widths of all three Ca lines, or just the sum of the two strongest ones. Since the weakest line at 8498\AA\ is significantly fainter than the other two, and therefore more difficult to fit in low S/N spectra, we chose to use the sum of the two broader lines at 8542 and 8662\AA, hereafter called $\sumew$.

We verified the quality of the equivalent widths derived from Voigt profile fits by comparing them to the results of a simple numerical integration of the lines within their respective bandpasses. A comparison between both is shown in Fig.~\ref{fig:integr_vs_model} for the sum $\sumew$ of the two strongest lines for all spectra. Apparently, the equivalent widths from the numerical integration are slightly but systematically higher than those from Voigt profile fits, especially at larger widths. Presumably at higher metallicities not only the Ca lines broaden, but also fainter metal lines get stronger, which affects the numerical integration more than the Voigt profiles. However, the correlation is linear as expected.

\begin{figure}
 \includegraphics{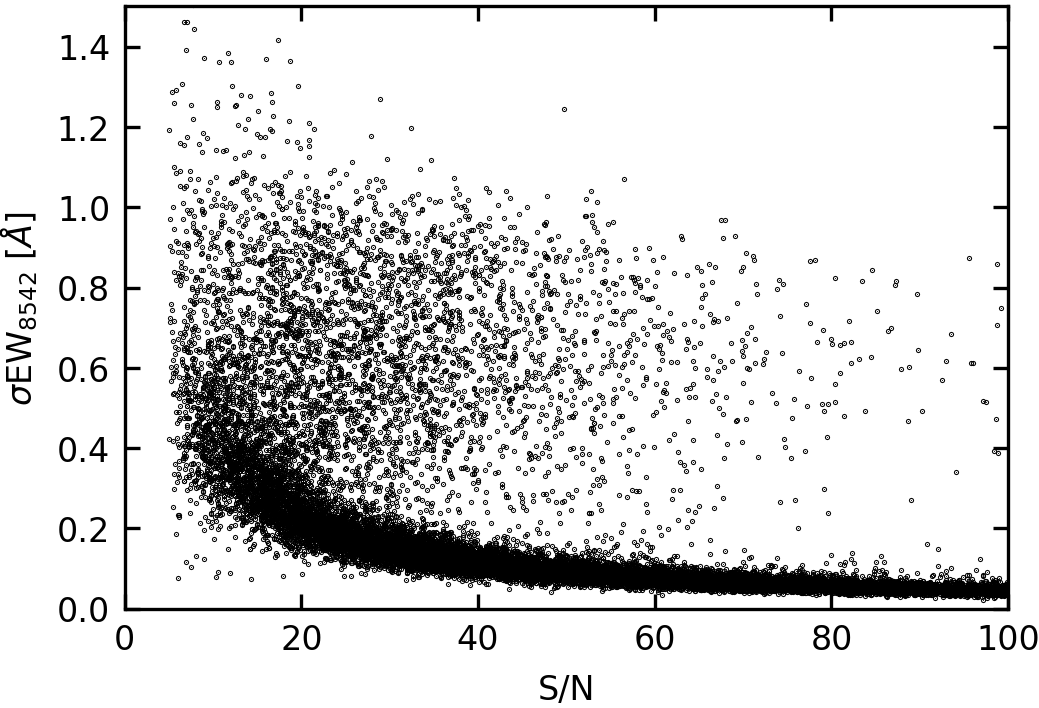}
 \caption{The uncertainties for the equivalent width of the $\mathrm{Ca}_{8542}$ line as function of S/N.}
 \label{fig:snr_vs_uncertainties}
\end{figure}

In order to obtain uncertainties for our equivalent widths, we took the full covariance matrix from the Voigt profile fits and used it to draw 10,000 combinations of parameters for each fitted line. We evaluated and integrated the Voigt profiles as before and use the standard deviation of all results as the uncertainty for the EW of the single line. Figure~\ref{fig:snr_vs_uncertainties} shows those uncertainties for the $\mathrm{Ca}_{8542}$ line as a function of S/N. Unfortunately, we cannot use the raw spectra for calibrating the uncertainties \citep[as in, e.g.,][]{2008MNRAS.383..183B} because we have significant variations in S/N between all spectra for a single star.

\begin{figure}
 \includegraphics{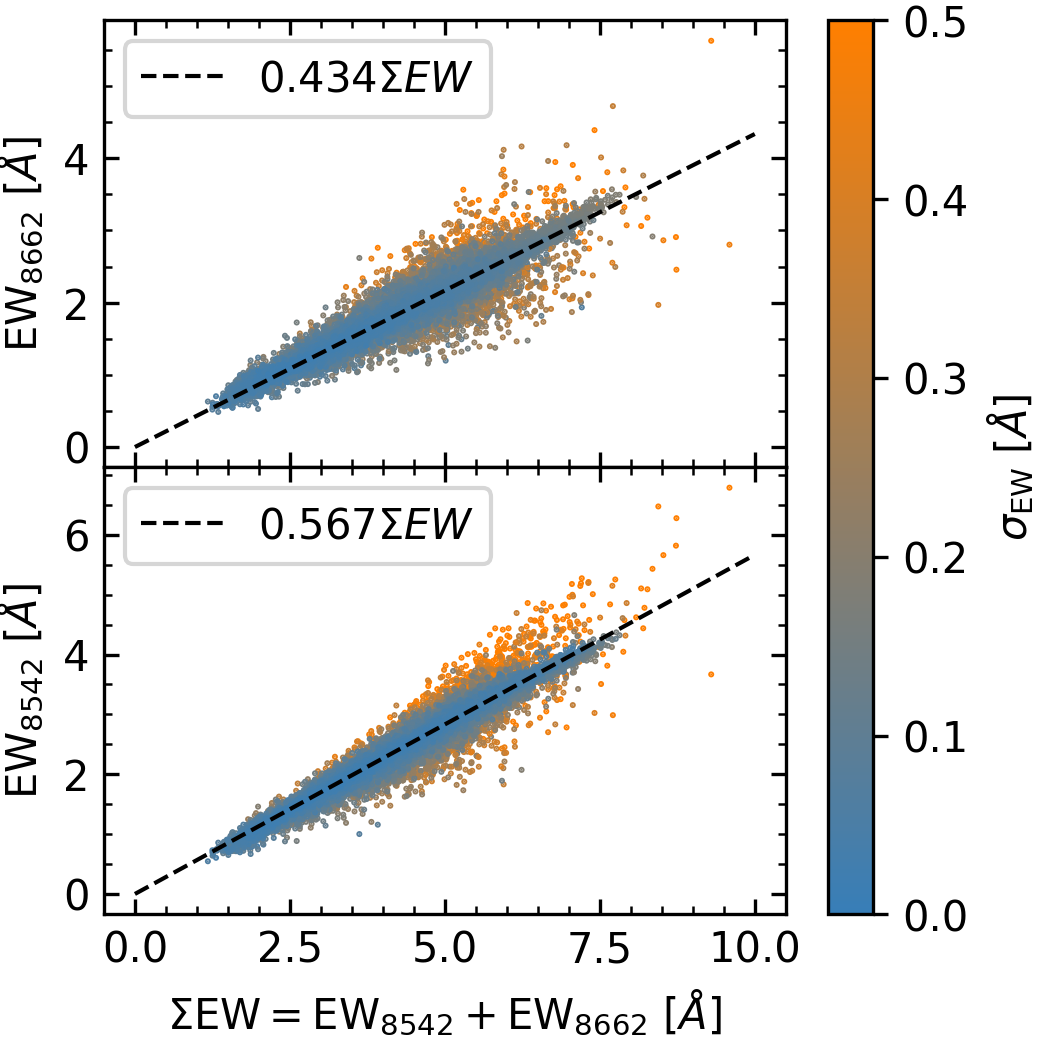}
 \caption{Measured equivalent widths of the two strongest Ca lines plotted as a function of their sum. The uncertainties of the EW measurements on the single line are color-coded.}
 \label{fig:single_vs_sum}
\end{figure}
The quality of the fit on a single spectrum can also be derived from the ratio of the equivalent widths of the two strongest lines, which should be constant. In Fig.~\ref{fig:single_vs_sum} we show the equivalent widths of those two lines as a function of their sum. Fitting a line to the data using the inverse square of the uncertainties as weights on both axes yields a negligible error for the slope. We found $EW_{8542}=0.567 \Sigma EW$ and $EW_{8662}=0.434 \Sigma EW$, which is in perfect agreement with \citet{2015A&A...580A.121V}, who determined the slopes to be $0.57$ and $0.43$, respectively. Written as a ratio of line strengths, we find $W_{8542}/W_{8662}=1.31\pm0.20$, which, again, agrees with the value of $1.32\pm0.09$ derived by \citet{2013MNRAS.434.1681C}. 

\begin{figure}
 \includegraphics{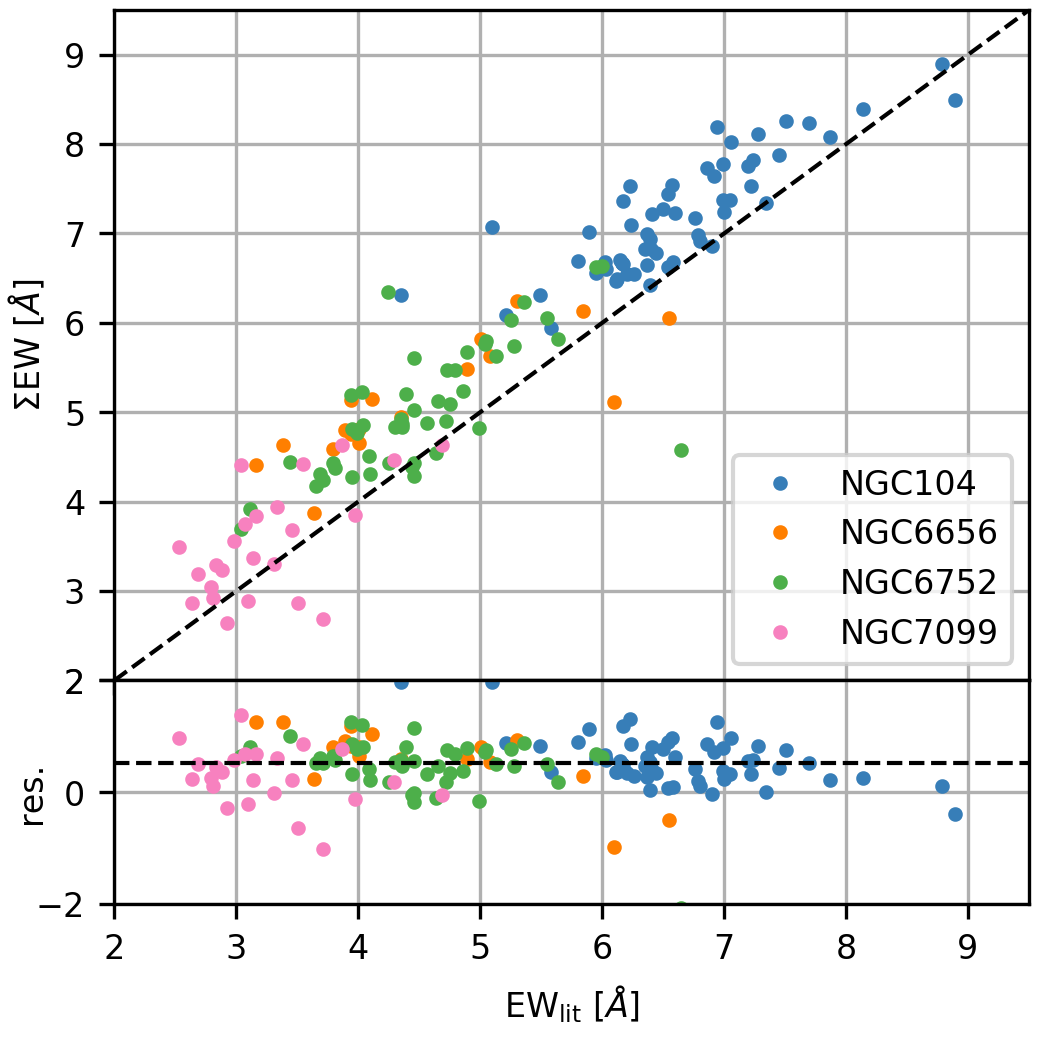}
 \caption{Comparison of our derived equivalent widths with those of \citet{2011A&A...530A..31L}. For the purpose of this plot, $\Sigma EW$ denotes the sum of all three lines in the Ca triplet. In the upper plot the dashed black line shows the identity, while in the lower plot it indicates the mean offset of $\sim 0.52\,\AA$.}
 \label{fig:compare_ew}
\end{figure}
A comparison of single results with literature is a bit more difficult, since the very dense centers of the clusters that we observed are usually too crowded for other observation techniques. One exception that we found are the AAOmega observations performed by \citet{2011A&A...530A..31L}, with which we have 155 stars in four different GCs in common, covering almost our full range of metallicities from $-2.28$ (\object{NGC\,7099}) to $-0.69\,\dex$ (\object{NGC\,104}). They published the sum of the EWs for all three CaT lines, so we did the same and compared the results in Fig.~\ref{fig:compare_ew} after removing three outliers that have unusually high EW measurements in the literature. Our equivalent widths show a constant offset of $0.52\,\AA$, which is most probably due to a different method used for determining the widths (e.g., different integration intervals), with a scatter of $0.51\,\AA$.

\begin{table}
 \caption{Derived mean-reduced equivalent widths from different methods and cluster metallicities.} 
 \begin{center}
 \begin{tabular}{rcccccc}
  \hline \hline
  NGC & $\left< W'_\mathrm{HB} \right>$  & $\left< W'_\mathrm{all} \right>$ & $\left< W'_\mathrm{M} \right>$ & $\left< W'_\mathrm{lum} \right>$ & $\feh$ \\ \hline
       104 & $ 5.59$ & $ 5.66$ & $ 5.60$ & $ 5.63$ & $-0.69$ \\
     362 & $ 4.84$ & $ 4.82$ & $ 4.76$ & $ 4.78$ & $-1.05$ \\     
    1851 & $ 4.99$ & $ 4.97$ & $ 4.91$ & $ 4.94$ & $-1.19$ \\
    1904 & -- & -- & -- & $ 3.80$ & $-1.61$ \\
    2808 & $ 5.05$ & $ 5.04$ & $ 4.99$ & $ 5.02$ & $-1.13$ \\
    3201 & $ 4.39$ & $ 4.29$ & $ 4.22$ & $ 4.24$ & $-1.46$ \\
    5139 & -- & -- & $ 3.74$ & $ 3.77$ & $-1.56$ \\
    5286 & $ 3.57$ & $ 3.64$ & $ 3.52$ & $ 3.55$ & $-1.63$ \\
    5904 & $ 4.79$ & $ 4.76$ & $ 4.67$ & $ 4.69$ & $-1.15$ \\
    6093 & $ 3.46$ & $ 3.51$ & $ 3.53$ & $ 3.55$ & $-1.73$ \\
    6218 & -- & -- & $ 4.58$ & $ 4.60$ & $-1.35$ \\
    6254 & $ 4.17$ & $ 4.06$ & $ 4.09$ & $ 4.11$ & $-1.65$ \\
    6266 & -- & -- & -- & $ 5.07$ & $-1.11$ \\
    6293 & -- & -- & -- & $ 2.37$ & $-1.86$ \\
    6388 & $ 5.67$ & $ 5.70$ & $ 5.81$ & $ 5.82$ & $-0.57$ \\
    6441 & $ 5.64$ & $ 5.66$ & $ 5.79$ & $ 5.78$ & $-0.49$ \\
    6522 & -- & -- & -- & $ 4.83$ & $-1.35$ \\
    6541 & $ 3.48$ & $ 3.52$ & $ 3.41$ & $ 3.44$ & $-1.80$ \\
    6624 & $ 5.71$ & $ 5.72$ & $ 5.68$ & $ 5.70$ & $-0.36$ \\
    6656 & $ 3.45$ & $ 3.51$ & $ 3.50$ & $ 3.52$ & $-1.91$ \\
    6681 & $ 4.18$ & $ 4.21$ & $ 4.20$ & $ 4.23$ & $-1.54$ \\
    6752 & $ 4.09$ & $ 4.10$ & $ 4.06$ & $ 4.08$ & $-1.44$ \\
    7078 & $ 1.99$ & $ 2.15$ & $ 2.02$ & $ 2.05$ & $-2.28$ \\
    7089 & $ 3.95$ & $ 3.98$ & $ 3.90$ & $ 3.93$ & $-1.58$ \\
    7099 & $ 2.32$ & $ 2.46$ & $ 2.38$ & $ 2.40$ & $-2.28$ \\

  \hline
 \end{tabular}
 \end{center}
 Notes. The index ``HB'' denotes results from the linear relation using stars with $V-\vhb<+0.2$ as discussed in Sect.~\ref{sec:rew}, while ``all'' uses the results for all RGB stars using a quadratic relation from Sect.~\ref{sec:extending_below}. Finally, ``M'' uses the absolute magnitude in F606W instead of $V-\vhb$ and ``lum'' the luminosity, both as presented in Sect.~\ref{sec:calib_lum}. The uncertainties for the reduced EWs are usually of the order $0.1\mbox{--}0.4\dex$, the metallicities are taken from \citet{2016A&A...590A...9D}.
 \label{table:reduced_rw}
\end{table}

\subsection{Reduced equivalent widths}
\label{sec:rew}
In general, the strength of an atomic absorption line is driven by the effective temperature of the star, its surface gravity, and the abundance of the element in question. \citet{2004AJ....127..840P} showed theoretically that, for a fixed metallicity, the strength of the CaT lines for stars on the RGB increases with increasing $\teff$ and decreases for increasing $\logg$, and that both effects roughly cancel each other. Therefore, the strength of the lines is a function of luminosity alone.
Because luminosity is a parameter that is not easily obtained, other indicators can replace it such as the (extinction corrected) absolute magnitudes in $V$ or $I$, or the brightness difference to the horizontal branch $V-\vhb$. These indicators are all independent of reddening, distance, and photometric zero-point.

As discussed above, we use the HST photometry to extract spectra from the observed MUSE data\-cubes. The available F606W photometry in these catalogs can also be used to calculate the brightness difference to the HB ($F606W-F606W_\mathrm{HB}$) using the HB brightnesses in this filter available from \citet{2010ApJ...708..698D}, with which we have 19 clusters in common. However, we continue calling it $V-\vhb$.

\begin{figure*}
 \includegraphics{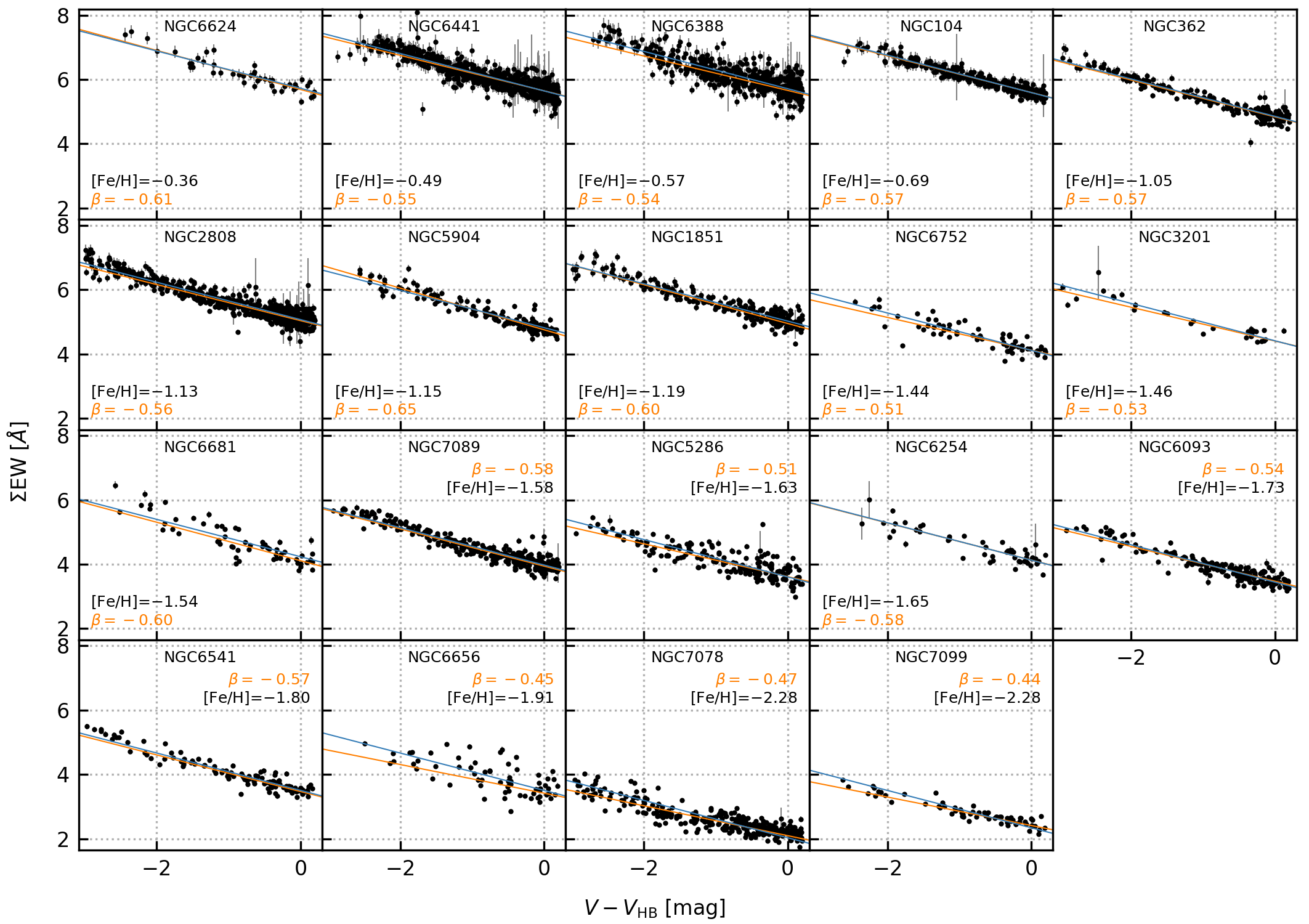}
 \caption{Sum of the equivalent widths $\Sigma \mathrm{EW}$ of the two strongest Ca lines plotted over $V-\vhb$ for all 20 clusters in the sample, sorted by mean metallicity. Only stars are included with $V-\vhb<+0.2$ as described in the text. A linear fit to the data for each individual cluster is plotted in orange, for which the slope $\beta$ is given in the title of each panel. The blue lines show the result of a global fit, where the same slope has been used for all clusters, yielding $\beta=-0.581 \pm 0.004$. In some cases, the individual and global fits are indistinguishable.}
 \label{fig:calib_rew_above_hb}
\end{figure*}
\begin{figure}
 \includegraphics{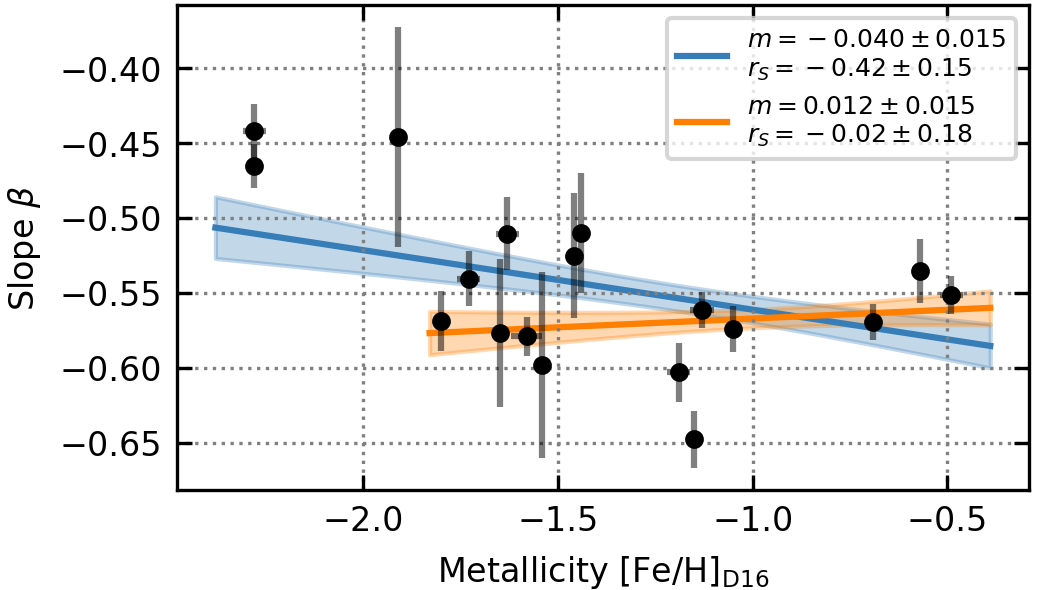}
 \caption{Slopes $\beta$ from individual fits for each cluster are shown as a function of metallicity. The blue line indicates a linear fit to the data with the shaded area representing its $1 \sigma$ uncertainty band, while the orange line is the equivalent when ignoring the three most metal-poor clusters. As discussed in the text, there might be a real trend, but for the further analysis we assume $\beta$ to be constant.}
 \label{fig:beta_over_feh}
\end{figure}

\citet{2009ApJ...705.1481D} observed that the relation between $V-\vhb$ and $\sumew$ flattens for $V-\vhb>+0.2\,\mathrm{mag}$. This change of slope was confirmed theoretically by \citet{2010A&A...513A..34S}. Consequently, for the rest of this subsection we proceed only with stars having $V<\vhb+0.2\,\mathrm{mag}$ (assuming that $V$ and $F606W$ magnitudes -- or at least the brightness differences to the HB in these bands -- are similar enough). The numbers and percentages (relative to all RGB stars) of stars fulfilling this criterion are listed in column (5) of Table~\ref{table:numstars}. The brightnesses of the horizontal branches $\vhb$ are taken from \citet{2010ApJ...708..698D} -- if none are given, that column is marked with a dash.

Using the assumption of the strengths of the Ca lines being a function of $V-\vhb$ alone, we define the reduced equivalent $W'$ as
\begin{equation}
 \sumew = \beta \cdot (V - \vhb) + W'.
 \label{eq:rew_linear}
\end{equation}

We performed a linear fit for $\sumew$ as a function of $V-\vhb$ for each cluster, yielding a slope $b$ and a reduced EW $W'$.. In addition, a global function was fitted to all the data, deriving individual $W'$ for each cluster, but using the same slope $\beta$ for all of them. The data itself and the results for both approaches (blue and orange lines) are shown in Fig.~\ref{fig:calib_rew_above_hb}.

The global fit with all clusters yielded a slope of $\beta=-0.581  \pm 0.004$. The slopes from the individual fits are given in brackets in each panel and are usually similar to the global slope. In the literature, we find values of $-0.55$ \citep{2018A&A...619A..13V}, $-0.627$ \citep{2012A&A...540A..27S}, with both using V magnitudes, and $-0.74\pm0.01$ and $-0.60\pm0.01$ \citep{2007AJ....134.1298C} when using V and I magnitudes, respectively.

\begin{figure*}
 \includegraphics{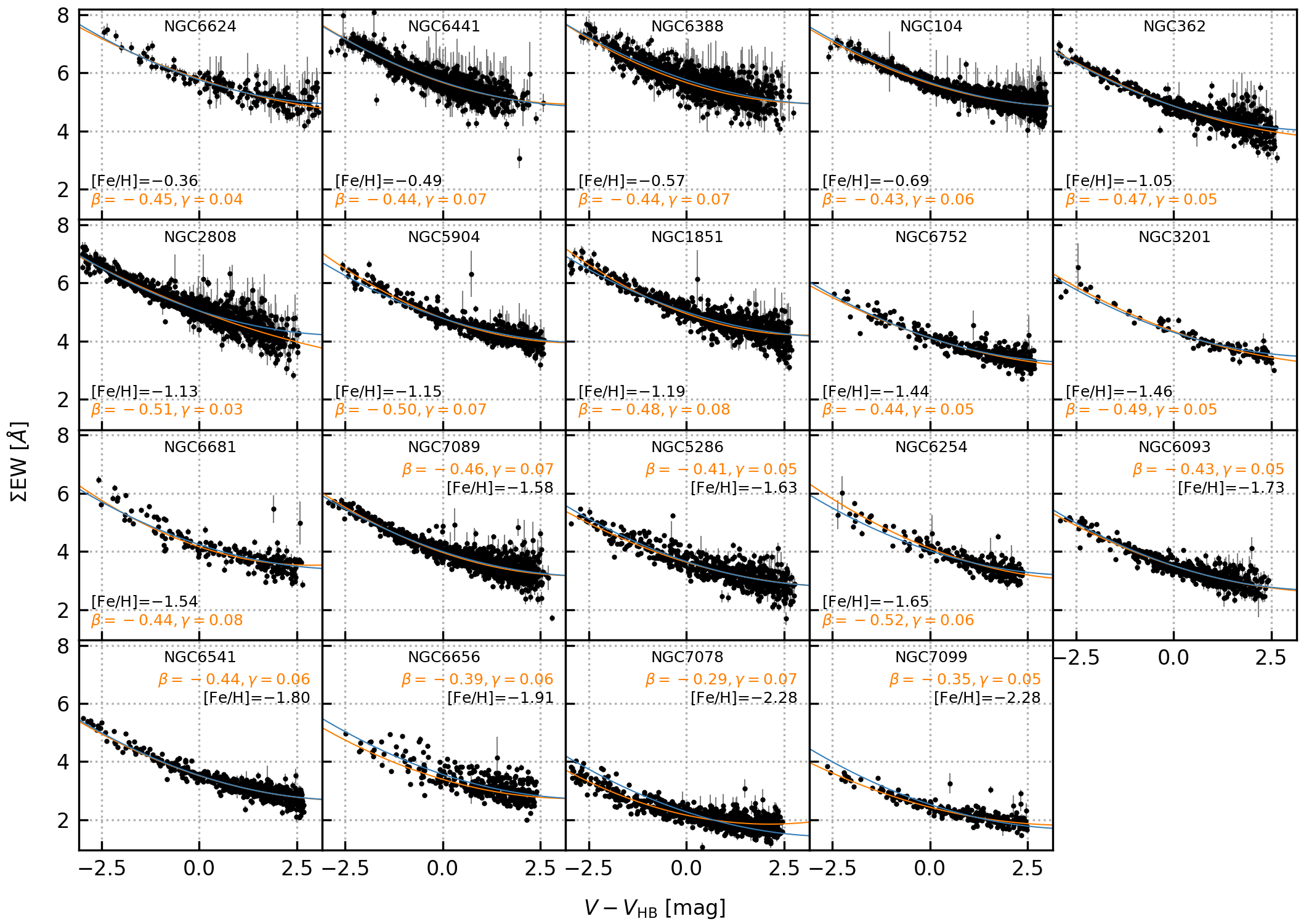}
 \caption{Similarly to Fig.~\ref{fig:calib_rew_above_hb}, the sum of the equivalent widths $\Sigma \mathrm{EW}$ of the two strongest Ca lines is plotted over $V - \vhb$ for all RGB stars. Quadratic fits to each individual cluster are shown in orange, while a global fit, where the same values for $\beta$ and $\gamma$ are used (giving $\beta=-0.442 \pm 0.002$ and $\gamma=0.058 \pm 0.001$), is plotted for each cluster in blue.}
 \label{fig:calib_rew_all}
\end{figure*}

The intercept of the fitted lines correspond to the reduced equivalent width $W'$ for each cluster. From these fits, the uncertainties are unrealistically small due to the large number of points. Therefore, we derived $W'$ for each spectrum using the global slope $\beta$ and calculated the mean reduced equivalent widths $\left<W'\right>$ for all clusters \citep[see][]{2014A&A...563A..76M}, which are given in Table~\ref{table:reduced_rw} with the index ``HB.'' This approach also yields uncertainties for the reduced equivalent widths (i.e.,\ the standard deviation of all results per cluster) and has been used by \citet{2014A&A...563A..76M}.
As a result, we get the following calibration when using a linear relation on all RGB stars brighter than the HB:
\begin{equation}
 W' = \sumew + 0.581 \cdot (V - \vhb).
\end{equation}
With the negligible statistical uncertainty for the slope $\beta$, the error on the reduced equivalent width $W'$ is just equal to the error on the sum of equivalent widths, i.e.\ $\sigma_{W'} = \sigma_{\sumew}$.

Using theoretical models from \citet{1992A&A...254..258J}, a prediction was made by \citet{2004AJ....127..840P} that there should be an increasing slope $\beta$ with increasing metallicity. In Fig.~\ref{fig:beta_over_feh} we show the slopes from the individual fits for each cluster as a function of metallicity. Two lines have been fitted to the data, one to all the clusters (orange) and one without our three most metal-poor ones (blue), both using the uncertainties as weights. The slopes $m$ and Spearman correlation coefficients $r_S$ are given in the legend. While with all data there might be some trend, it completely disappears when ignoring the three metal-poor clusters. Therefore, we chose to ignore any trend and use the same slope, $\beta=-0.581$ from the global fit, for all clusters.

\citet{2007AJ....134.1298C} reported seeing a trend of slope with metallicity, but within the uncertainties, while \citet{2013MNRAS.434.1681C} detected a significant trend and suggest to add more terms to the final CaT-metallicity (see also Sect.~\ref{sec:calib_lum}) relation as described by \citet{2010A&A...513A..34S}.

\subsection{Extending the calibration below the HB}
\label{sec:extending_below}

\citet{2007AJ....134.1298C} state that from theoretical predictions there is no reason why the relation between $V-\vhb$ and $\sumew$ must be linear, so they suggested adding a quadratic term: 
\begin{equation}
 \sumew = W' + \beta (V - \vhb) + \gamma (V - \vhb)^2.
 \label{eq:rew_quadratic}
\end{equation}
This approach removes the necessity for using only stars brighter than the HB, and thus, following Table~\ref{table:numstars}, we actually increase our sample size by a factor of five. The results of the fits with the quadratic equation are shown in Fig.~\ref{fig:calib_rew_all}. As in the previous subsection, we performed a fit for each individual cluster as well as a global fit, for which we forced the same values for $\beta$ and $\gamma$ for all clusters. As before, the two results only differ significantly for those clusters with very few RGB stars. The global fit yields values of $\beta=-0.442 \pm 0.002$ and $\gamma=0.058 \pm 0.001$.
This yields the final calibration for the reduced equivalent width using a quadratic relation on all RGB stars, even extending below the HB:
\begin{equation}
 W' = \sumew + 0.442 (V - \vhb) - 0.058 (V - \vhb)^2. 
 \label{eq:calib_rew_all}
\end{equation}
As we did for the linear relation, we calculated the reduced equivalent widths for all stars using this equation and derived a mean width for each cluster. The results are listed in Table~\ref{table:reduced_rw} with the index ``all.''

\begin{figure}
 \includegraphics{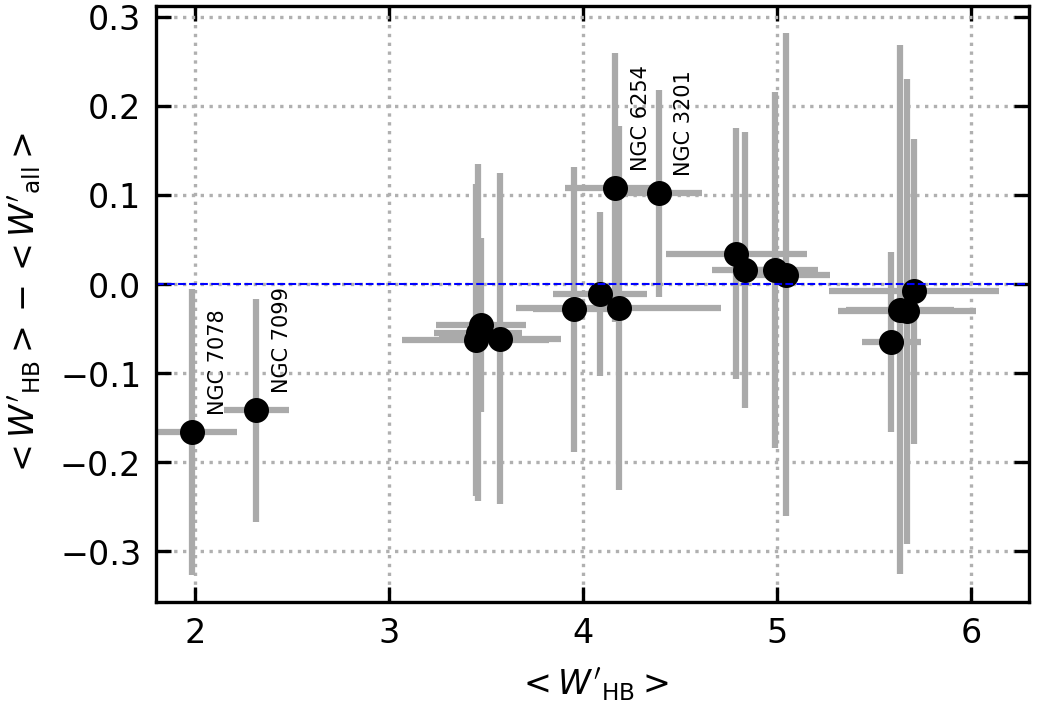}
 \caption{Comparison of average reduced equivalent widths $\left< W' \right>$ obtained from a linear fit to stars with $V-\vhb<+0.2$ (``HB'') and from a quadratic fit to the full sample (``all''). The error bars on the y axis are those of $\left< W'_\mathrm{all} \right>$.}
 \label{fig:calib_offsets}
\end{figure}
\begin{figure}
 \includegraphics{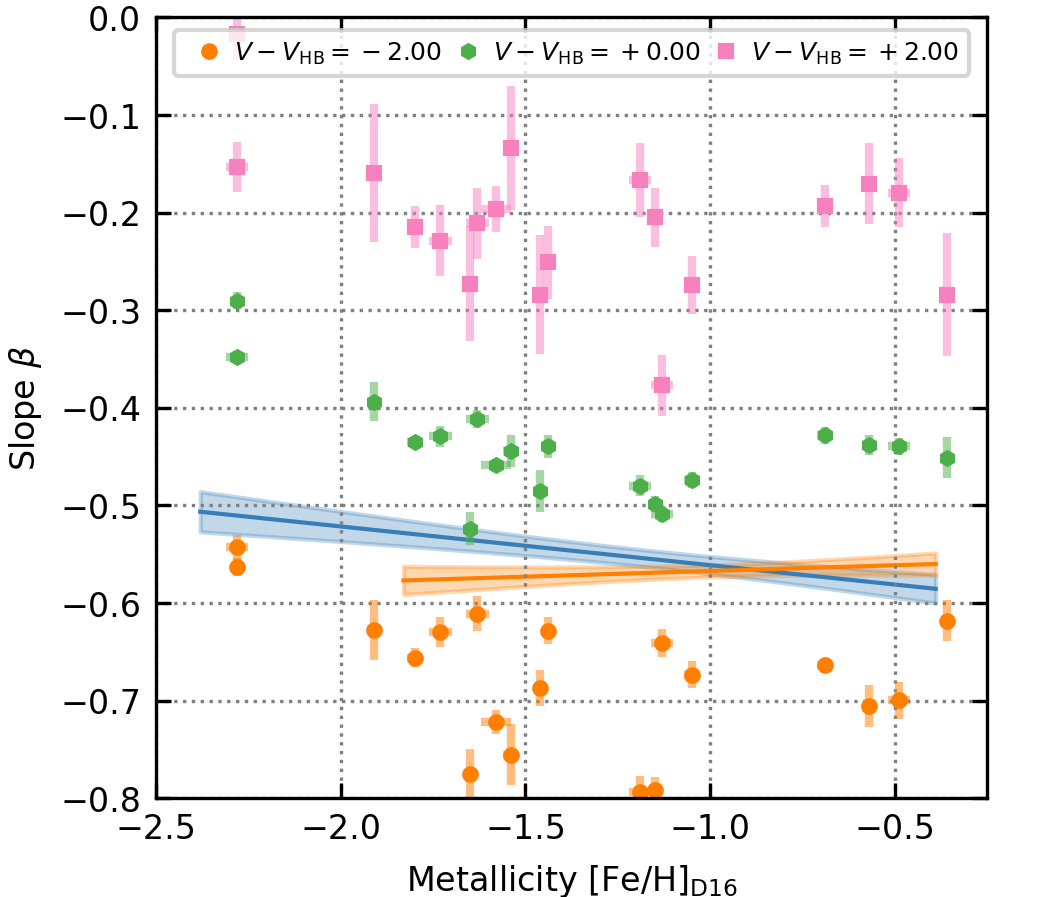}
 \caption{As in Fig.~\ref{fig:beta_over_feh}, slopes are plotted as function of metallicity, but here for the quadratic fit from Fig.~\ref{fig:calib_rew_all} at three different values for $V-\vhb$. The blue and orange lines show the fitted relations from Fig.~\ref{fig:beta_over_feh}.}
 \label{fig:calib_over_feh2}
\end{figure}
We expected this method, using a quadratic equation and including all stars on the RGB, to produce the same reduced equivalent widths as the classical approach, where a linear relation is fitted to only stars brighter than the HB. To verify that we compare the results of both methods in Fig.~\ref{fig:calib_offsets}. We find large deviations for only four clusters, of which two are the ones with the lowest metallicity in our sample, that is \object{NGC\,7078} and \object{NGC\,7099}. This might indicate problems with the calibration for very low metallicities. The other two mild outliers are \object{NGC\,3201} and \object{NGC\,6254}, which both have a relatively low number of RGB stars in our sample. Otherwise, the reduced equivalent widths derived from both methods are within the error bars.

We also repeated the analysis concerning the slope of the relation as it was done for the linear relation in Sect.~\ref{sec:rew} (see Fig.~\ref{fig:beta_over_feh}). Since the slope is not constant for a quadratic relation, we plotted it for three different values of $V-\vhb$ in Fig.~\ref{fig:calib_over_feh2}. For stars brighter than the HB (orange), the trend of this relation is similar to the linear case, indicated by the blue and orange lines and only shows some offset. For stars at the HB brightness or fainter, the blue trend line seems to fit better, meaning a larger absolute slope for larger metallicities, which agrees better with the predictions by \citet{2004AJ....127..840P}.

\subsection{Using absolute magnitude and luminosity instead of $V-\vhb$}
\label{sec:calib_lum}
The CaT metallicity relation as presented in this paper requires the brightness difference between a star and the horizontal branch. While this can be obtained easily in stellar populations like globular clusters, it is next to impossible for field stars. However, with $V-\vhb$ just being a proxy for the luminosity, we can use the luminosity directly, or -- a quantity easier to measure -- the absolute brightness in any given filter.

As before, we used HST magnitudes measured in the F606W filter, which we corrected for distance and extinction as given by \citet[][2010 edition]{1996AJ....112.1487H} to obtain absolute magnitudes $M_\mathrm{F606W}$. For deriving luminosities, we need bolometric corrections, which we calculated from our grid of PHOENIX spectra \citep{2013A&A...553A...6H}. The stars' effective temperature and surface gravity needed to apply the corrections to the data come from our analysis pipeline as described in \citet{2016A&A...588A.148H}. These two approaches open up new windows especially for investigating the CaT-metallicity in field stars, for which, in the era of Gaia \citep{2016A&A...595A...1G}, distances and spectral types are now easily available. Although we have F606W photometry available for most of our clusters, there were some exceptions for which we had to use different filters for deriving luminosities: for \object{NGC\,1904}, \object{NGC\,6266}, and \object{NGC\,6293} we use F555W, and for \object{NGC\,6522} we use F625W.

\begin{figure}
 \includegraphics{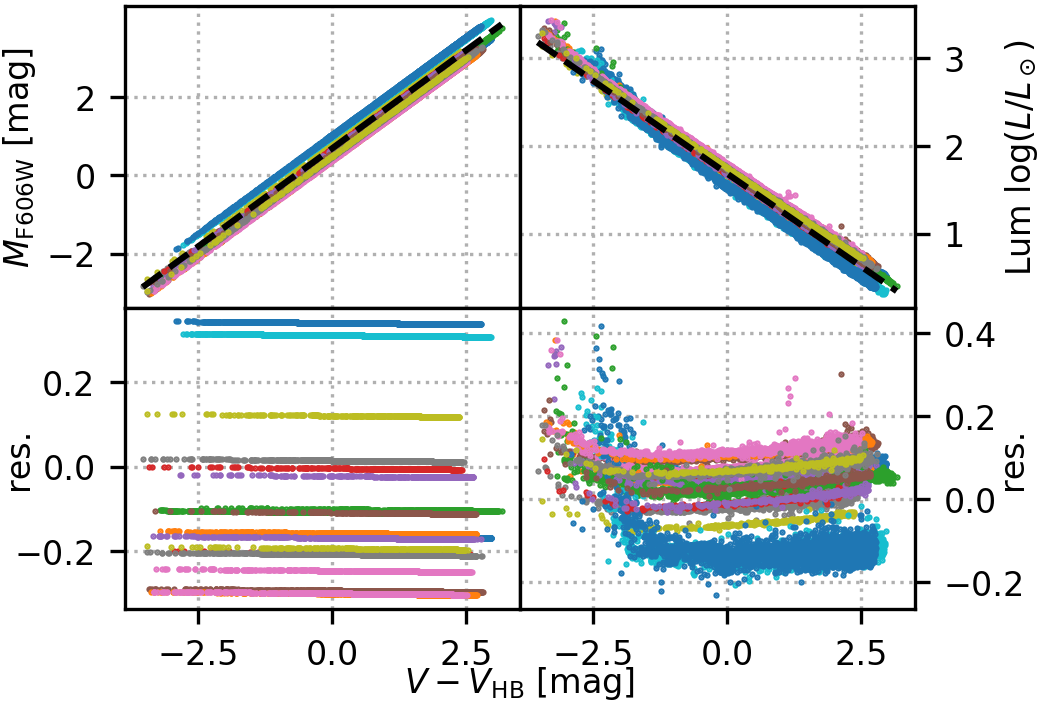}
 \caption{Absolute magnitude in F606W (left panels) and luminosity (right panels) over brightness difference to the HB for all stars in the sample. Different colors belong to different clusters. The dashed black lines indicate linear fits to the data, to which the differences are shown in the lower panels.}
 \label{fig:v-vhb_lum}
\end{figure}
Figure~\ref{fig:v-vhb_lum} shows a comparison between $V-\vhb$ magnitudes and both the derived absolute magnitudes and the logarithm of the derived luminosities. There is a linear relation for both parameters as expected (dashed black line). The offset between individual clusters (different colors) might stem from the difficulty of determining HB brightnesses, especially in clusters where the HB is not really horizontal.

We repeated the analysis done in Section~\ref{sec:extending_below} with absolute magnitudes and luminosities instead of the brightness differences to the HB. For the case of absolute brightness, we obtain
$\beta=-0.426 \pm 0.002$ and $\gamma=0.054 \pm 0.001$, and therefore, equivalent to Eq.~\ref{eq:calib_rew_all}:
\begin{equation} 
 W' = \sumew + 0.426 M_\mathrm{F606W}' - 0.054 M_\mathrm{F606W}'^2,
 \label{eq:calib_rew_all_M}
\end{equation}
with $M_\mathrm{F606W}' = M_\mathrm{F606W} - 0.687$ from the $y$-intercept of the linear relation in Fig.~\ref{fig:v-vhb_lum}. We apply this correction to get similar reduced equivalent widths as from the method using $V-\vhb$.
In the same way, we obtain a calibration for the luminosities (see Fig.~\ref{fig:calib_rew_lum}) and get $\beta=1.006 \pm 0.005$ and $\gamma=0.259 \pm 0.007$, and therefore:
\begin{equation}
 W' =  \sumew + 1.006 L' - 0.259 L'^2,
 \label{eq:calib_rew_all_lum}
\end{equation}
with $L'=\log(L/L_\odot) - 1.687$.

The calibrations for both $M_\mathrm{F606W}$ and $L$ are shown in Figs.~\ref{fig:calib_rew_M} and \ref{fig:calib_rew_lum}, respectively. When using the absolute magnitude, we have \object{NGC\,5139} (\object{$\omega$~Centauri}) as an additional cluster. We could not use this cluster with the previous $V-\vhb$ calibration due to its missing HB brightness. With its many populations and broad range of metallicities (see Sect.~\ref{sec:ngc5139}), it does not show the same narrow trend as the other clusters. However, since the global fit was not significantly affected by this, we kept it for the calibration. For all the other clusters, we see again some deviations between the global and the individual fits of the quadratic relation, but they are mostly minor.

The full sample of 25 cluster can only be calibrated using the relation based on luminosities $L$ (see Fig.~\ref{fig:calib_rew_lum}). In addition to the previously discussed clusters, we can now include, among others, \object{NGC\,6293}, for which we have only very few RGB stars. Therefore the individual fit for this cluster is significantly different from the global one, which, however, also reproduces reasonably well the distribution of W'.

\begin{figure}
 \includegraphics{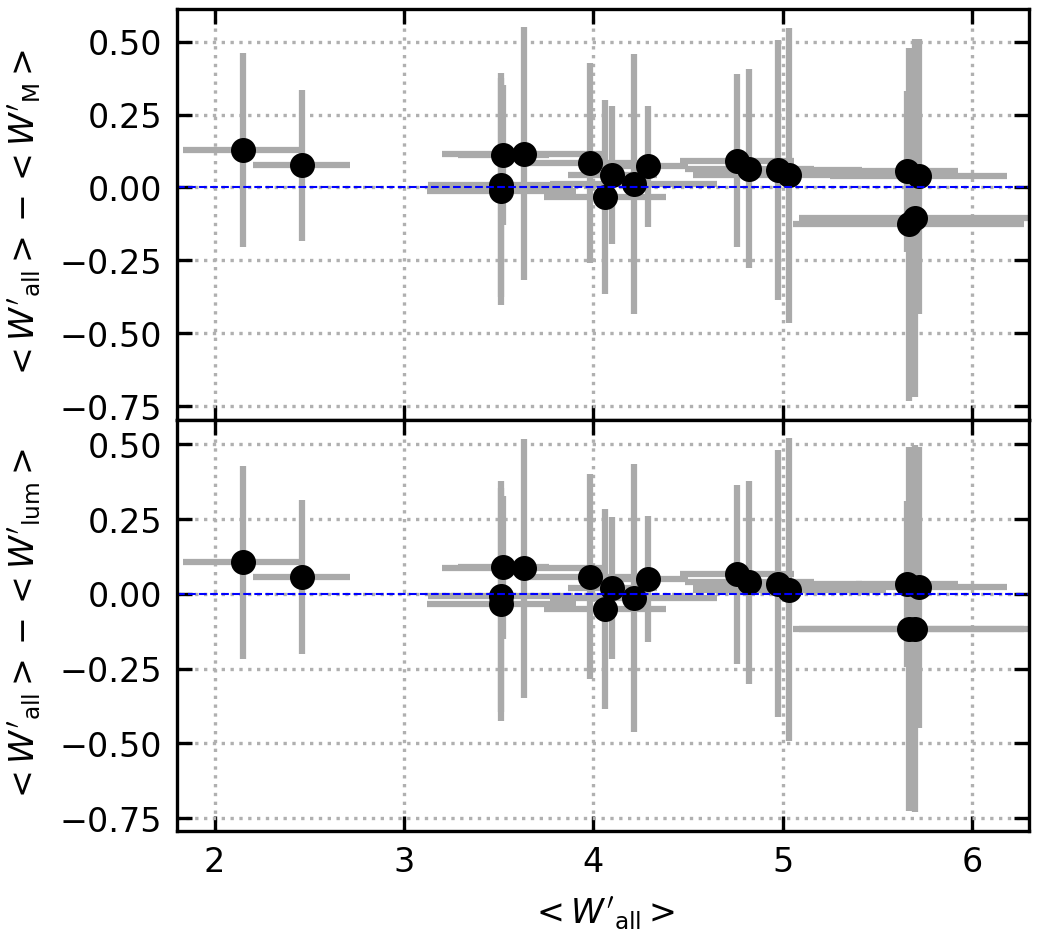}
 \caption{A comparison of average reduced equivalent widths $\left< W'_\mathrm{M} \right>$ and $\left< W'_\mathrm{lum} \right>$ based on absolute magnitudes and luminisities with those obtained using the brightness difference to the HB $\left< W'_\mathrm{all} \right>$.}
 \label{fig:calib_offsets_lum}
\end{figure}
The average reduced equivalent widths for all clusters in the sample are given in Table~\ref{table:reduced_rw} with the indices ``M'' for absolute magnitudes in F606W and ``lum'' for luminosities, respectively. Figure~\ref{fig:calib_offsets_lum} compares the derived average reduced equivalent widths per cluster with those from the analysis as described in Sect.~\ref{sec:extending_below}. As one can see, they agree well within the error bars.

\subsection{Metallicity calibration}
\begin{figure}
 \center
 \includegraphics{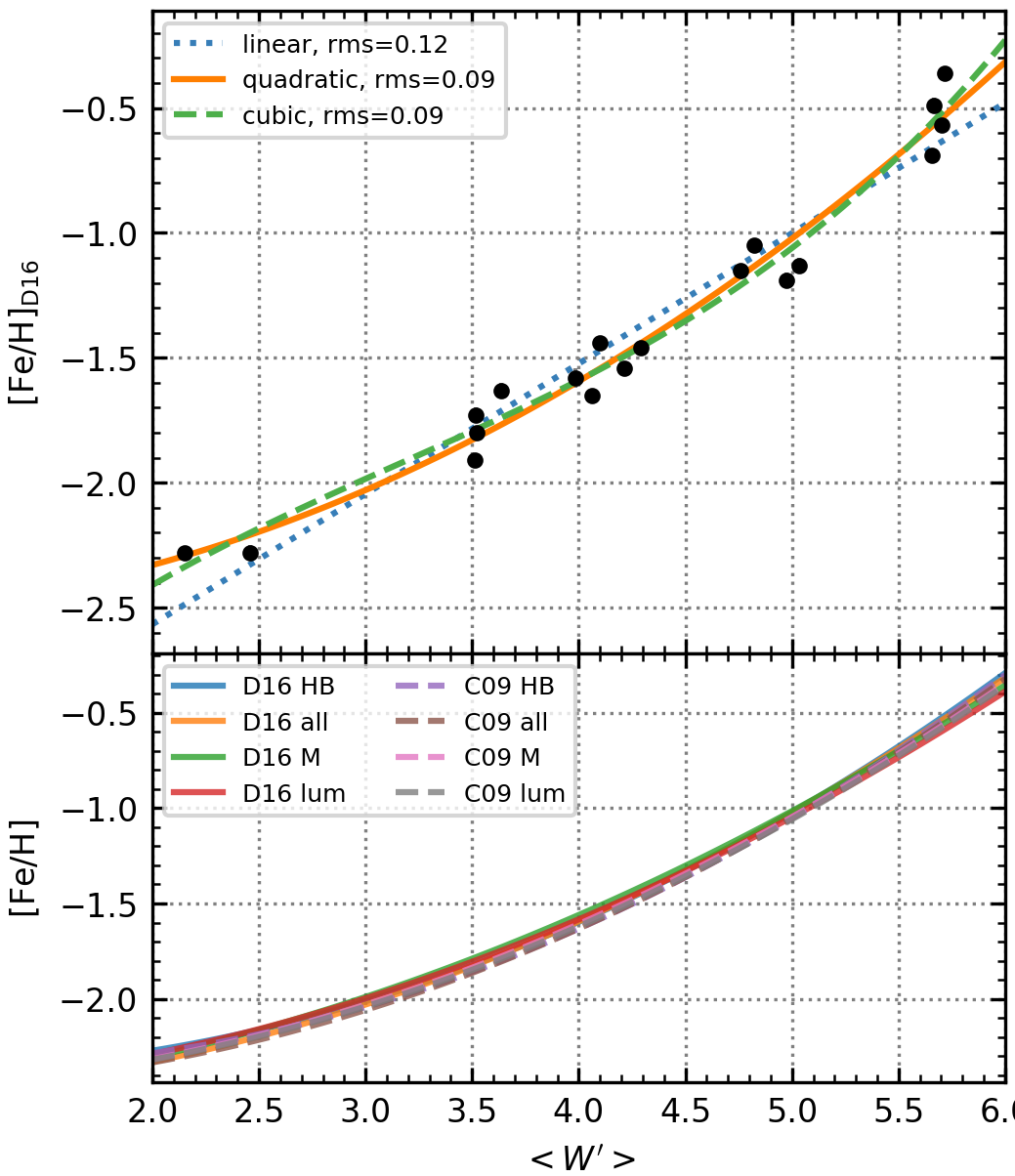}
 \caption{In the upper panel the metallicity from the literature \protect\citep{2016A&A...590A...9D} is plotted over the mean reduced equivalent width of each cluster, derived from the quadratic relation on all RGB stars based on $V-\vhb$. Three polynomials of different degree are fitted to the data and the RMS for each is given in the legend. In the lower panel, quadratic fits to three more calibrations using the same metallicity scale (only brighter than HB, and based on $M$ and $L$) are provided for comparison, together with all four on a different metallicity scale \protect\citep[C09, from][]{2009A&A...508..695C}.}
 \label{fig:calib_feh}
\end{figure}

Up to this point, we have presented four different methods for calculating reduced EWs ($W'$), identified by the following indices in plots and tables:
\begin{itemize}
  \item \textbf{HB:} Using only stars with $V>\vhb+0.2$ and a linear relation based on brightness differences to the HB $V-\vhb$.
  \item \textbf{all:} Using all RGB stars and a quadratic relation based on $V-\vhb$.
  \item \textbf{M:} Same as \textbf{all}, but based on the absolute magnitudes $M_\mathrm{F606W}$.
  \item \textbf{lum:} Same as \textbf{all}, but based on luminosities $L$.
\end{itemize}

Although the calibrations based on luminosity $L$ should be the method of choice in most scenarios, it also depends heavily on model assumptions -- not only for deriving $\teff$ and $\logg$ for each star, but also for calculating the bolometric corrections. Calculating the luminosity also requires an absolute magnitude $M$, which, in turn, can only be derived using good values for distance and extinction. However, when using $L$ or $M$ for the calibration it is not necessary to derive brightness differences to the HB $V-\vhb$, which is complicated even in some globular clusters and impossible for field stars. 
Nevertheless, we do have reliable HB brightnesses for 19 of our clusters from \citet{2010ApJ...708..698D} and we prefer not to depend on any model assumptions, so we will use the calibration based on $V-\vhb$ for these 19 clusters. For the remaining clusters, the absolute magnitude calibration would be the best choice, but we only have F606W photometry available for two additional clusters. So instead of presenting results from three different calibrations, the metallicities for all the remaining six clusters are derived from the luminosity approach. When necessary, these six clusters are marked in plots and tables with an asterisk.

Using the set of average reduced equivalent widths as given in Table~\ref{table:reduced_rw}, we now calibrate them with mean cluster metallicities from the literature \citep[from][]{2016A&A...590A...9D}. The upper plot in Fig.~\ref{fig:calib_feh} shows the metallicities from the literature as a function of reduced equivalent width. The three lines show a linear, quadratic, and cubic fit to the data, taking into account the errors on both axes (which are small), together with their corresponding root mean squares (RMS). The coefficients for all relations are given in Table~\ref{table:feh_calib} for the following equation:
\begin{equation}
 \feh = p0 + p_1 \cdot W' + p_2 \cdot {W'}^2 + p_3 \cdot {W'}^3.
 \label{eq:feh_calib}
\end{equation}
Since both the Bayesian (BIC) and the Akaike (AIC) information criteria give the best results for the quadratic relation, we chose this for further analyses. Therefore, the relation between reduced equivalent width and metallicity is given by:
\begin{equation}
 \feh = -2.52  - 0.04 W' - 0.07 {W'}^2.
\end{equation}

\begin{table}
 \caption{Coefficients for CaT-metallicity calibration as given by Eq.~\ref{eq:feh_calib} for three different polynomial degrees.}
 \begin{center}
 \begin{tabular}{cccc}
 \hline \hline
 $p_0$ & $p_1$ & $p_2$ & $p_3$ \\ \hline
$-3.61 \pm 0.13$ & $0.52 \pm 0.03$ & -- & -- \\
$-2.52 \pm 0.32$ & $-0.04 \pm 0.16$ & $0.07 \pm 0.02$ & -- \\
$-4.02 \pm 1.29$ & $1.22 \pm 1.07$ & $-0.27 \pm 0.28$ & $0.03 \pm 0.02$ \\

 \hline
 \end{tabular}
 \end{center} 
 \label{table:feh_calib}
\end{table}
\begin{table}
 \caption{Coefficients for CaT-metallicity calibration as given by Eq.~\ref{eq:3d_calib}.}
 \begin{center}
 \begin{tabular}{cccccc}
 \hline \hline
 a & b & c & d & e & f \\\hline
$-3.456$ & -0.074 & -0.100 & 0.540 & 2.101 & -0.011\\
$\pm0.050$ & $\pm0.017$ & $\pm0.005$ & $\pm0.009$ & $\pm0.117$ & $\pm0.004$\\
 \hline
 \end{tabular}
 \end{center} 
 \label{table:3d_calib}
\end{table}

The metallicities derived from this relation show two systematic errors: the mean metallicity of a cluster will be the value of the relation given above at the mean reduced EW of the cluster and, therefore, will have a small offset to the literature value for most clusters. Furthermore, the slope of the relation at any given point defines the metallicity spread.

Following previous studies \citep[see, e.g.,][]{2018A&A...619A..13V}, the uncertainties are calculated as the quadratic sum of the uncertainty $\sigma_{W'}$ and the RMS for the used relation as given in Fig.~\ref{fig:calib_feh}. For the quadratic relation this yields:
\begin{equation}
 \sigma_{\feh} = \sqrt{\sigma_r^2 + \mathrm{RMS}^2},
 \label{eq:feh_calib_err}
\end{equation}
where $\mathrm{RMS}$ is the root mean square for the used relation and $\sigma_r$ is the propagated uncertainty for $\sigma_{W'}$:
\begin{equation}
 \sigma_r = (p_1 + 2 \cdot p_2 \cdot W') \sigma_{W'} = (-0.03 + 0.14 W') \sigma_{W'}.
\end{equation}

We already demonstrated that all previously discussed approaches yield a similar $W'$. The lower panel in Fig.~\ref{fig:calib_feh} shows that the metallicity calibrations are comparable, no matter what method for calculating the reduced EWs has been used. There is also no significant variation when using different metallicity scales, in this case taken from \citet[D16]{2016A&A...590A...9D} and \citet[C09]{2009A&A...508..695C}.

The use of absolute magnitudes or luminosities allows us to use a different approach and to get rid of the intermediate step of calculating reduced equivalent widths completely. A CaT-metallicity relation based on absolute magnitudes was described by \citet{2010A&A...513A..34S}. They suggest a direct relation between the equivalent widths of the Ca lines and the metallicity:
\begin{equation}
    \feh = a + b \cdot M + c \cdot \sumew + d \cdot \sumew^{-1.5} + e \cdot \sumew \cdot M,
\end{equation}
where M is the absolute magnitude in an arbitrary filter and the term for $\sumew^{-1.5}$ was introduced to account for variations at low metallicities. The limits for this calibration were given as $-3<\vhb<0$ and $-3<M_V<0.8$, i.e.\ for stars brighter than the HB only.

Instead of the absolute magnitude $M$ we are using the luminosity $\log L/L_\odot$ and extend the relation to stars fainter than the HB in the same way as before by introducing a quadratic term for the luminosity:

\begin{align}
  \feh = a &+ b \cdot \log (L/L_\odot) + c \cdot \log (L/L_\odot)^2 + d \cdot \sumew \nonumber\\
           &+ e \cdot \sumew^{-1.5} + f \cdot \sumew \cdot \log (L/L_\odot),
  \label{eq:3d_calib}
\end{align}

Unfortunately, we do not have metallicities for all the stars in our sample available for the calibration, so we assume it to be the same for all stars in each cluster. The coefficients for the best fitting polynomial are given in Table~\ref{table:3d_calib}. This calibration will be referenced to using the index \emph{poly} and is treated more as an experimental approach for comparison.

\section{Metallicity distributions}
\label{sec:feh_dists}
\begin{figure}
 \includegraphics{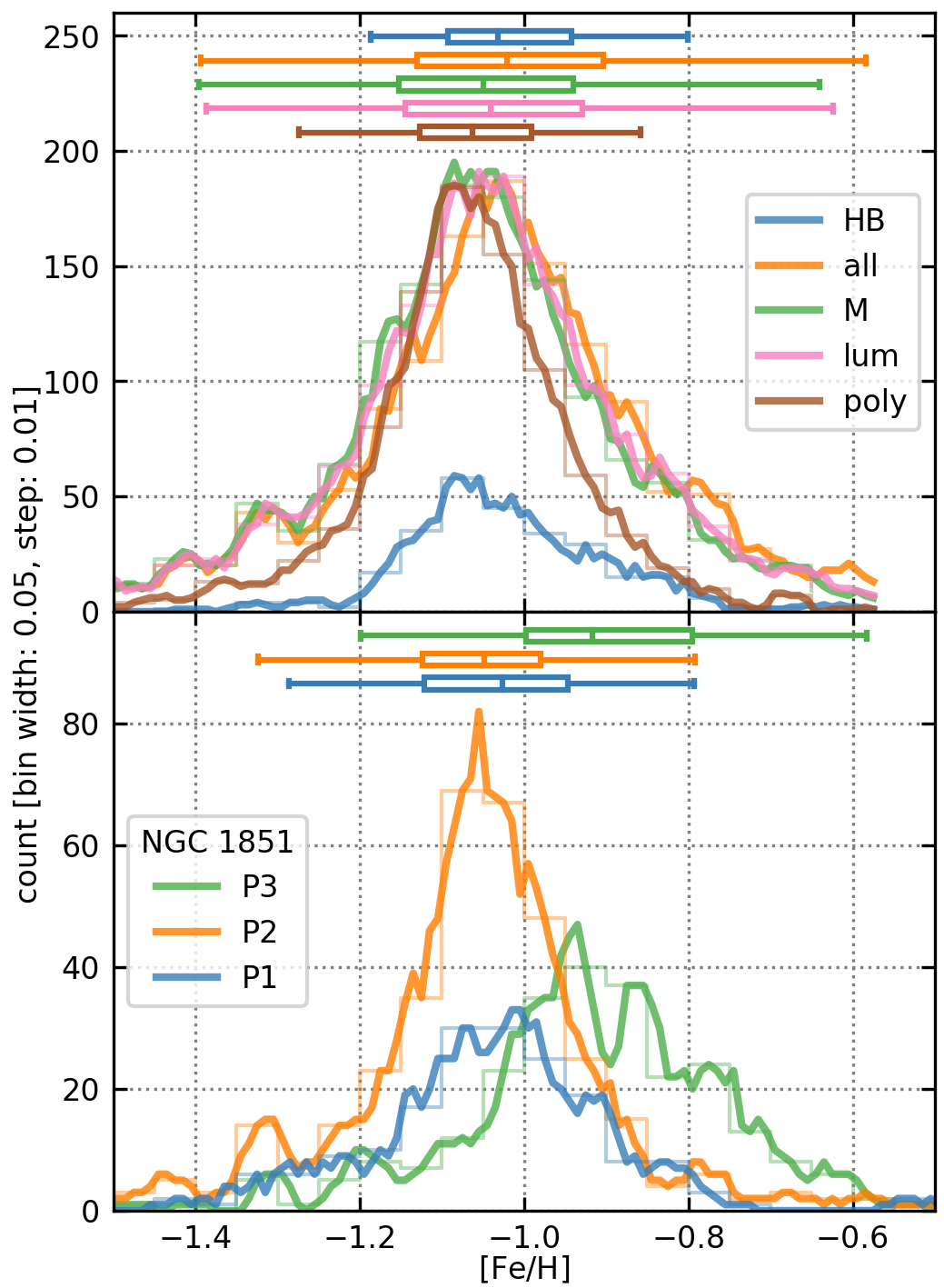}
 \caption{Metallicity distributions for \object{NGC\,1851} as derived from all four CaT-metallicity relations discussed in the text are shown in the upper panel, while distributions for the three populations we obtained from the chromosome map in Fig.~\ref{fig:ngc1851_cmap} for the ``all'' calibration from above are shown in the lower panel. As for all the following, similar plots, the bars at the top of the plots show the 5--95\% range of the data (lines with caps), the interquartile range ($Q_1$--$Q_2$, boxes) and medians (vertical line) for all distributions.}
 \label{fig:ngc1851_dist}
\end{figure}
\begin{figure*}
 \includegraphics{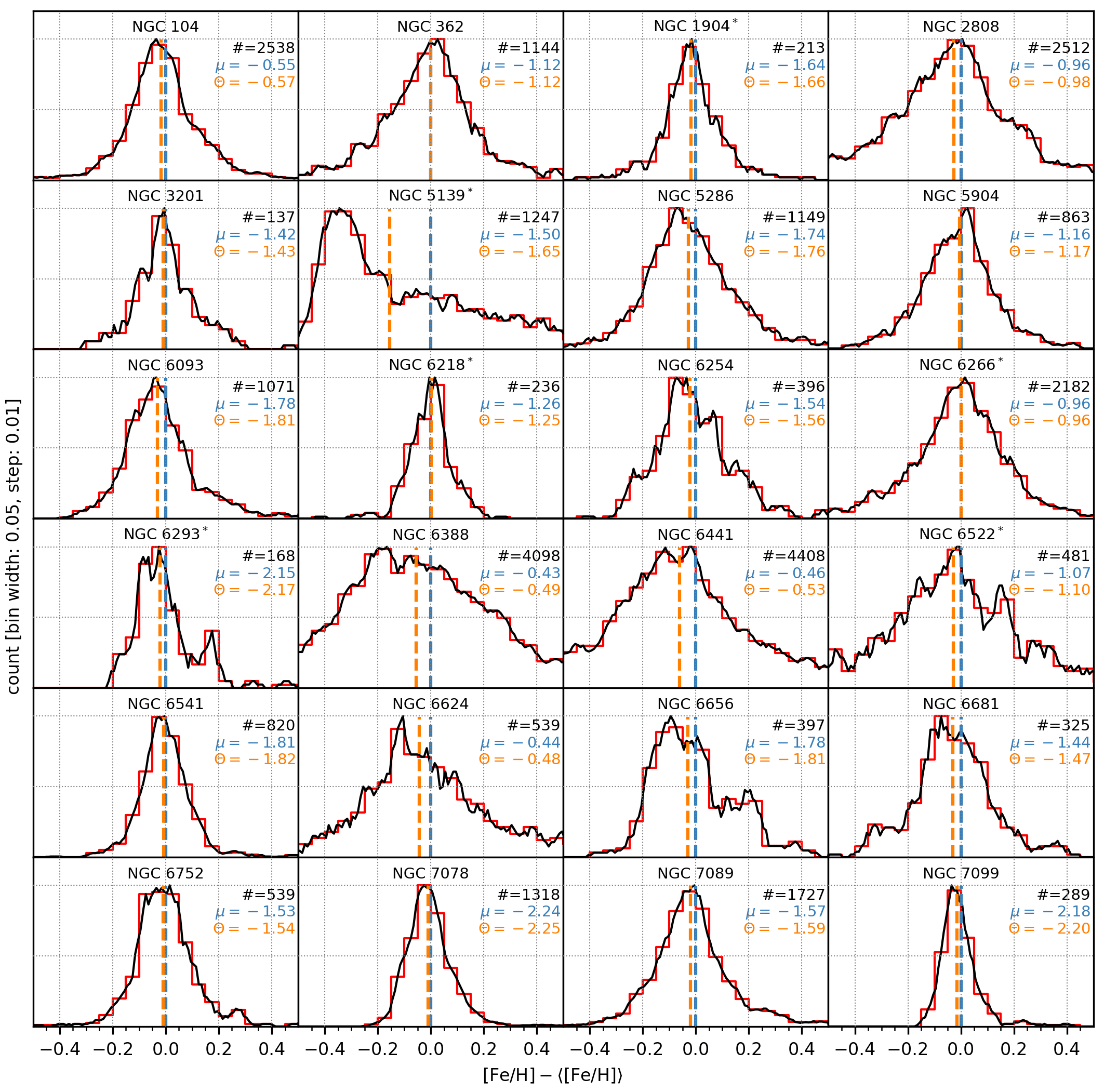}
 \caption{Classic (red) and rolling (black) histograms for the $\feh$ distributions for all clusters (except \object{NGC\,1851}, which is shown in Fig.~\ref{fig:ngc1851_dist})} as derived from the CaT relation. The metallicities for all clusters are shifted by their respective means so that they peak around 0. The x axes all show the same range within $\pm0.5\mathrm{dex}$, while the y axes are scaled to the peak values of each clusters. Horizontal grid lines are shown at peak height and at half that height, while vertical lines are located at $0$, $\pm0.2$, and $\pm0.4\mathrm{dex}$. For each cluster, the number of stars is given (\#) that have been used for calculating the distribution, as well as the mean $\mu=\left< \feh \right>$ and the median $\Theta$. All the numbers are also given in Table~\ref{table:results}.
 \label{fig:feh_dists}
\end{figure*}
Having presented five different approaches for deriving metallicities from reduced EWs, we can now apply them to all stars in our sample. The upper panel of Fig.~\ref{fig:ngc1851_dist} shows the resulting metallicity distributions for \object{NGC\,1851}. The thinner lines show classical histograms with a bin size of $0.05\mathrm{dex}$. Since the shape of a histogram not only depends on the bin size, but also on the starting value, we decided to include a rolling histogram (also known as convolved frequency), which was obtained by shifting the positions of the bins in steps of $0.1$ and connecting the points with a solid line. This way, smaller structures in the shape of the distribution show up more prominently.

In the case of \object{NGC\,1851}, all calibrations produce similar metallicity distributions, although the one based only on stars brighter than the HB, apart from including fewer stars, shows a significant tail towards higher metallicities that does not exist in the other distributions. The distributions derived from the other methods based on the intermediate step of calculating reduced EWs (\emph{all}, \emph{M}, and \emph{lum}) are all very similar, and also show the same small features, for example the little bump at $\sim -1.3\dex$. On the other hand, the metallicities derived from the polynomial fit (\emph{poly}) are systematically lower, and the distribution is a little narrower and does not exhibit the smaller features that exist in the others.

\begin{figure}
 \includegraphics{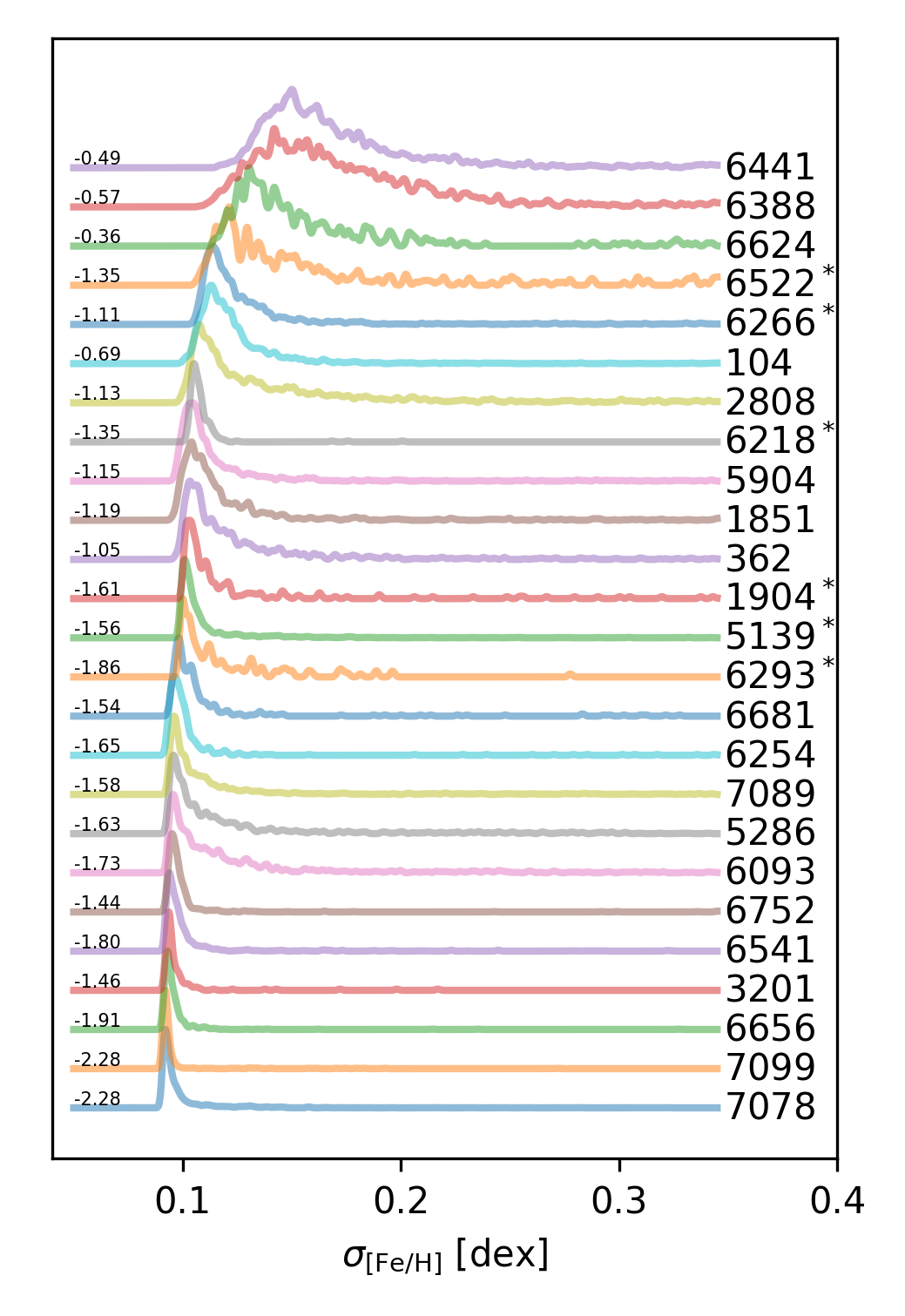}
 \caption{Distribution for the uncertainties of the derived metallicities for all RGB stars in all clusters, convolved with a Gaussian with $\sigma=0.001\dex$. The NGC numbers of the clusters are given at the right end of the distributions, while the mean metallicities from \protect\citet{2016A&A...590A...9D} are listed on the left.}
 \label{fig:uncertainties_feh}
\end{figure}
\begin{figure}
 \includegraphics{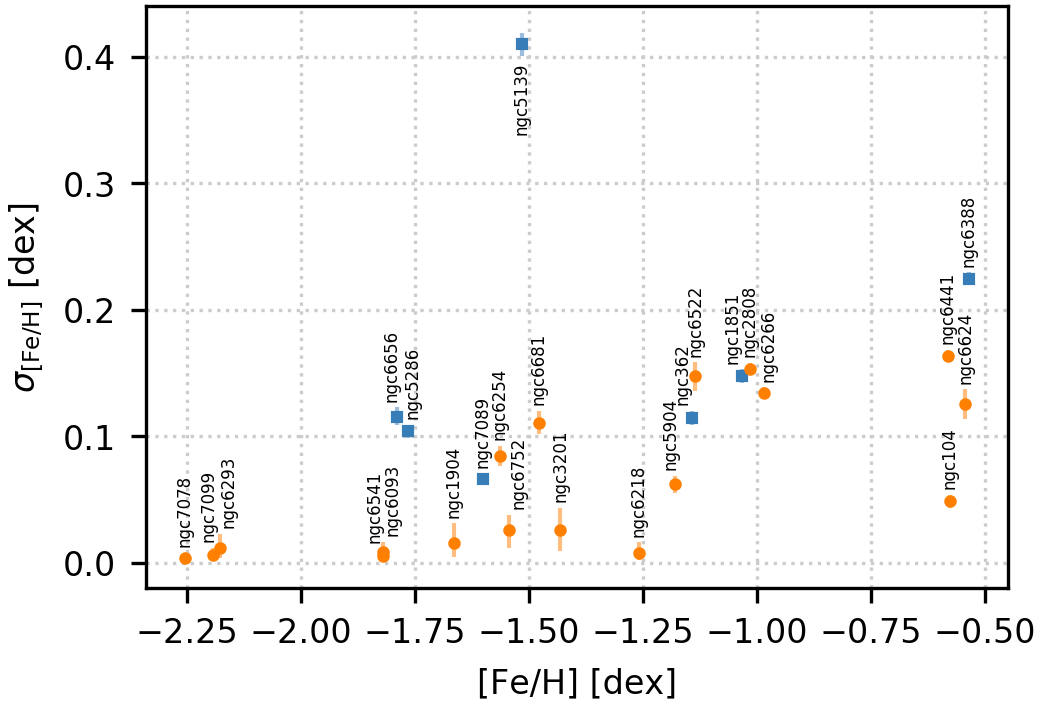}
 \caption{Intrinsic metallicity distributions for all clusters as derived from a maximum-likelihood analysis. Metal-complex type II clusters are marked with blue squares.}
 \label{fig:err_dists_lm}
\end{figure}
\begin{figure*}
  \center
  \includegraphics{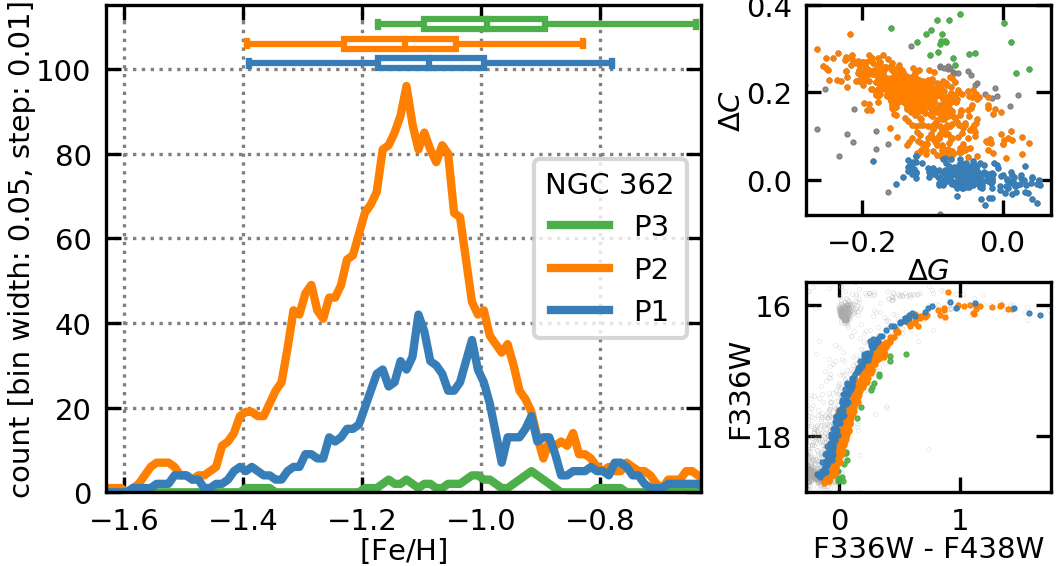}          
  \includegraphics{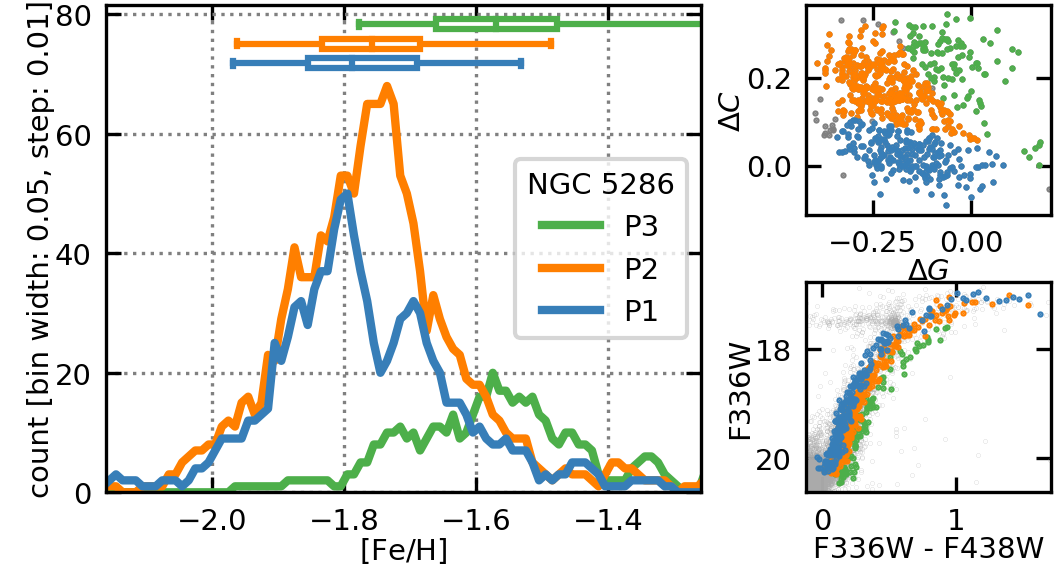} \\
  \includegraphics{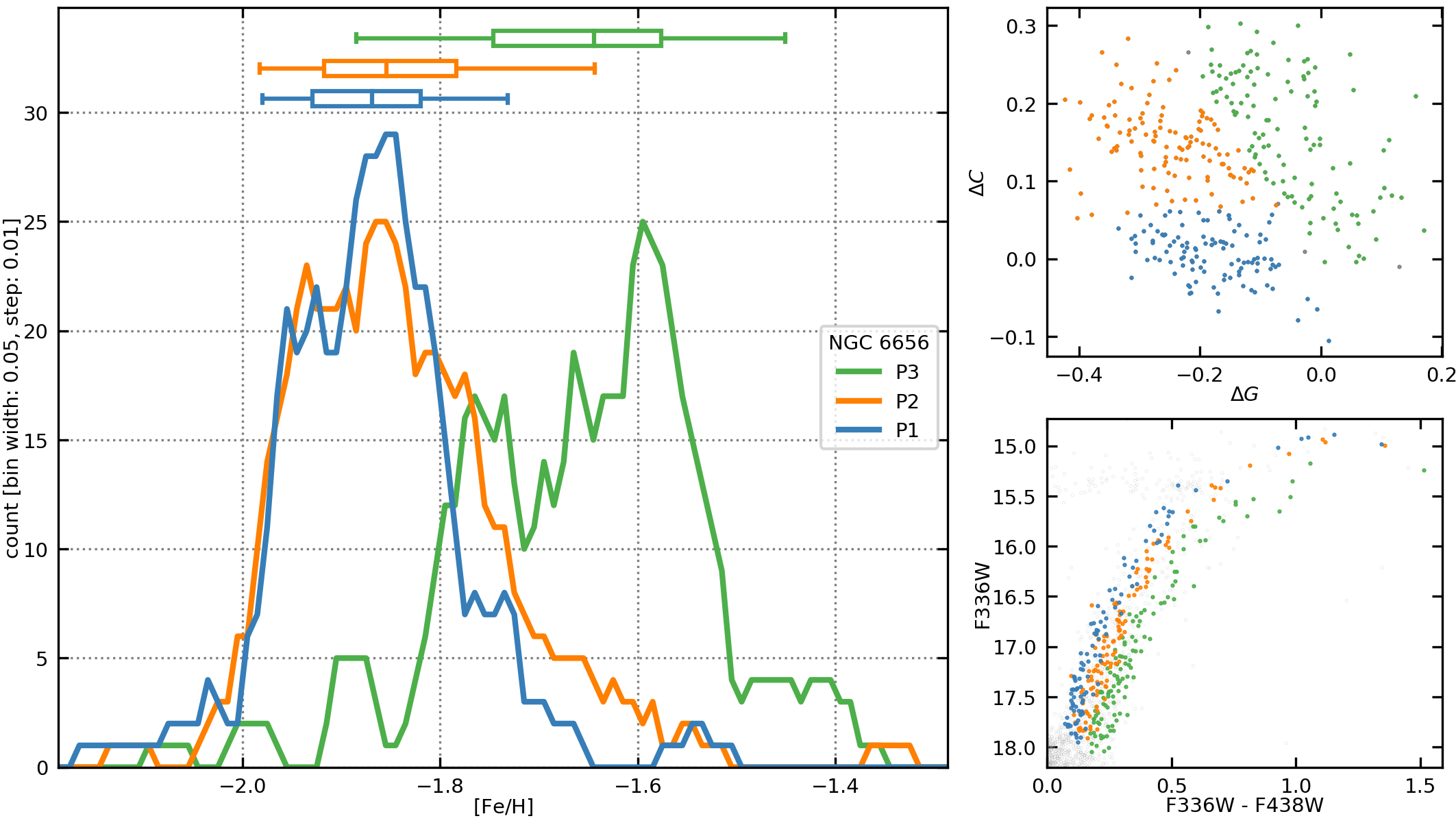} \\
  \includegraphics{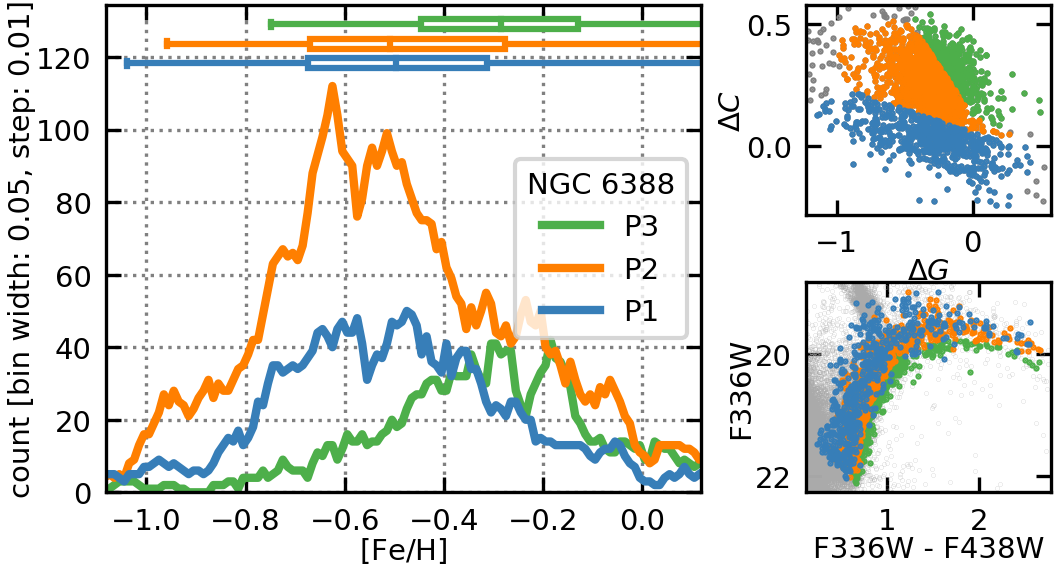} 
  \includegraphics{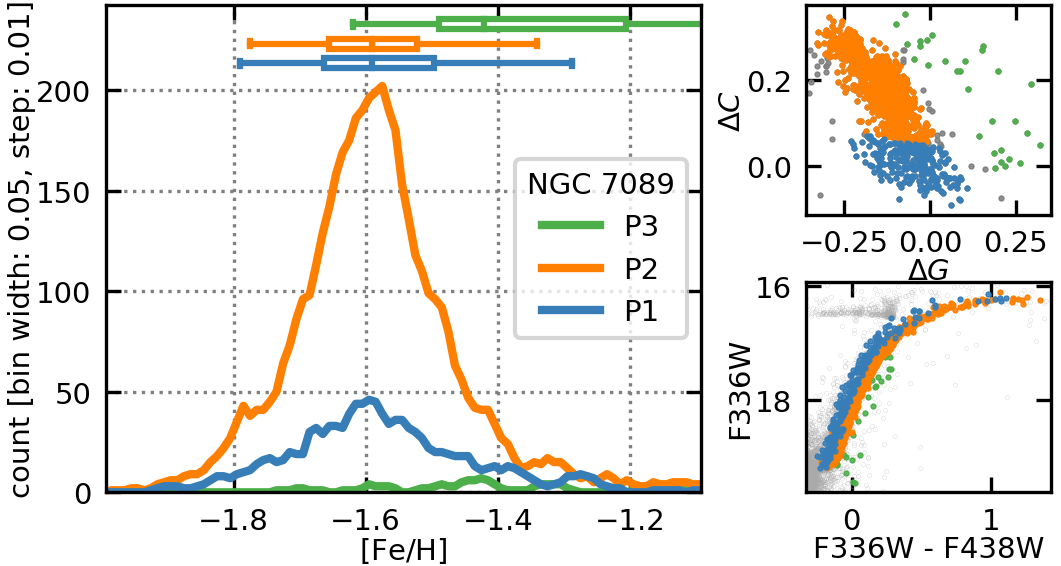} 
  \caption{Metallicity distributions for the stellar populations of five more Type~II clusters are shown in the big panels (as in Fig.~\ref{fig:ngc1851_dist} for \object{NGC\,1851}). The upper smaller panels show the chromosome maps and the lower panels the CMDs of the RGBs of the respective clusters (both as in Fig.~\ref{fig:ngc1851_cmap}). The color-coding is the same in all plots for a single cluster.}
  \label{fig:multipop}
\end{figure*}
While the metallicity distribution for \object{NGC\,1851} is shown in Fig.~\ref{fig:ngc1851_dist}, those of the remaining 24 clusters in our sample are given in Fig.~\ref{fig:feh_dists}. The results for clusters marked with a star have been obtained using the luminosity calibration presented in Sect.~\ref{sec:calib_lum}, since for those clusters there is no HB brightness available from \citet{2010ApJ...708..698D}. The medians, means, standard deviations, and first and third quartiles for the metallicities of all clusters are listed in Table~\ref{table:results}. In order to assess whether the width of the distribution is dominated by the errors on the individual measurements, the distribution of metallicity uncertainties is shown for every cluster in Fig.~\ref{fig:uncertainties_feh} -- the median uncertainty for the whole sample is $\sim 0.12\,\mathrm{dex}$. With the mean metallicities of the clusters given on the left side of the distributions, we see a clear trend of the uncertainties with metallicity. The uncertainty distributions for high-metallicity clusters are significantly broader and peak at higher values compared to the low-metallicity clusters. This result confirms our suspicion that for high metallicities, our EW measurements (or at least their uncertainties) are affected by smaller lines in the wings of the CaT lines, likely more as a systematic than a random error.

We also investigated the reliability of our metallicity measurements using a maximum likelihood approach. Under the assumption that a cluster has a mean metallicity of $\mu_{\rm [Fe/H]}$ and an intrinsic metallicity spread of $\sigma_{\rm [Fe/H]}$, the probability of measuring a value $m$ with uncertainty $\delta_{\rm m}$ can be approximated as
\begin{equation}
p(m,\,\delta_{\rm m})=\frac{1}{2\pi\sqrt{\sigma_{\rm [Fe/H]}^2 + \delta_{\rm m}^2}}\exp\left(-\frac{(m-\mu_{\rm [Fe/H]})^2}{2(\sigma_{\rm [Fe/H]}^2 + \delta_{\rm m}^2)}\right)\,.
\end{equation}
For each cluster, we determined the intrinsic parameters $\mu_{\rm [Fe/H]}$ and $\sigma_{\rm [Fe/H]}$ of the metallicity distribution using the affine-invariant MCMC sampler \textsc{emcee} \citep{2013PASP..125..306F}. Most clusters in our sample do not show an intrinsic metallicity spread, hence we expect $\sigma_{\rm [Fe/H]}$ to be consistent with zero in such objects. On the other hand, if significant spreads are found, they can be attributed to residual trends in our metallicity measurements (e.g., with luminosity) or systematic effects in the analysis that are not accounted for by our formal uncertainties.

Figure~\ref{fig:err_dists_lm} shows the results of this analysis for all clusters. For all our Type~I clusters (in orange), we expect an intrinsic spread of $\sigma_{\rm [Fe/H]}=0$. While this is true for most clusters with $\feh \lessapprox -1.7$, we see values significantly larger than this for higher metallicities, which indicates that at least for those clusters we under-estimate the uncertainties for the metallicities. At the same time, all Type~II clusters (in blue) show intrinsic spreads, which we would expect from these metal-complex clusters.

Looking at the distributions for both the metallicities and their uncertainties, we identified some problematic cases: those with erratic metallicity distributions and those with high average uncertainties. While for \object{NGC\,6293} and \object{NGC\,6522} this is certainly due to low number statistics, this does not apply to \object{NGC\,6388} and \object{NGC\,6441}, since they both include more than 2,000 RGB stars. For these two clusters, the available photometry is difficult to handle due to extremely broadened main sequences and giant branches (see Sect.~\ref{sec:observations}). This directly affects the extraction process of the raw spectra from the data cubes and therefore the quality of the extracted spectra. However, two other high-metallicity clusters, namely \object{NGC\,6624} and \object{NGC\,6522}, show a similar behavior, so this is presumably connected to systematic errors as discussed before.

We combined the derived metallicities for all stars with the chromosome maps that we created for all clusters with available UV HST photometry and plot metallicity distributions for the different populations. The CMD and the chromosome map for \object{NGC\,1851} is shown in Fig.~\ref{fig:ngc1851_cmap}, revealing three different, clearly separated populations. The lower panel in Fig.~\ref{fig:ngc1851_dist} shows the metallicity distributions for all three populations, using the same color-coding. The bars at the top of the plot show the 5--95\% range of the data (lines with caps), the interquartile range ($Q_1$--$Q_2$, boxes) and medians (vertical line) for all distributions.
The parameters derived from the metallicity distributions for all populations are listed in Table~\ref{table:results}. Please note that the clusters \object{NGC\,1904}, \object{NGC\,6266}, \object{NGC\,6293}, and \object{NGC\,6522} were not included in the HUGS survey \citep{2018MNRAS.tmp.2405N}, so we did not create chromosome maps for them.

\begin{figure*}
 \includegraphics{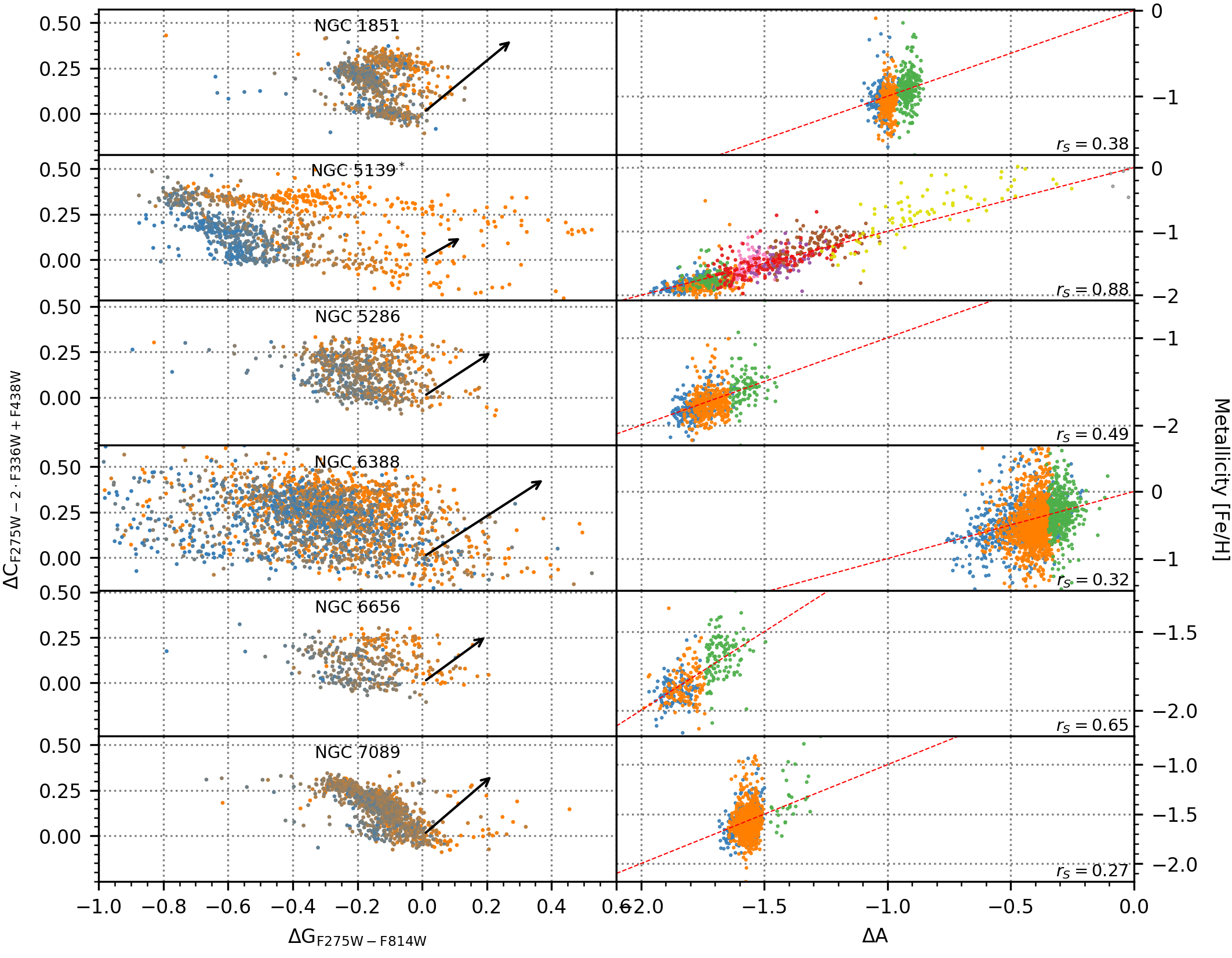}
 \caption{In the panels on the left, the chromosome maps are plotted for all our Type~II clusters, color-coded with our derived metallicity, ranging from the median minus $0.3$ (blue) to median plus $0.3\dex$ (orange). The arrows show the directions of the metallicity slopes and their lengths indicate a change of $0.15\dex$. In the panels on the right, the metallicities are plotted as function of a pseudo-color $\Delta \mathrm{A}$ along the arrows on the left, color-coded by population. A linear fit to the data is shown as dashed red line. The Spearman correlation coefficient $r_S$ is given for every cluster.}
 \label{fig:chm_trends}
\end{figure*}

\subsection{Type I/II clusters}
\label{sec:typeI_II}
For the clusters for which we created chromosome maps, we investigated the metallicity distributions of the different populations in each cluster. For Type~I cluster we do not expect any variations in metallicity, and for almost all of them in our sample this seems to be true. Figure~\ref{fig:multipop_all} shows the distributions for all Type~I clusters, and they are very similar for all clusters only showing two populations. For each cluster the distributions are shown in the larger panel on the left, while the chromosome map and the CMD are shown next to it on the right. For two of the Type~I clusters, more than two populations have been found: \object{NGC\,2808} and \object{NGC\,7078}. The latter has recently been re-labeled as Type~II by \citet{2018MNRAS.477.2004N}. They will be discussed in detail in Sections \ref{sec:ngc2808} and \ref{sec:ngc7078}, respectively.

On the other hand, \object{NGC\,1851}  is one of those clusters that \citet{2017MNRAS.464.3636M} classified as Type~II (or metal-complex) clusters. These clusters do not simply show two populations, but contain a third population -- or even more. Previous studies have shown that these populations also show a significant difference in their chemical compositions, showing up as a split in metallicity. 
Figures~\ref{fig:multipop}~and~\ref{fig:multipop_ngc5139} show the metallicity distributions for the populations of the six remaining Type~II clusters in our sample. For all of them, we see a difference of $\sim 0.2\dex$ between the mean and median metallicities of populations P1/P2, and P3. Only \object{NGC\,362} and \object{NGC\,1851} show a smaller variation of $\sim 0.12\dex$. The populations P3 of all clusters usually contain a hundred or more stars (besides \object{NGC\,362} and \object{NGC\,7089}). Table~\ref{table:type2_kstest} shows the results of Two-sample Kolmogorov-Smirnov tests comparing the metallicities of the clusters' P3 stars with that of all their other stars. With $D>>D_\mathrm{crit}$ and $p<<1$ for all clusters, we assume the splits in the metallicities to be significant.

\begin{table}
 \caption{Results from a Two-sample Kolmogorov-Smirnov test for all Type~II clusters comparing the metallicities of their P3 stars with that of all the other stars.}
 \begin{center}
 \begin{tabular}{lccc}
 \hline \hline
 Cluster & $D$ & $D_\mathrm{crit}$ & $p$-value \\ \hline
 NGC 362 & 0.435 & 0.050 & $0.00032$ \\
NGC 1851 & 0.403 & 0.005 & $3.1\cdot10^{-15}$ \\
NGC 5286 & 0.534 & 0.011 & $2.1\cdot10^{-15}$ \\
NGC 6388 & 0.340 & 0.003 & $1.2\cdot10^{-37}$ \\
NGC 6656 & 0.615 & 0.013 & $1.3\cdot10^{-15}$ \\
NGC 7089 & 0.162 & 0.008 & $0.0018$ \\

 \hline
 \end{tabular}
 \end{center} 
 Notes. $D$ denotes the result of the Two-sample KS test. $D_{crit}=c(\alpha ){\sqrt{(n+m)/(nm)}}$ is the critical value (with sample sizes $n$ and $m$) for $\alpha=0.1$ and $c(\alpha)=1.073$. Finally, the last column gives the two-tailed $p$-value.
 \label{table:type2_kstest}
\end{table}

With the large number of stars in our sample, we can then investigate whether the Type~II clusters show a real bimodality in metallicity or whether it is a continuous trend. In the left panels of Fig.~\ref{fig:chm_trends}, the chromosome maps of all our Type~II clusters are plotted, color-coded with metallicity. An arrow in each panel shows the direction of the metallicity gradient and its length indicates a change of $0.15\dex$. Interestingly, these arrows all point in the same direction, indicating a global trend. We note that \object{NGC\,362} is missing in this overview because we could not determine a metallicity gradient due to the small number of stars in its P3 population.

\begin{figure*}
  \includegraphics{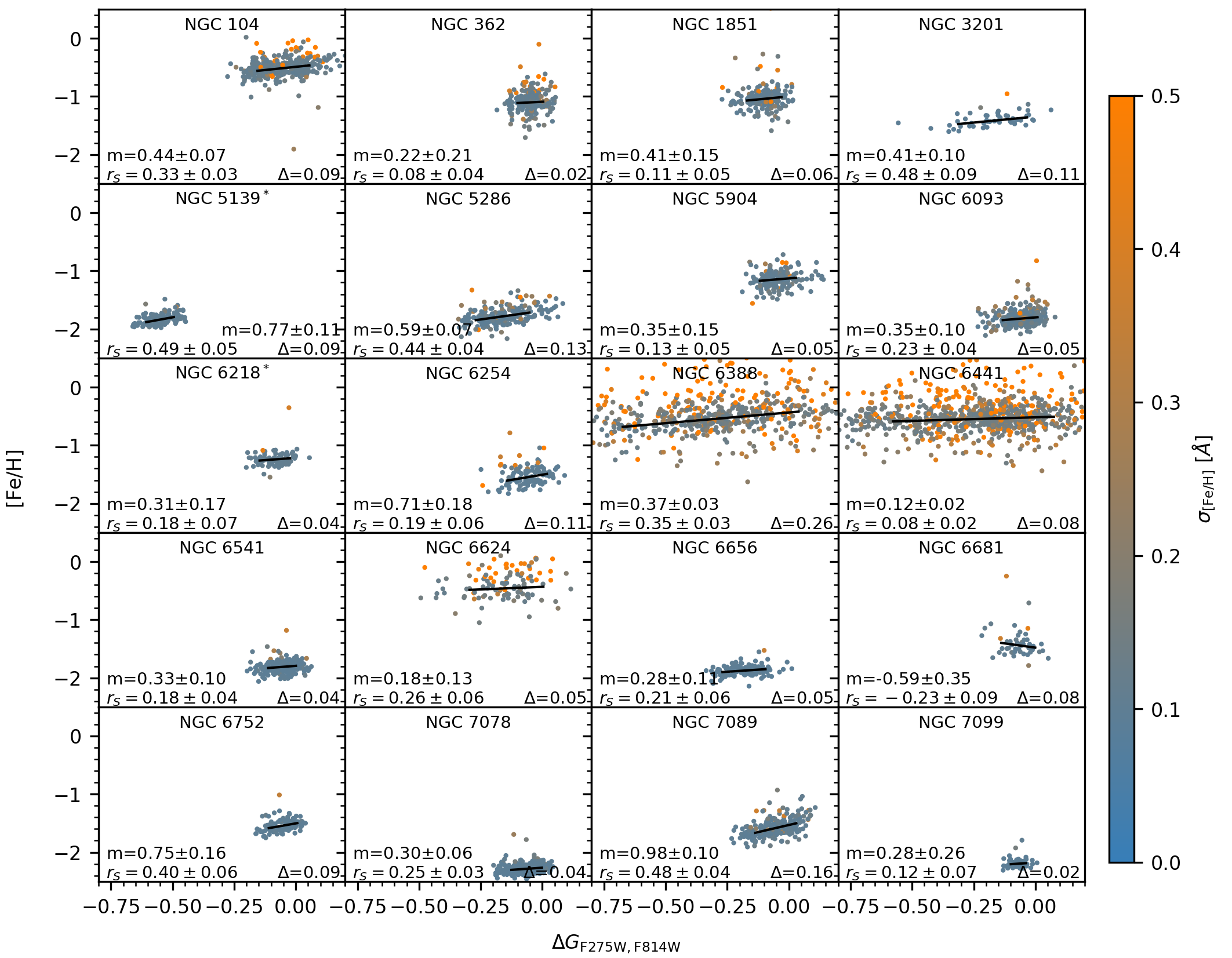}          
  \caption{Metallicity as a function of $\Delta \mathrm{G}=\Delta \mathrm{F275W,F814W}$ is plotted for the primordial populations P1 in all clusters in our sample with available UV photometry. The black lines show a linear fit (with the length being the FWHM of the distribution) and values are given for the fitted slopes $m$, the Spearman correlation coefficients $r_S$, and the differences $\Delta$ in metallicity between the extremities of the black lines. Note that for \object{NGC\,6388} and \object{NGC\,6441} not the full ranges are shown.}
  \label{fig:1g_variations}
\end{figure*}

In the right panels of Fig.~\ref{fig:chm_trends}, the metallicity is plotted as a function of the pseudo-color $\Delta \mathrm{A}$ along the arrows on the left. The stars are color-coded by population, using the same colors as in the other figures. We fitted a line through all datapoints  (in red) and the Spearman correlation coefficient $r_S$ is given for all clusters. For \object{NGC\,7089}, the number of stars in P3 is probably too low (30) for this kind of analysis, but for all other clusters we see a clear, narrow, and continuous trend of metallicity with $\Delta \mathrm{A}$. Even \object{NGC\,5139} with its complicated structure follows this relation very nicely with all stars and all populations. The other clusters also do not show a clear separation of P3 in these diagrams. These results hint towards a continuous trend and not a bimodality.

\section{Intrinsic abundance variations in the primordial populations}
\label{sec:1g_variations}
In the chromosome maps, the primordial P1 population is often extended along the $\Delta \mathrm{G}_\mathrm{F275W-F814W}$ axis, indicating some variations in the chemical composition. \citet{2019MNRAS.487.3815M} discussed two possible explanations for this color spread: either a variation in He content or in [Fe/H] and [O/Fe]. They also state that a spread in metallicity would result in a positive correlation with $\Delta \mathrm{G}$, although they could not find strong evidences supporting this. While \citet{2018A&A...616A.168L} assume the cause to be a spread in the initial helium and possibly nitrogen abundance, \citet{2019MNRAS.486.5895T} find no conclusive explanation for the spread in P1.

In Fig.~\ref{fig:1g_variations}, the metallicities of all P1 stars are plotted as a function of $\Delta \mathrm{G}$ for all clusters in our sample with available UV photometry. The black lines represent linear fits to the data with their lengths indicating the FWHM of the metallicity distribution in $\Delta \mathrm{G}$. For each cluster, the slope $m$ of the linear fit is indicated as well as the Spearman correlation coefficient $r_S$ and the difference in metallicity between the extremities of the black lines. The latter value can be used as an estimator for the total change in metallicity within the P1 stars.
Although the Spearman correlation coefficient barely reaches $+0.5$, except for a few clusters, it is positive for all besides \object{NGC\,6681}, where the slope is dominated by some outliers -- removing them yields a correlation of about zero. Except for \object{NGC\,6681}, the slope is positive for all clusters and the $1\sigma$ error interval excludes a flat line, with the possible exceptions of \object{NGC\,362} and \object{NGC\,7099}.
For \object{NGC\,3201}, the slope of the metallicity as a function of $\Delta \mathrm{G}$ has been determined by \citet{2019arXiv191002892M}. They find a value of $0.5$, which is within the error bars of our result of $0.41 \pm 0.10$.

The total variation in metallicity (given as $\Delta$ in the plots) is typically about $0.04\dex$, but goes up to about $0.1\dex$ or more for some clusters. The cases of \object{NGC\,6388} and \object{NGC\,6441} might be explained by large uncertainties (see Fig.~\ref{fig:uncertainties_feh}), but due to the large number of stars the trends are significant. As expected, the trend is more pronounced for wide distributions in $\Delta \mathrm{G}$, but some of the narrower ones also show a clear increase of $\feh$ with $\Delta \mathrm{G}$. Surprisingly, the trend does not seem to be affected by the Type~I/II classification.

\begin{figure*}
  \includegraphics{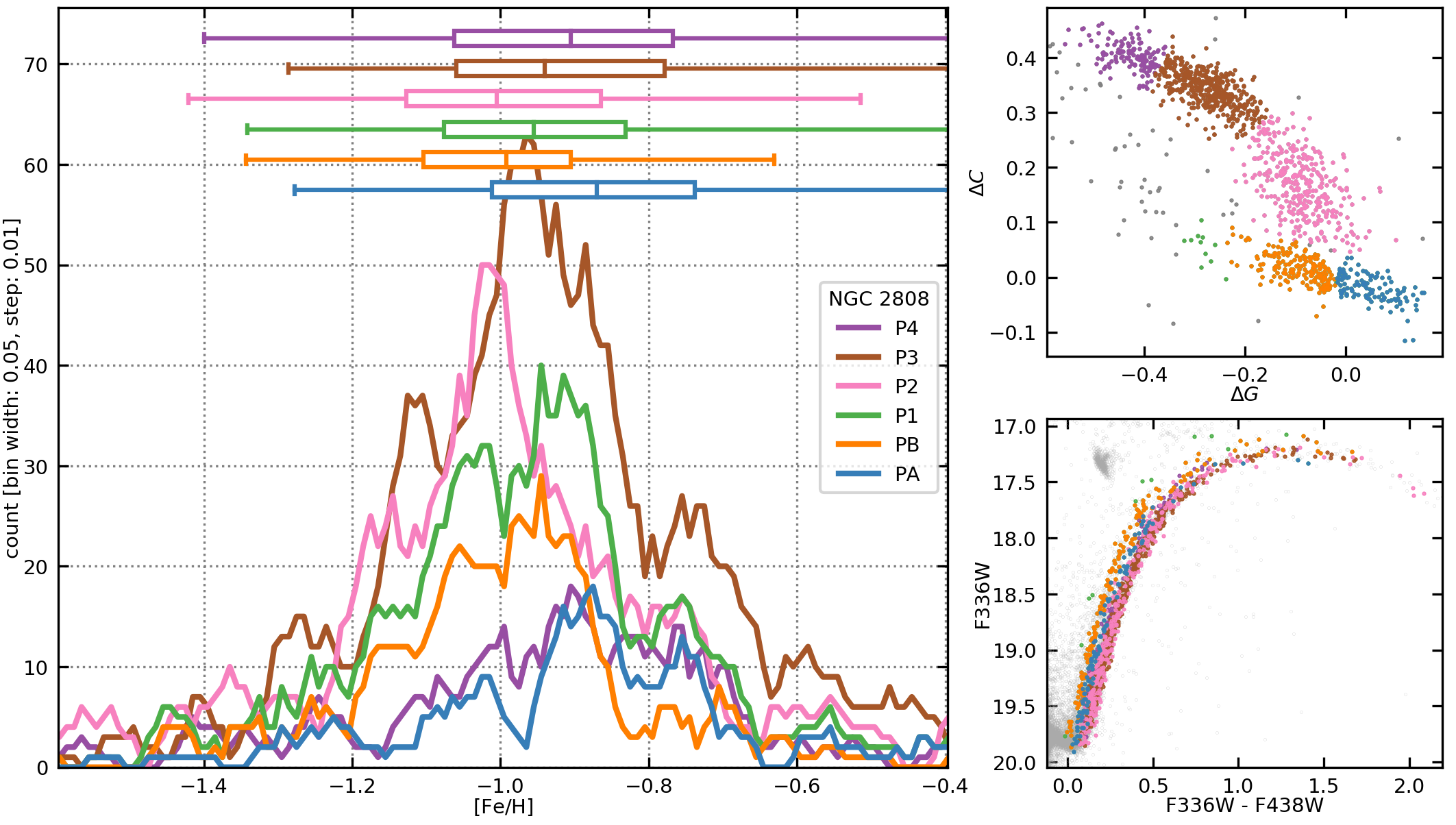}          
  \caption{Metallicity distributions for \object{NGC\,2808}. Note that population P1 is a combination of PA and PB.}
  \label{fig:multipop_ngc2808}
\end{figure*}

When using model spectra for deriving element abundances, an error in the determination of the effective temperature can cause variations in metallicity. Although we derive our results using a different method, we might also see a trend with temperature. However, we do not see any significant change in $\teff$ and $\logg$ (from our full-spectrum fits) with the pseudo-color $\Delta \mathrm{G}$, so the trends we see in Fig.~\ref{fig:uncertainties_feh} are probably not temperature related.
Although our results cannot give strong evidence on the variation of metallicity within the primordial populations of globular clusters, we cannot exclude this possibility. The special case of \object{NGC\,2808} will be discussed in more detail in Sect.~\ref{sec:ngc2808}.

\section{Individual clusters}
\label{sec:individual}
In this section, we discuss in detail the results for all individual clusters. While some of them show peculiarities and are therefore of interest in other regards, we will concentrate only on their metallicity distributions -- both for the whole cluster (mainly Fig.~\ref{fig:feh_dists}) and for its different populations (Figs.~\ref{fig:multipop_all} and \ref{fig:multipop}, and for some individual clusters), when available. 

\subsection{\object{NGC\,104} / \object{47~Tuc}}
The metal-rich, nearby, and well-studied globular cluster \object{NGC\,104} harbours two known populations. The Na-O anti-correlation has been observed, among others, by \citet{2009A&A...505..139C, 2013A&A...550A..34C} and \citet{2013A&A...549A..41G}. The presence of two populations has been shown photometrically by \citet{2012ApJ...744...58M}. Although \citet{2018MNRAS.476..496F} do not find a split in $\feh$, they report different values for the alpha element abundance $[\alpha/\mathrm{Fe}]$ for the two populations of $0.41$ and $0.23\dex$, respectively. With our method being based on CaT equivalent width, we are presumably biased by different alpha element abundances, so the difference in metallicity as visible in the 47\,Tuc panel of Fig.~\ref{fig:multipop_all} of $0.07\dex$ presumably corresponds to the split from the literature.

\subsection{\object{NGC\,362}}
Being classified as a Type~II cluster, \object{NGC\,362} shows a small P3 population which, unfortunately, in our sample only consists of 17 stars. The distribution is also very broad, but its median of $-0.99\,\mathrm{dex}$ differs significantly from that of P1 ($-1.09\,\mathrm{dex}$) and P2 ($-1.13\,\mathrm{dex}$).

\citet{2013A&A...557A.138C} found a split in the RGB of this cluster with a secondary sequence that consists of about $6\%$ of all RGB stars, which most likely corresponds to our population P3. They found an enrichment in Ba and probably all s-process elements.

\subsection{\object{NGC\,1851}}
\label{sec:ngc1851}
In this well-studied Type~II cluster, multiple RGBs have already been found by \citet{2009ApJ...695L..78L} and \citet{ 2009ApJ...707L.190H}. An actual split in the metallicity was suggested by \citet{2010ApJ...722L...1C} and estimated to be in the range of $0.06 \mbox{--} 0.08\,\mathrm{dex}$. \citet{2015ApJS..216...19L} found a split of $\sim 0.14\,\mathrm{dex}$. A split in the SGB was reported by \citet{2008ApJ...673..241M}, and they suggested that this could be explained by a difference of $0.2\dex$ in $\feh$, which, however, could be ruled out from the narrowness of the MS. On the other hand, \citet{2008ApJ...672L..29Y} found this to be consistent with a higher He content. Other studies like \citet{2012MSAIS..19..173M} and \citet{2010ApJ...722L..18V} did not find any variation in metallicity.

In our results, we see a separation in metallicity for \object{NGC\,1851} of $\sim 0.12\,\mathrm{dex}$ between stars in P3 and P1/P2. Furthermore, there is only a slight overlap of the $Q_1$--$Q_3$ intervals.

\begin{figure}
 \includegraphics{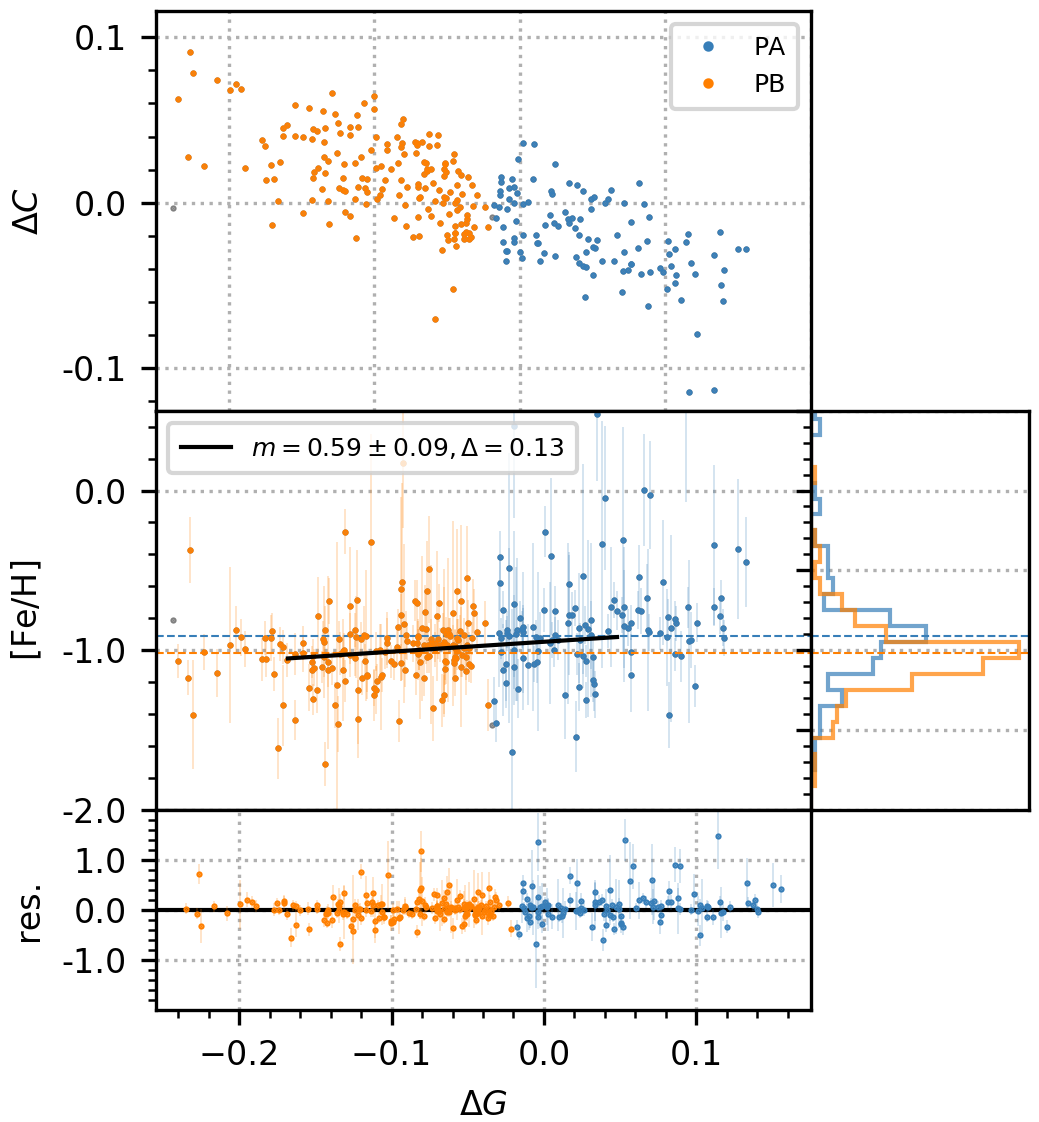}
 \caption{Splitting population P1 of \object{NGC\,2808} into two sub-populations A and B.}
 \label{fig:ngc2808_1g_var}
\end{figure}

\begin{table}
 \caption{Parameters of the metallicity distributions, for the whole clusters and single populations (given in 2nd column). Given are the number of stars, the medians $\Theta$, means $\mu$, and standard deviations $\sigma$ of the distributions, as well as the 1st and 3rd quartiles $Q_1$ and $Q_3$.}
 \begin{center}
 \begin{tabular}{lcrrrrrr}
 \hline \hline
   NGC  & P   & \#    & $\Theta$ & $\mu$ & $\sigma$ & $Q_1$ & $Q_3$ \\ \hline 
    104 &  -- &  2538 &    -0.57 &    -0.56 &     0.21 &    -0.65 &    -0.49 \\
    104 &  P1 &   340 &    -0.51 &    -0.50 &     0.17 &    -0.58 &    -0.42 \\
    104 &  P2 &  1270 &    -0.58 &    -0.56 &     0.18 &    -0.64 &    -0.51 \\
  \hline
    362 &  -- &  1144 &    -1.12 &    -1.12 &     0.21 &    -1.23 &    -1.03 \\
    362 &  P1 &   218 &    -1.09 &    -1.08 &     0.19 &    -1.17 &    -0.99 \\
    362 &  P2 &   579 &    -1.13 &    -1.12 &     0.18 &    -1.23 &    -1.04 \\
    362 &  P3 &    22 &    -0.99 &    -0.97 &     0.28 &    -1.10 &    -0.89 \\
  \hline
   1851 &  -- &  1358 &    -1.02 &    -0.99 &     0.32 &    -1.13 &    -0.90 \\
   1851 &  P1 &   184 &    -1.03 &    -0.99 &     0.41 &    -1.12 &    -0.95 \\
   1851 &  P2 &   353 &    -1.05 &    -1.04 &     0.23 &    -1.12 &    -0.98 \\
   1851 &  P3 &   265 &    -0.92 &    -0.89 &     0.22 &    -1.00 &    -0.80 \\
  \hline
  1904$^*$ &  -- &   213 &    -1.66 &    -1.64 &     0.20 &    -1.72 &    -1.59 \\
  \hline
   2808 &  -- &  2512 &    -0.98 &    -0.96 &     0.35 &    -1.13 &    -0.84 \\
   2808 &  P1 &   297 &    -0.95 &    -0.93 &     0.31 &    -1.08 &    -0.83 \\
   2808 &  P2 &   336 &    -1.00 &    -0.97 &     0.33 &    -1.13 &    -0.86 \\
   2808 &  P3 &   481 &    -0.94 &    -0.88 &     0.35 &    -1.06 &    -0.78 \\
   2808 &  P4 &   144 &    -0.91 &    -0.88 &     0.37 &    -1.06 &    -0.76 \\
   2808 &  PA &   114 &    -0.87 &    -0.83 &     0.36 &    -1.01 &    -0.73 \\
   2808 &  PB &   171 &    -0.99 &    -0.99 &     0.23 &    -1.10 &    -0.90 \\
  \hline
   3201 &  -- &   137 &    -1.43 &    -1.42 &     0.12 &    -1.50 &    -1.35 \\
   3201 &  P1 &    52 &    -1.42 &    -1.40 &     0.12 &    -1.46 &    -1.33 \\
   3201 &  P2 &    66 &    -1.44 &    -1.42 &     0.10 &    -1.48 &    -1.37 \\
  \hline
  5139$^*$ &  -- &  1247 &    -1.65 &    -1.50 &     0.45 &    -1.82 &    -1.31 \\
  5139$^*$ &  P1 &   174 &    -1.84 &    -1.83 &     0.08 &    -1.90 &    -1.79 \\
  5139$^*$ &  P2 &   216 &    -1.83 &    -1.80 &     0.15 &    -1.89 &    -1.76 \\
  5139$^*$ &  P3 &    96 &    -1.74 &    -1.72 &     0.12 &    -1.80 &    -1.67 \\
  5139$^*$ &  P4 &    87 &    -1.53 &    -1.50 &     0.16 &    -1.60 &    -1.40 \\
  5139$^*$ &  P5 &   127 &    -1.21 &    -1.24 &     0.19 &    -1.35 &    -1.09 \\
  5139$^*$ &  P6 &    59 &    -1.48 &    -1.47 &     0.18 &    -1.59 &    -1.35 \\
  5139$^*$ &  P7 &    28 &    -0.15 &    -0.18 &     0.19 &    -0.28 &    -0.03 \\
  5139$^*$ &  P8 &   144 &    -1.51 &    -1.50 &     0.22 &    -1.66 &    -1.39 \\
  5139$^*$ &  P9 &    78 &    -0.69 &    -0.72 &     0.39 &    -1.02 &    -0.45 \\
  \hline
   5286 &  -- &  1149 &    -1.76 &    -1.74 &     0.22 &    -1.86 &    -1.65 \\
   5286 &  P1 &   226 &    -1.79 &    -1.77 &     0.14 &    -1.86 &    -1.69 \\
   5286 &  P2 &   332 &    -1.76 &    -1.75 &     0.15 &    -1.83 &    -1.68 \\
   5286 &  P3 &   104 &    -1.57 &    -1.53 &     0.27 &    -1.66 &    -1.48 \\
  \hline
   5904 &  -- &   863 &    -1.17 &    -1.16 &     0.20 &    -1.26 &    -1.08 \\
   5904 &  P1 &   167 &    -1.14 &    -1.14 &     0.14 &    -1.23 &    -1.06 \\
   5904 &  P2 &   506 &    -1.18 &    -1.17 &     0.15 &    -1.25 &    -1.11 \\
  \hline
   6093 &  -- &  1071 &    -1.81 &    -1.78 &     0.20 &    -1.89 &    -1.73 \\
   6093 &  P1 &   269 &    -1.81 &    -1.80 &     0.14 &    -1.88 &    -1.74 \\
   6093 &  P2 &   437 &    -1.80 &    -1.77 &     0.19 &    -1.87 &    -1.73 \\
  \hline
  6218$^*$ &  -- &   236 &    -1.25 &    -1.26 &     0.10 &    -1.31 &    -1.21 \\
  6218$^*$ &  P1 &    83 &    -1.24 &    -1.23 &     0.13 &    -1.31 &    -1.20 \\
  6218$^*$ &  P2 &   120 &    -1.26 &    -1.27 &     0.07 &    -1.31 &    -1.23 \\
  \hline
   6254 &  -- &   396 &    -1.56 &    -1.54 &     0.18 &    -1.64 &    -1.47 \\
   6254 &  P1 &   109 &    -1.52 &    -1.51 &     0.17 &    -1.63 &    -1.42 \\
   6254 &  P2 &   178 &    -1.57 &    -1.54 &     0.14 &    -1.63 &    -1.49 \\
  \hline
  6266$^*$ &  -- &  2182 &    -0.96 &    -0.96 &     0.25 &    -1.07 &    -0.85 \\
  \hline
  6293$^*$ &  -- &   168 &    -2.17 &    -2.15 &     0.12 &    -2.23 &    -2.10 \\
  \hline
   6388 &  -- &  4098 &    -0.48 &    -0.43 &     0.48 &    -0.69 &    -0.24 \\
   6388 &  P1 &   579 &    -0.50 &    -0.45 &     0.45 &    -0.68 &    -0.31 \\
   6388 &  P2 &  1203 &    -0.51 &    -0.44 &     0.42 &    -0.67 &    -0.28 \\
   6388 &  P3 &   411 &    -0.28 &    -0.25 &     0.39 &    -0.45 &    -0.13 \\
 \hline
 \end{tabular}
 \end{center} 
 \label{table:results}
\end{table}

\renewcommand{\thetable}{\arabic{table} (Cont.)}
\addtocounter{table}{-1}
\begin{table}
 \caption{Parameters of the metallicity distributions.}
 \begin{center}
 \begin{tabular}{lcrrrrrr}
 \hline \hline
    NGC  & P   & \#    & $\Theta$ & $\mu$ & $\sigma$ & $Q_1$ & $Q_3$ \\ \hline 
   6441 &  -- &  4408 &    -0.53 &    -0.46 &     0.48 &    -0.71 &    -0.32 \\
   6441 &  P1 &   826 &    -0.52 &    -0.46 &     0.42 &    -0.65 &    -0.35 \\
   6441 &  P2 &  1546 &    -0.53 &    -0.48 &     0.40 &    -0.68 &    -0.36 \\
  \hline
  6522$^*$ &  -- &   481 &    -1.10 &    -1.07 &     0.36 &    -1.25 &    -0.93 \\
  \hline
   6541 &  -- &   820 &    -1.82 &    -1.81 &     0.11 &    -1.87 &    -1.76 \\
   6541 &  P1 &   274 &    -1.81 &    -1.80 &     0.09 &    -1.85 &    -1.75 \\
   6541 &  P2 &   396 &    -1.81 &    -1.80 &     0.10 &    -1.86 &    -1.76 \\
  \hline
   6624 &  -- &   539 &    -0.48 &    -0.44 &     0.39 &    -0.62 &    -0.32 \\
   6624 &  P1 &   119 &    -0.39 &    -0.34 &     0.33 &    -0.56 &    -0.20 \\
   6624 &  P2 &   286 &    -0.52 &    -0.50 &     0.29 &    -0.66 &    -0.39 \\
  \hline
   6656 &  -- &   397 &    -1.81 &    -1.78 &     0.20 &    -1.89 &    -1.68 \\
   6656 &  P1 &   107 &    -1.87 &    -1.87 &     0.09 &    -1.93 &    -1.82 \\
   6656 &  P2 &   119 &    -1.85 &    -1.84 &     0.11 &    -1.92 &    -1.78 \\
   6656 &  P3 &   116 &    -1.65 &    -1.65 &     0.13 &    -1.75 &    -1.57 \\
  \hline
   6681 &  -- &   325 &    -1.47 &    -1.44 &     0.27 &    -1.56 &    -1.38 \\
   6681 &  P1 &    50 &    -1.45 &    -1.36 &     0.37 &    -1.55 &    -1.34 \\
   6681 &  P2 &   208 &    -1.47 &    -1.45 &     0.24 &    -1.55 &    -1.38 \\
  \hline
   6752 &  -- &   539 &    -1.54 &    -1.53 &     0.12 &    -1.61 &    -1.48 \\
  6752$^*$ &  P1 &   114 &    -1.52 &    -1.52 &     0.10 &    -1.59 &    -1.47 \\
  6752$^*$ &  P2 &   264 &    -1.53 &    -1.51 &     0.12 &    -1.58 &    -1.46 \\
  \hline
   7078 &  -- &  1318 &    -2.25 &    -2.24 &     0.10 &    -2.30 &    -2.20 \\
   7078 &  P1 &   331 &    -2.28 &    -2.27 &     0.08 &    -2.32 &    -2.23 \\
   7078 &  P2 &   259 &    -2.26 &    -2.26 &     0.09 &    -2.31 &    -2.22 \\
   7078 &  P3 &   292 &    -2.24 &    -2.23 &     0.07 &    -2.27 &    -2.19 \\
  \hline
   7089 &  -- &  1727 &    -1.59 &    -1.57 &     0.19 &    -1.67 &    -1.51 \\
   7089 &  P1 &   238 &    -1.59 &    -1.57 &     0.15 &    -1.66 &    -1.50 \\
   7089 &  P2 &   930 &    -1.59 &    -1.58 &     0.15 &    -1.66 &    -1.52 \\
   7089 &  P3 &    30 &    -1.42 &    -1.34 &     0.25 &    -1.50 &    -1.18 \\
  \hline
   7099 &  -- &   289 &    -2.20 &    -2.18 &     0.09 &    -2.23 &    -2.16 \\
  \hline
 \end{tabular}
 \end{center} 
\end{table}
\renewcommand{\thetable}{\arabic{table}}

\subsection{\object{NGC\,1904} / M\,79}
The metallicity distribution for \object{NGC\,1904} shows no spread in metallicity, and without UV photometry we cannot create chromosome maps and investigate the different populations of this cluster. No anomalous metallicity distribution could be found in the literature.

\subsection{\object{NGC\,2808}}
\label{sec:ngc2808}
For the Type~I cluster \object{NGC\,2808}, \citet{2015ApJ...808...51M} reported five different populations from an analysis of the HUGS photometry, and, assuming a constant metallicity \citep[see][]{2006A&A...450..523C}, found four of those populations (our P2-P4) to be enhanced in He when compared to the primordial population (our P1). According to \citet{2011A&A...534A...9S} and \citet{2018A&A...616A.168L}, a change in He also produces a change in luminosity and effective temperature. 

\begin{figure*}
  \includegraphics{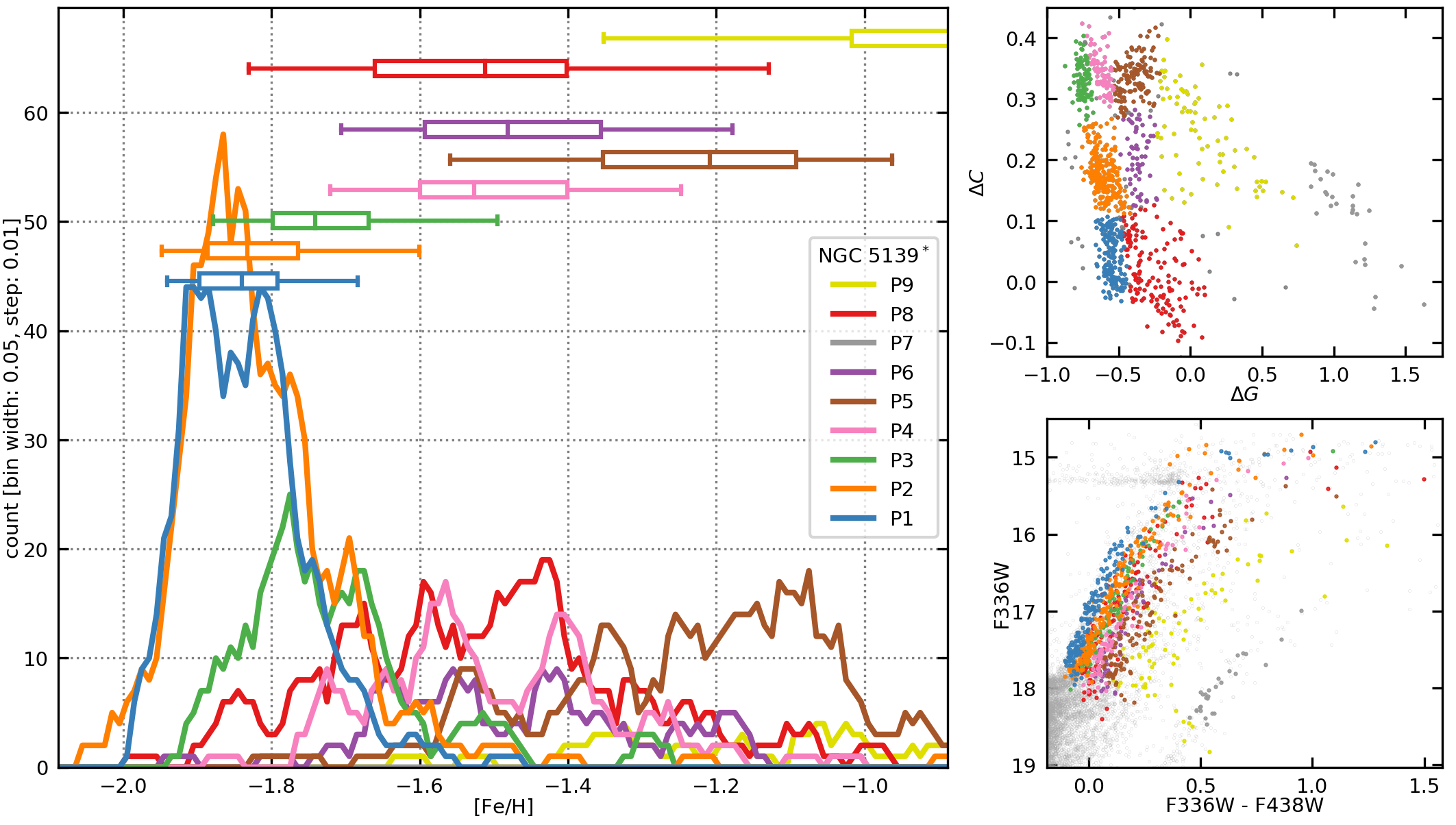}          
  \caption{Metallicity distributions for the type~II cluster \object{NGC\,5139}. Note that the median values for populations P7 and P9 are outside the plotted metallicity range.}
  \label{fig:multipop_ngc5139}
\end{figure*}
\begin{figure}
  \includegraphics{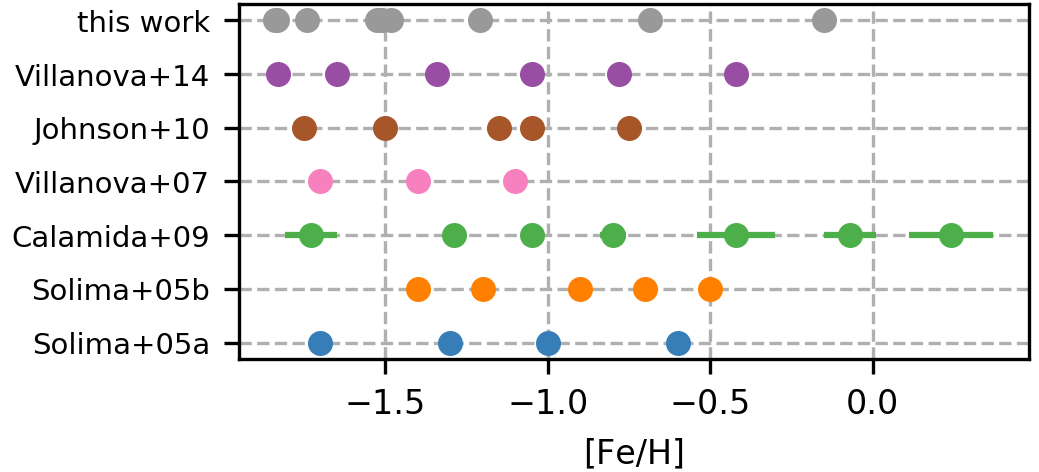}          
  \caption{Comparison of metallicities for different populations in \object{NGC\,5139} as reported in the literature with the results from this work.}
  \label{fig:ngc5139_lit}
\end{figure}

We applied the same grouping in the chromosome map of \object{NGC\,2808} into five populations and see no significant split in metallicity (see Fig.~\ref{fig:multipop_ngc2808}). However, the metallicity seems to be increasing from P2 to P4, that is with decreasing $\Delta \mathrm{G}$. This trend is opposite to what we see in Type~II clusters where metallicity increases with increasing $\Delta \mathrm{G}$. In Figure~\ref{fig:chm_trends}, we showed that at least for Type~II cluster the metallicity also increases with $\Delta \mathrm{C}$, so we might see the same effect in \object{NGC\,2808}.

For \object{NGC\,2808}, we investigate in more detail the metallicity variations in the P1 populations as discussed in Sect.~\ref{sec:1g_variations}. The chromosome map for the two sub-populations in P1, called PA and PB, is shown in Fig.~\ref{fig:ngc2808_1g_var}. When looking at the metallicity histograms for the two populations in the small panel on the right hand side, we see a little shift of about $0.12$ in the medians and even $0.16\dex$ in means. However, the middle panel shows a more continuous trend, as was already suggested from Fig.~\ref{fig:1g_variations} for all the other clusters. The black line shows a linear fit to the data and the residuals are shown in the lower panel, indicating that what we see is rather a continuous trend than an actual split.
We present a more in-depth analysis of the multiple populations chemistry of \object{NGC\,2808} in \citet{2019A&A...631A..14L}.

\subsection{\object{NGC\,3201}}
\object{NGC\,3201} is a halo cluster, for which \citet{2013ApJ...764L...7S} found an unusual intrinsic spread in iron abundance of $0.4\dex$. \citet{2015ApJ...801...69M} obtained the same result, but only when deriving the abundance from \ion{Fe}{i} lines.  For \ion{Fe}{ii} they reported no spread, so they argue that this is caused by NLTE effects driven by iron overionization. \citet{2013ApJ...764L...7S} also detected a metal-poor tail, although containing only 5 stars. In our results, we see neither a spread in metallicity nor a metal-poor tail. If at all, we see some stars with an excess metallicity.
The binary content of multiple populations in \object{NGC\,3201} is investigated in detail in \citet{2019arXiv191201627K}. 

\subsection{\object{NGC\,5139} / \object{$\omega$~Centauri}}
\label{sec:ngc5139}
For the peculiar cluster \object{$\omega$~Centauri}, a bimodal distribution of metallicities has been known for a long time \citep{1985ApJ...295..437H} and has been quantified by \citet{1996ApJ...462..241N} using calcium abundances, giving $\mathrm{[Ca/H]}=-1.4\dex$ for one and $-0.9$ for the other population. This bimodality has been confirmed later photometrically using HST \citep{1997PhDT.........8A,2004ApJ...605L.125B}, showing a split all along its CMD, from the MS to the RGB.

On the sub-giant branch (SGB), \citet{2005ApJ...634..332S} found four populations with $\feh=-1.7\dex$, $-1.3$, $-1.0$ (all with [$\alpha$/Fe$]=+0.3$), and $-0.6$ (with [$\alpha$/Fe$]=+0.1$) using CaT abundances. \citet{2007ApJ...663..296V} identified four populations using GIRAFFE spectra: two old populations with $-1.7$ and $-1.1\dex$, and two 1-2\,Gyrs younger populations with $-1.7$ and $-1.4\dex$, respectively. Six different SGBs have been identified by \citet{2014ApJ...791..107V}, with $\feh=-1.83$, $-1.65$, $-1.34$, $-1.05$, $-0.78$, and $-0.42$.

\citet{2005MNRAS.357..265S} found four different populations on the RGB using FORS1 photometry at the VLT and derived metallicities photometrically using the color distribution. The metallicities they obtained were $\feh=-1.4\dex$, $-1.2$, $-0.9$, $-0.7$, and $-0.5$, respectively. Str\"omgren photometry was used by \citet{2009ApJ...706.1277C} to find four major peaks in the metallicity distribution at $\feh=-1.73\dex$, $-1.29$, $-1.05$, and $-0.80$, and three minor ones at $-0.42$, $-0.07$, and $+0.24\dex$. High resolution spectroscopy of 855 red giants was obtained by \citet{2010ApJ...722.1373J}, who found five peaks in their metallicity distribution at $\feh \approx -1.75$, $-1.50$, $-1.15$, $-1.05$, and $-0.75$. 

\object{NGC\,5139} is one of our clusters without a $V-\vhb$ magnitude, so we rely on the luminosity calibration as presented in Sect.~\ref{sec:extending_below}. Due to the complex structure of \object{$\omega$~Centauri}, its $\sumew$-luminosity diagram shows a large spread (see Fig.~\ref{fig:calib_rew_lum}) and we expect some offset in our metallicities. In the metallicity calibration itself it is offset from the model by $\sim 0.11\dex$ towards lower metallicities, so we considered this as a systematic error. Furthermore, the reported variations in [Ca/Fe] and [$\alpha$/Fe] will have an effect on our results, presumably causing another systematic error.

Using chromosome maps created from HST UV photometry \citet{2017ApJ...844..164B} found at least 15 different populations, of which we identified nine, as shown in Fig.~\ref{fig:multipop_ngc5139}. We derived significantly different mean metallicities for most of these populations, which are all listed in Table~\ref{table:results}. In order to compare our results with the previously discussed literature values, they are all plotted in Fig.~\ref{fig:ngc5139_lit}. Note that our three most metal-rich clusters have metallicities of $-0.57$ (\object{NGC\,6388}), $-0.49$ (\object{NGC\,6441}), and $-0.36$ (\object{NGC\,6624}), so our CaT-metallicity is only valid up about these values. Therefore, the most metal-rich populations in \object{NGC\,5139} are either outside this limit (P7), or very close to it (P9), and must be treated with care. One of these, namely P7, is part of the bimodality that has been known for decades. Due to the limitation of our calibration at high metallicity, and the fact that Ca is enhanced in P7, our metallicity value is higher than expected.

For the more metal-poor populations, the comparison with literature values is better, although we did not try to match individual populations to those from the literature. The metallicity for our lowest-metallicity populations P1 and P2 is a little too low compared to all literature values except \citet{2014ApJ...791..107V}. The intermediate metal-rich populations all have a matching population in at least one previous study. But obviously, not even those agree well with each other.
Comparing the metallicity distributions with the chromosome map, which is also shown in Fig.~\ref{fig:ngc1851_dist}, we see that the metallicity increases steadily both with $\Delta \mathrm{G}$ and $\Delta \mathrm{C}$, as discussed in Sect.~\ref{sec:typeI_II} and shown in Fig.~\ref{fig:chm_trends}.

\subsection{\object{NGC\,5286}}
The poorly studied cluster \object{NGC\,5286} has been classified as Type\,II by \citet{2017MNRAS.464.3636M}. Three sub-populations have been found based on a CN index by \citet{2017ApJ...844...14L}, which they also group into two populations with different calcium HK' strengths that also differ in abundances of Fe and s-process elements. \citet{2015MNRAS.450..815M} called the cluster anomalous and found two populations with a metallicity split of $0.17\dex$. In our data we also see a clear split in metallicity, with $-1.72$ and $-1.71\dex$ for the populations P1 and P2, respectively, and $-1.60\dex$ for population P3.

\subsection{\object{NGC\,5904} / \object{M\,5}}
\citet{2017ApJ...844...77L} found bimodal CN and [N/Fe] distributions in \object{NGC\,5904} and \citet{2009A&A...505..117C} also confirmed the existence of the well-known Na-O anticorrelation. They found it to be homogeneous in [Fe/H] at a level below 6\%, so we do not expect to see any split. 

\subsection{\object{NGC\,6093} / M\,80}
Even in the title of their paper, \citet{2015A&A...578A.116C} call \object{NGC\,6093} a cluster with a ``normal chemistry'', which we can confirm with the inconspicuous, Gaussian-shaped metallicity distribution derived from our results.

\subsection{\object{NGC\,6218} / \object{M\,12}}
\citet{2007A&A...464..939C} found no star-to-star scatter of metallicity in \object{NGC\,6218}, which we can confirm from our results, both with the overall distribution as well as with the non-existence of any separation of metallicities between the two populations (see Fig.~\ref{fig:multipop_all}).

\begin{figure*}
  \includegraphics{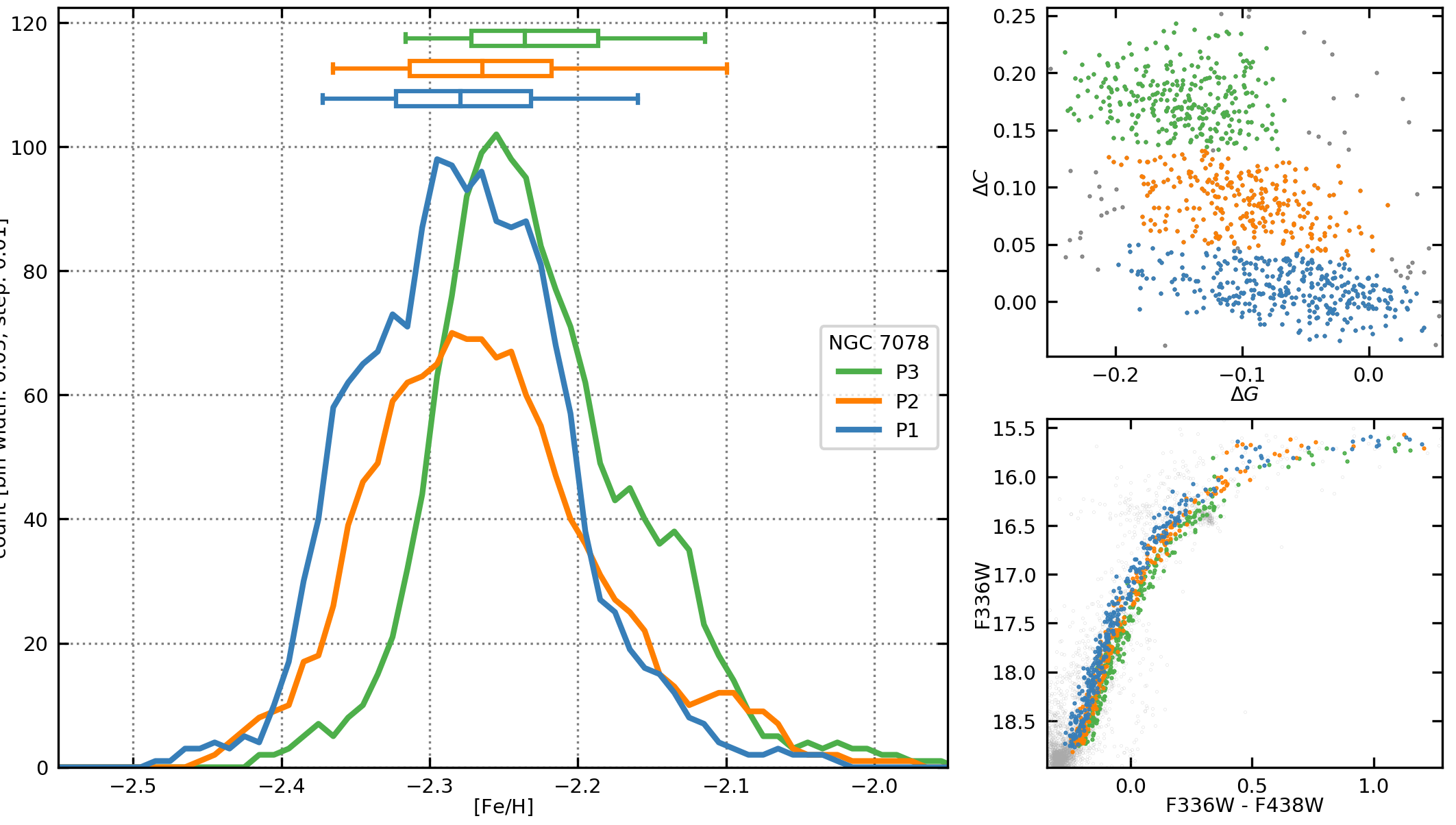}          
  \caption{Metallicity distributions for \object{NGC\,7078}.}
  \label{fig:multipop_ngc7078}
\end{figure*}

\subsection{\object{NGC\,6254} / \object{M\,10}}
\object{NGC\,6254} shows no sign of any metallicity spread and no significant separation can be found for its two populations, which is consistent with no report of abnormalities in the literature.

\subsection{\object{NGC\,6266} / \object{M\,62}}
For \object{NGC\,6266}, we see a very broad metallicity distribution, similar to that of \object{NGC\,104} or \object{NGC\,1851}. Also the typical uncertainties are similar, although the distribution for \object{NGC\,6266} extends to higher values. Unfortunately, due to missing UV photometry we could not create a chromosome map, so we were unable to investigate, whether this is caused by a split in metallicity between populations. However, \citet{2014MNRAS.439.2638Y} found no evidence for a dispersion in metallicity but with data for only seven bright giants.

\subsection{\object{NGC\,6293}}
With only 168 stars, our sample for \object{NGC\,6293} is very small, which is reflected in the metallicity distribution in Fig.~\ref{fig:feh_dists}. There may be a second peak in metallicity at $\feh - \left< \feh \right> \approx +0.2\dex$. No chromosome map is available for this cluster.

\subsection{\object{NGC\,6388}}
\begin{figure*}
  \centering
  \includegraphics{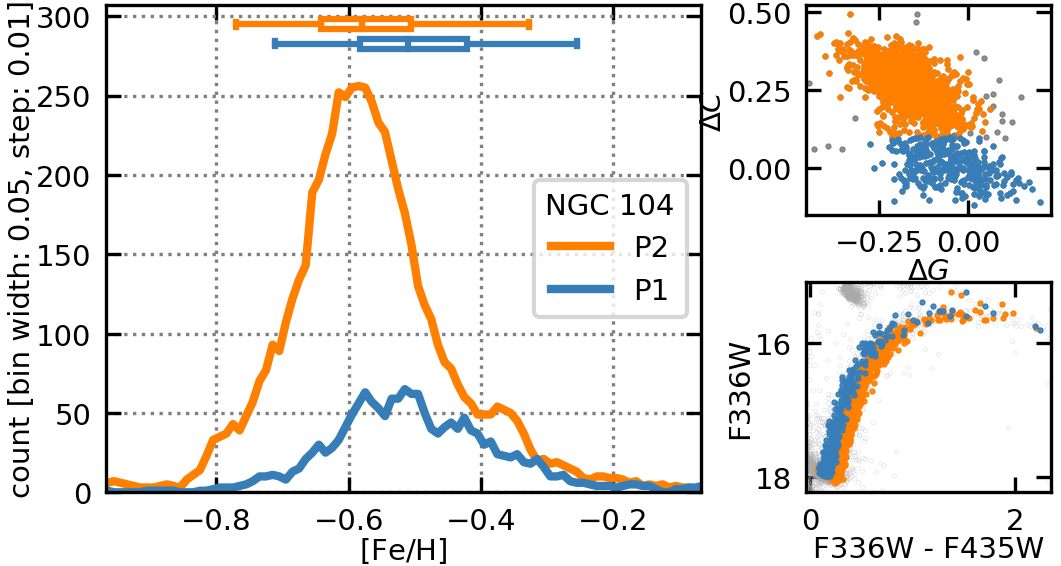}
  \includegraphics{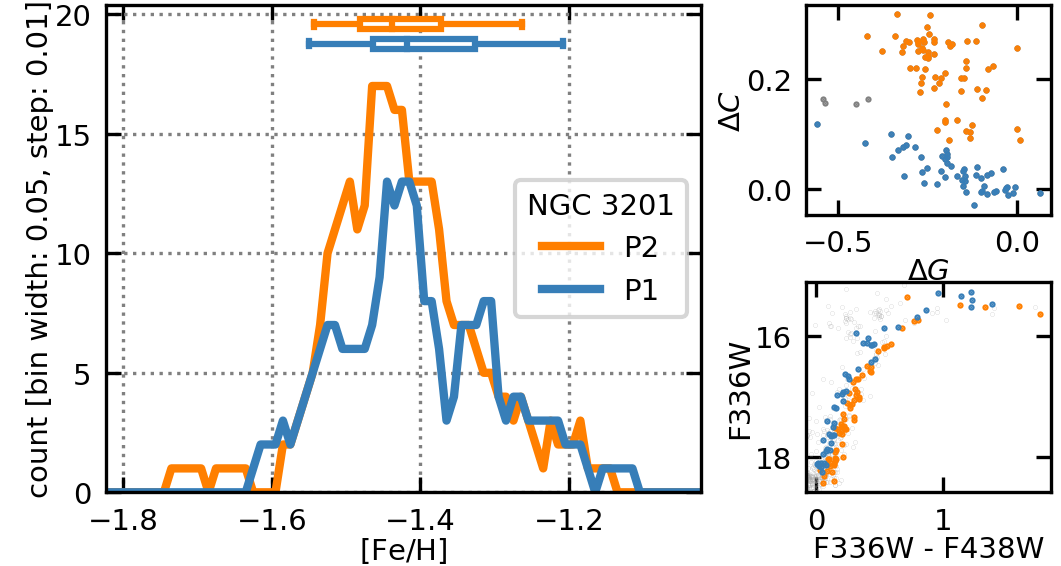} \\
  \includegraphics{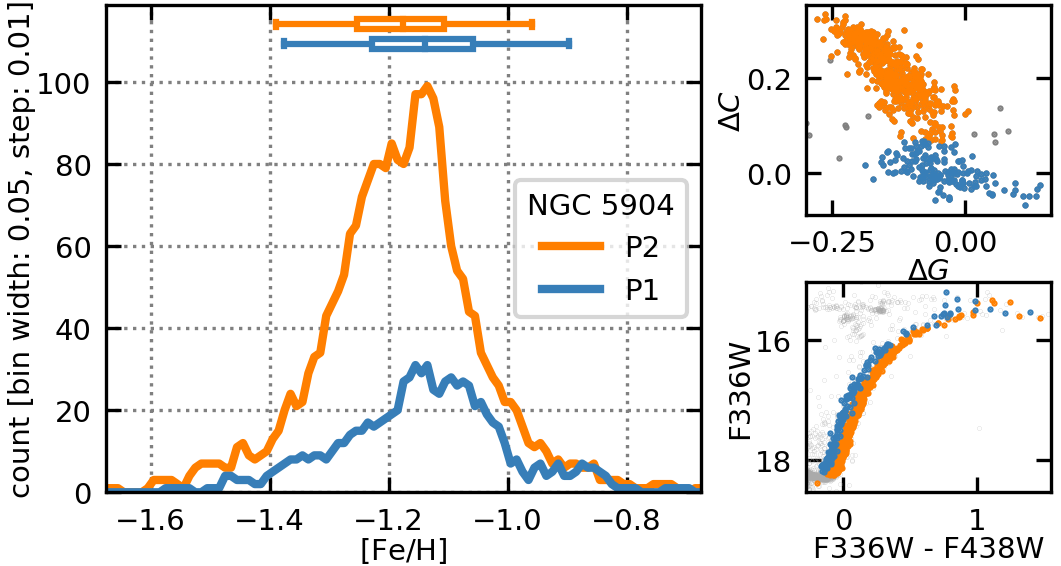} 
  \includegraphics{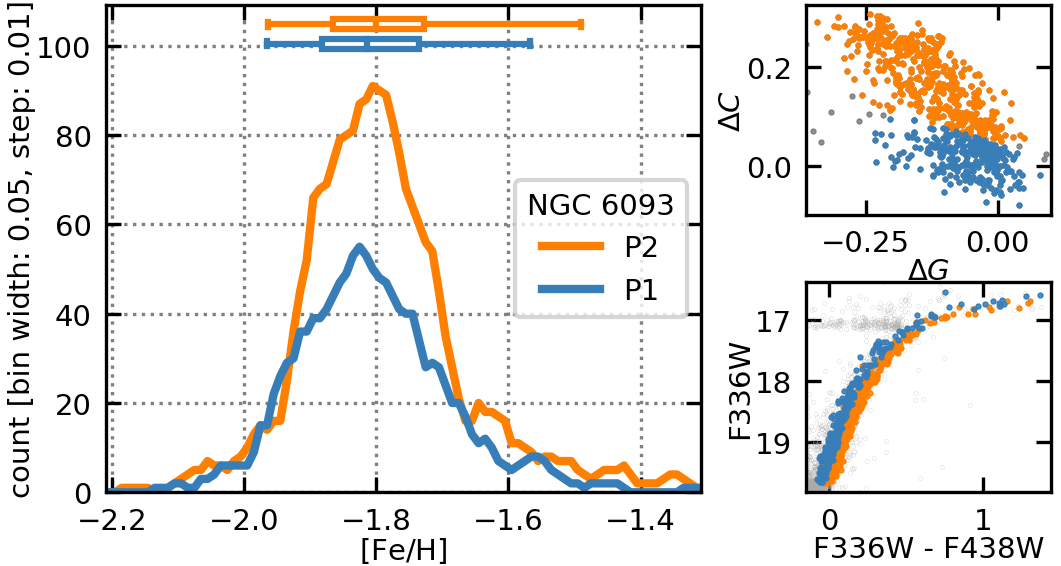} \\
  \includegraphics{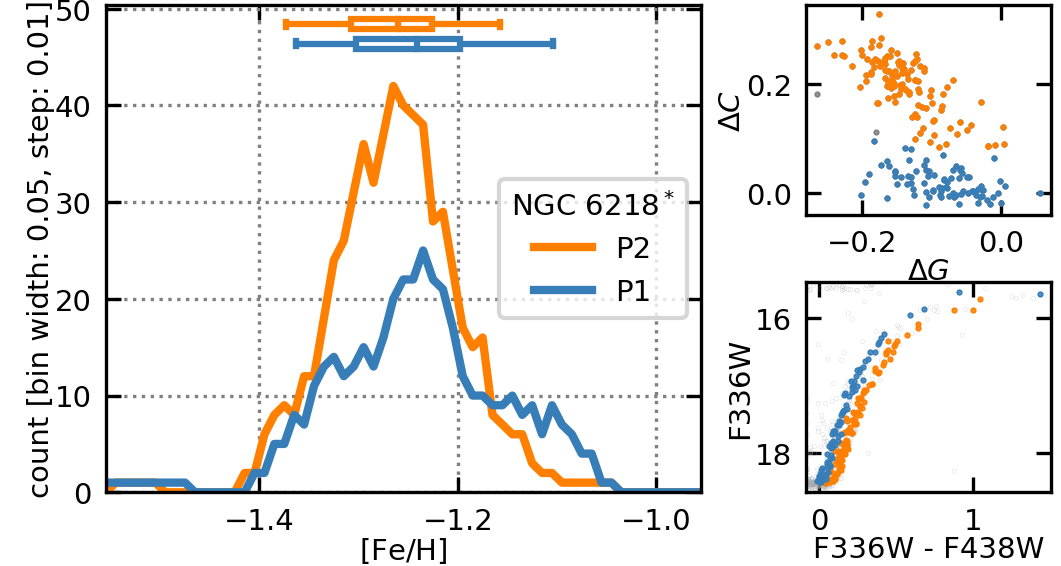} 
  \includegraphics{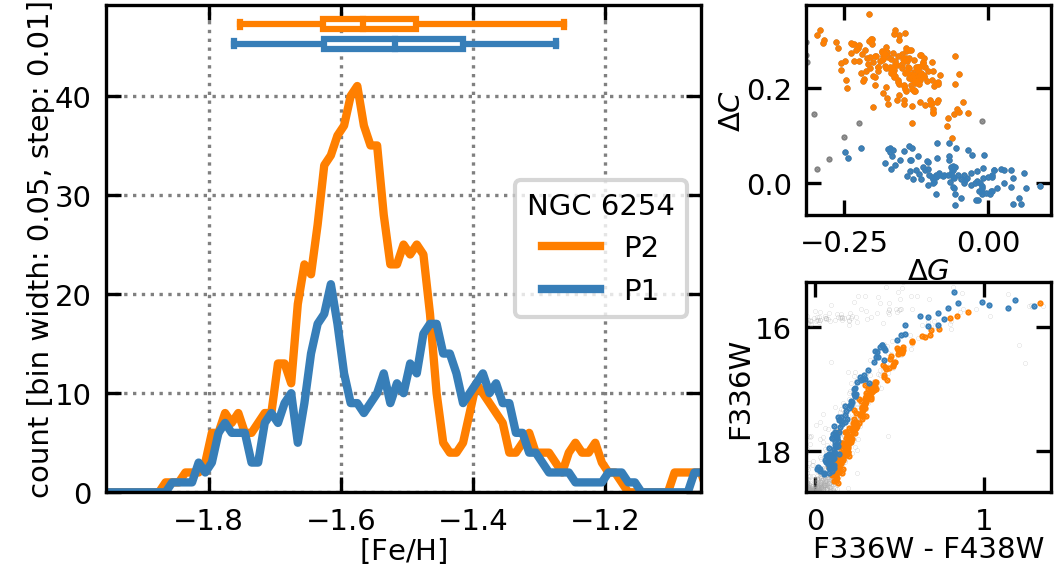} \\
  \includegraphics{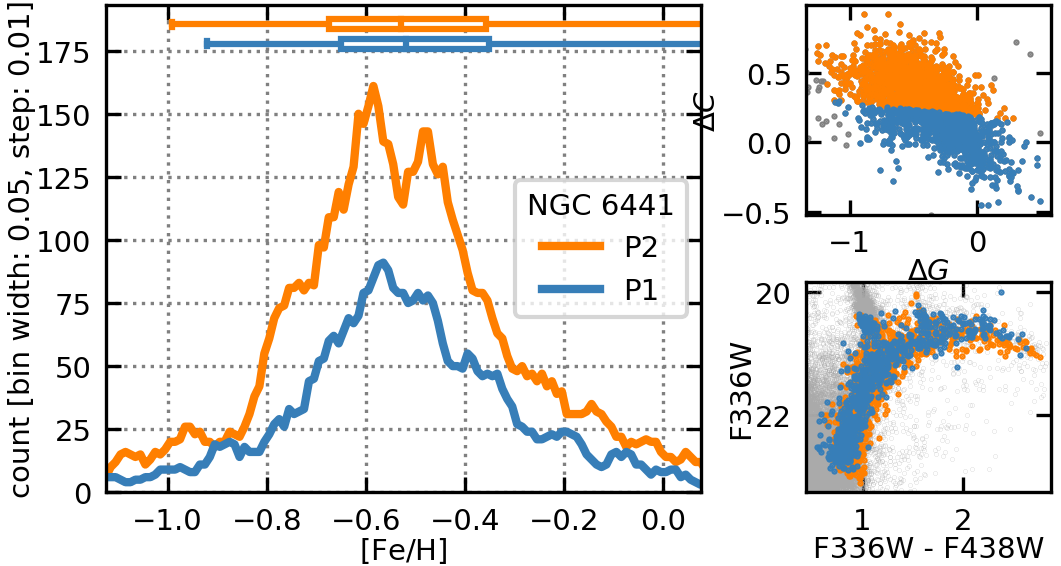} 
  \includegraphics{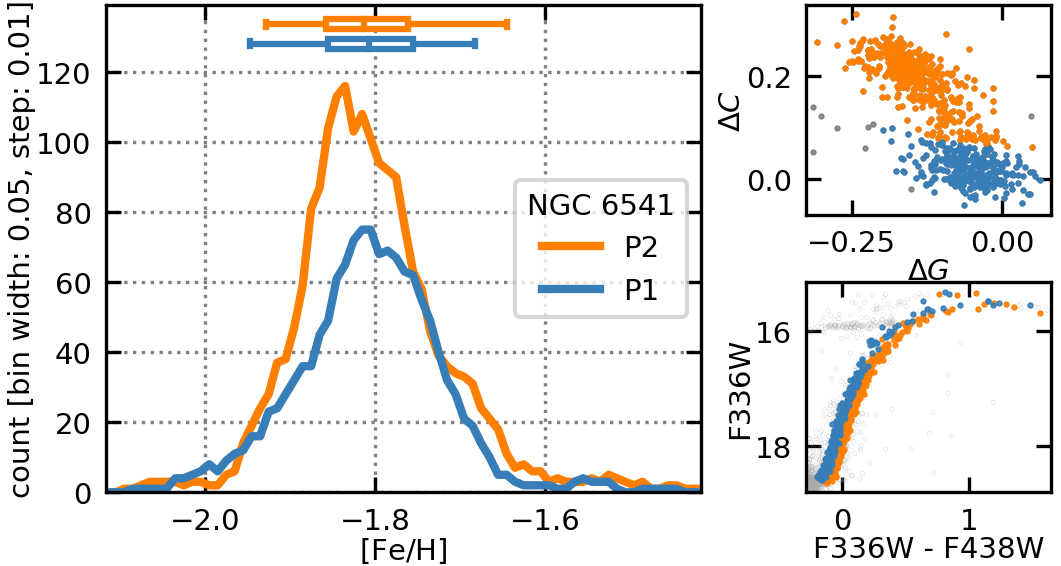} \\
  \includegraphics{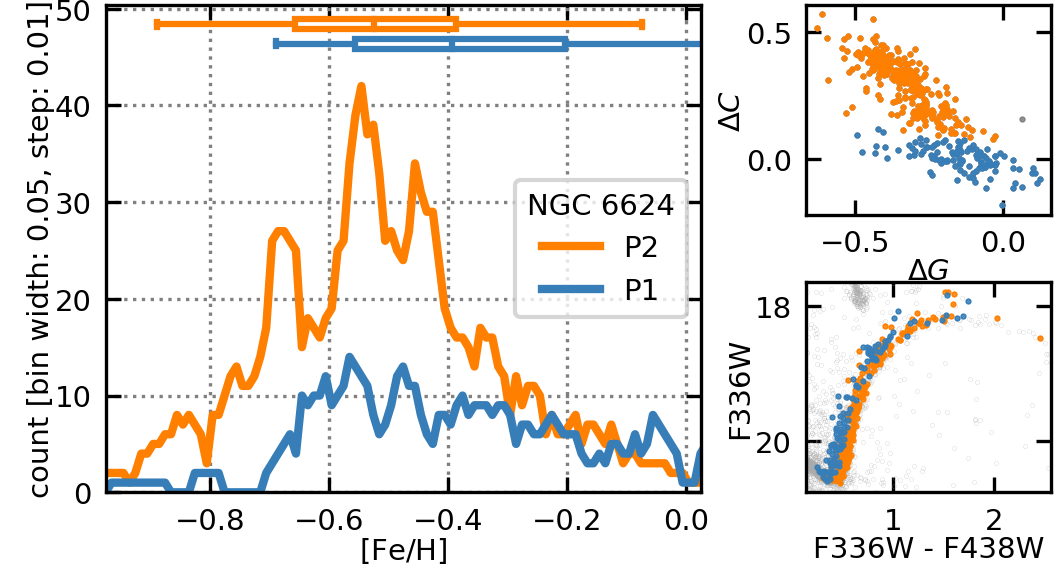} 
  \includegraphics{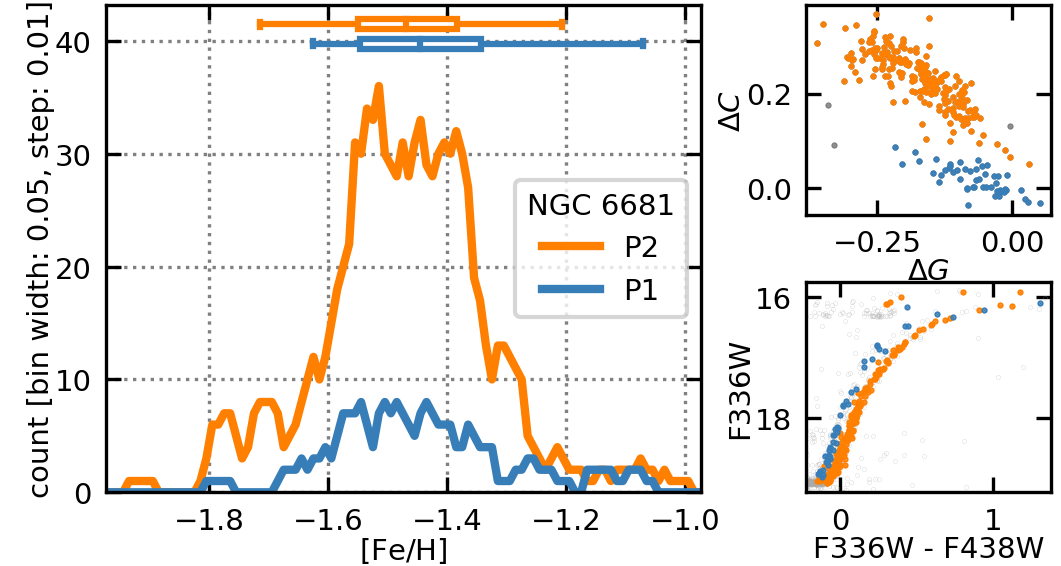} 
  \caption{Metallicity distributions for the different stellar populations in our Type~I clusters, equivalent to Fig.~\ref{fig:multipop}.}
  \label{fig:multipop_all}
\end{figure*}

Similarly to \object{NGC\,6441}, the analysis of the results for \object{NGC\,6388} is a bit cumbersome due to the large uncertainties we get for the metallicities, yielding a very broad and asymmetric distribution of metallicities. Furthermore, as a result of the uncertain photometry (see discussion in Sect.~\ref{sec:observations}) the chromosome map for this cluster is a little messy, so separating populations becomes difficult. Nevertheless, as shown in Fig.~\ref{fig:multipop_all}, we still see a significant split of about $0.22\dex$ in metallicity between the populations P1/P2 and P3, as expected for a Type~II cluster. However, \citet{2018A&A...614A.109C} excluded the existence of any intrinsic Fe dispersion in \object{NGC\,6388}.

\subsection{\object{NGC\,6441}}
While \object{NGC\,6388} and \object{NGC\,6441} are pretty similar with both being massive metal-rich bulge clusters, \citet{2017MNRAS.464.3636M} classify the former as Type~II and the latter as Type~I cluster. Both also share the problematic photometry, so the quality of the spectra seems to be degraded from the extraction process (see Sect.~\ref{sec:observations}), affecting the metallicity distribution we obtain. However, lacking an obvious P3 population in the chromosome map, we can only confirm the status of \object{NGC\,6441} as a simple Type~I.

\renewcommand{\thefigure}{\arabic{figure} (Cont.)}
\addtocounter{figure}{-1}
\begin{figure*}
  \centering
  \includegraphics{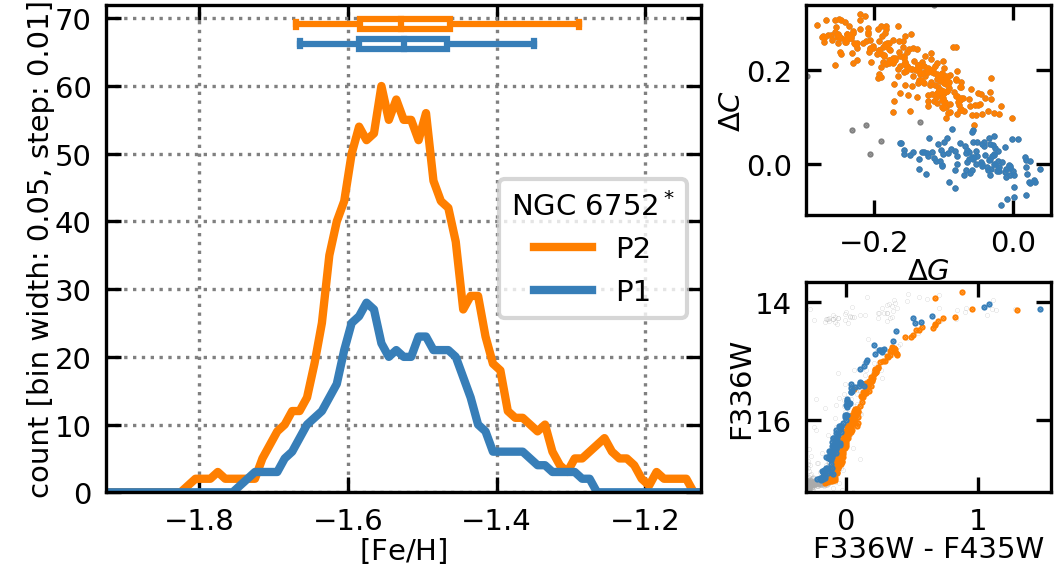}
  \includegraphics{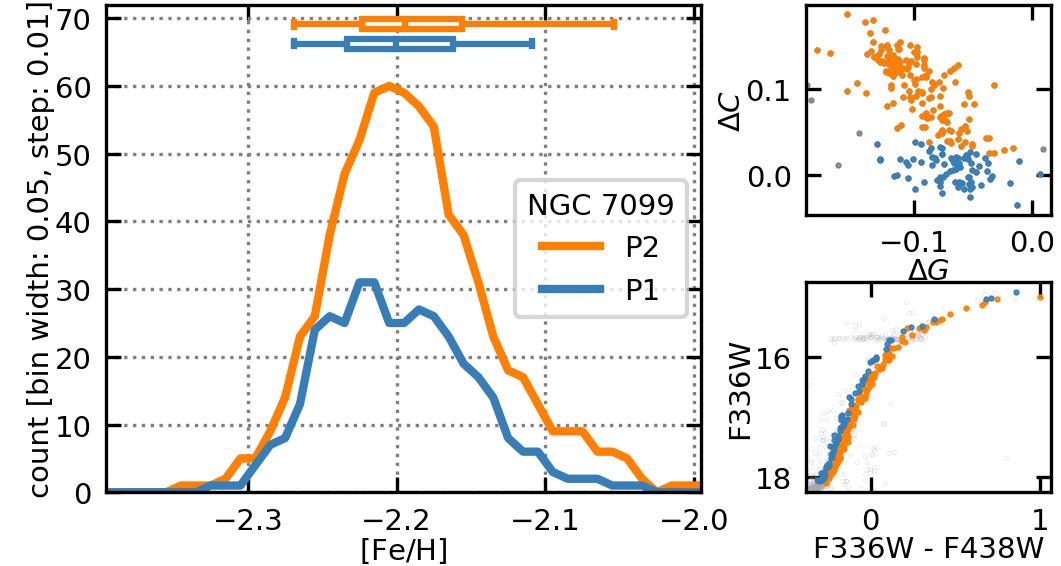}
  \caption{Metallicity distributions.}
\end{figure*}
\renewcommand{\thefigure}{\arabic{figure}}

\subsection{\object{NGC\,6522}}
Our sample of stars in the oldest known globular cluster in the Milky Way is very small, but the derived metallicity distribution does not show any abnormalities. No chromosome map is available for this cluster.

\subsection{\object{NGC\,6541}}
The metallicity distribution of the poorly studied cluster \object{NGC\,6541} shows no sign of any metallicity spread and no variation in metallicity between its two populations.

\subsection{\object{NGC\,6624}}
\object{NGC\,6624} is one of the clusters for which we only have a very small sample of RGB stars and obtain relatively large uncertainties for the derived metallicities. Nevertheless, we might see a bimodal distribution for its two populations in Fig.~\ref{fig:multipop}, although with both being very broad.

\subsection{\object{NGC\,6656} / \object{\object{M\,22}}}
\citet{2014A&A...563A..76M} found a split in $\feh$ for this Type~II cluster using CaT spectroscopy -- although the separation they found is quite large with $0.3 \mbox{--} 0.4\,\mathrm{dex}$. A separation of populations was also observed in Ca \emph{by} photometry by \citet{2015ApJS..219....7L}, where they suggest a merger scenario. \citet{2012MSAIS..19..173M} found a bimodality of metallicities with $\mathrm{\feh}_\mathrm{rich}=(-1.68\pm0.02)\,\mathrm{dex}$ and $\mathrm{\feh}_\mathrm{poor}=(-1.82\pm0.02)\,\mathrm{dex}$, i.e.\ $\sigma_{\feh}=0.14\,\mathrm{dex}$. An abundance difference of $(0.15\pm0.02)\dex$ was reported by \citet{2015MNRAS.450..815M}. \citet{2011A&A...532A...8M} also found a difference of about 0.1 dex in Ca between their $s$-rich and $s$-poor stars. From our data we get a separation of $\sim 0.2\,\mathrm{dex}$, which might be overestimated if the populations indeed have a different Ca abundance.

\subsection{\object{NGC\,6681} / \object{M\,70}}
\citet{2017ApJ...846...23O} confirmed the existence of the Na-O anti-correlation and found no sign of an intrinsic metallicity dispersion. Although having only few stars in our sample, their results are of good quality and we also see no sign of any spread in metallicity.

\subsection{\object{NGC\,6752}}
Although \citet{2013ApJ...767..120M} identified three different populations along the whole evolutionary sequence in \object{NGC\,6752}, they could only find two of them using their chromosome maps in \citet{2017MNRAS.464.3636M}, which we adopted for our analysis. Two stellar populations have also been reported by \citet{2015MNRAS.446.3319Y} and \citet{2018ApJS..238...24L}, based on their different C+N+O content, while \citet{2013MNRAS.434.3542Y} found no significant variation in iron-peak elements, even calling it one of the least complex clusters. We can confirm this conclusion with our results, which shows neither any kind of broadening in the overall distribution, nor any variation between the two populations.

\subsection{\object{NGC\,7078} / \object{M\,15}}
\label{sec:ngc7078}
Together with \object{NGC\,7099}, the massive cluster \object{NGC\,7078} is the most metal-poor one in our sample ($\feh_\mathrm{D16}=-2.28$). \object{M\,15} is a peculiar cluster with an unusual distribution of P1 stars being more centrally concentrated than P2 stars \citep{2015ApJ...804...71L}. It also shows no sign of a CN bimodality on the RGB \citep{2005AJ....130.1177C}, but, in contrast to other low-metallicity clusters, there is one on the MS \citep{2010A&A...524A..44P}.

Three populations have originally been identified in the chromosome map \citep{2015ApJ...804...71L}. \citet{2018MNRAS.477.2004N} even found five populations and argue that it might actually be a Type~II cluster. For our analysis, as shown in Fig.~\ref{fig:multipop_ngc7078}, we ignored the suggested split in the primordial population and we were unable to separate their population E that hints towards Type~II. However, we see a clear offset of population P3, which is probably caused by a contamination of this population. The difference in metallicity between P1/P2 and P3 is $\sim 0.03\dex$.
No spread in metallicity has been reported in the literature and \citet{2009A&A...505..117C} suggests that \object{NGC\,7078} may not contain a large number of stars with a different chemical composition. According to \citet{2018MNRAS.477.2004N}, only 5\% of stars should belong to this population.

\subsection{\object{NGC\,7089} / \object{M\,2}}
After a split on the SGB for this Type~II cluster has been reported by \citet{2012ApJ...760...39P}, a connection to the s-process bimodality along the RGB was made by \citet{2013MNRAS.433.1941L}, emphasizing also a similarity to the other Type~II clusters \object{NGC\,1851} and NGC\,6566. Seven populations have been found by \citet{2015MNRAS.447..927M} based on UV photometry.

Metallicities on the RGB have been measured from high-resolution spectra by \citet{2014MNRAS.441.3396Y}, finding three distinct populations: a main population with $\feh=-1.67\pm0.02\dex$ and two anomalous groups with $-1.51\pm0.04$ and $-1.03\pm0.03\dex$, respectively, yielding a split of $\sim 0.16\dex$ of the first one to the main group. They also adopted a CaT analysis like the one presented in this paper, and calculated metallicities from some medium-resolution spectra, obtaining $\feh=-1.58\pm0.08\dex$ for the normal and $-1.29\pm0.09\dex$ for the anomalous groups. They separate the latter into two sub-groups with five stars each and obtain similar results as for the high-resolution spectra, that is $-1.47\pm0.05$ and $-0.98\pm0.06\dex$, respectively.

Unfortunately, we only have 30 stars from population P3 in our data and its metallicity distribution is not well defined. We obtain a difference of about $0.17\dex$ from the medians and even $0.23\dex$ from the means. There is also a clear separation of the $Q_1$--$Q_3$ intervals.

\subsection{\object{NGC\,7099} / \object{M\,30}}
From our data, \object{NGC\,7099} shows neither a significant broadening of the overall metallicity distribution, nor a separation for its two populations, which is consistent with previous studies \citep[like, e.g.,][]{2018ApJ...856..130O} that found evidence for multiple populations, but no abnormalities in metallicity.

\section{Conclusions}
In this paper, we measure equivalent widths of the infrared Ca triplet for almost 34,000 red giants in 25 Galactic globular clusters. We follow a classical approach using a linear relation on stars brighter than the HB to obtain so-called reduced equivalent widths. However, with this approach, about 80\% of our RGB sample would be excluded, so we present an extension of this calibration to stars fainter than the HB and show that it is still valid when using a quadratic term in the relation.

Being aware that the brightness of the horizontal branch can not be used as a reference for field stars, we also present a calibration based on absolute magnitudes in F606W. Furthermore, we calculate bolometric corrections for all the stars in our sample and derive luminosities. That way, we are able to replace the commonly used calibration based on $V-\vhb$ by one using luminosities, which, in the era of Gaia, is now readily available for many stars, or will be in the near future.

Using our CaT-metallicity calibration, we derive metallicities for about 30,000 stars in 25 globular clusters with typical uncertainties of  $\sim 0.12\dex$. We create histograms for each of the clusters, showing the distribution of metallicities. We then use HST UV photometry, when available, to create 
chromosome maps of the clusters and investigate the metallicity distributions of the different populations. 
We find that the metallicity distribution of the P3 population is clearly different than that of P1 and P2 in the seven Type~II clusters in our sample. This adds further evidence to the idea that these clusters, also called metal-complex, harbor populations that have different metallicities. 
We discuss \object{NGC\,5139} (\object{$\omega$~Centauri}) in detail, presenting its complex metallicity structure, and the Type~I cluster \object{NGC\,2808}, which also shows some interesting substructure among its various populations. For \object{NGC\,7078}, we show more evidence that it probably also is a Type~II cluster, as discussed by \citet{2018MNRAS.477.2004N}. For all the other Type~I clusters, we find no significant split or spread in metallicity.

In this paper, we show that our unprecedented large sample of spectra from GC stars, even with a relatively low resolution of $R=2000\mbox{--}4000$, could be used to measure abundances in groups of stars to an accuracy formerly only reached with high-resolution spectroscopy.
This finding is in comfortable agreement with \citet{2008ApJ...682.1217K} who made a careful one-to-one comparison of metallicities obtained from medium and high resolution spectroscopy, however avoiding the CaT.
In the future, we plan to extend the equivalent width technique towards higher metallicities in partially resolved stellar populations of nearby galaxies \citep{2018A&A...618A...3R}.

\begin{acknowledgements}
Based on observations made with ESO Telescopes at the La Silla Paranal Observatory under programme IDs 094.D0142, 095.D-0629, 096.D-0175, 097.D-0295, 098.D-0148, 099.D-0019, 0100.D-0161, and 0101.D-0268. Also based on observations made with the NASA/ESA Hubble Space Telescope, obtained from the data archive at the Space Telescope Science Institute. STScI is operated by the Association of Universities for Research in Astronomy, Inc. under NASA contract NAS 5-26555. This research made use of Astropy,\footnote{http://www.astropy.org} a community-developed core Python package for Astronomy \citep{2013A&A...558A..33A,2018AJ....156..123A}.
We acknowledge funding from the Deutsche Forschungsgemeinschaft (grant DR 281/35-1 and KA 4537/2-1) and from the German Ministry for Education and Science (BMBF Verbundforschung) through grants 05A14MGA, 05A17MGA, 05A14BAC, and 05A17BAA. Acknowledgements: SK gratefully acknowledges financial support from the European Research Council (ERC-CoG-646928, Multi-Pop).
JB acknowledges support by FCT/MCTES through national funds (PIDDAC) by grant UID/FIS/04434/2019 and through Investigador FCT Contract No. IF/01654/2014/CP1215/CT0003.
\end{acknowledgements}



\bibliographystyle{aa}
\bibliography{muse_cat}



\begin{appendix}
\section{Additional figures}

\begin{figure*}
 \includegraphics{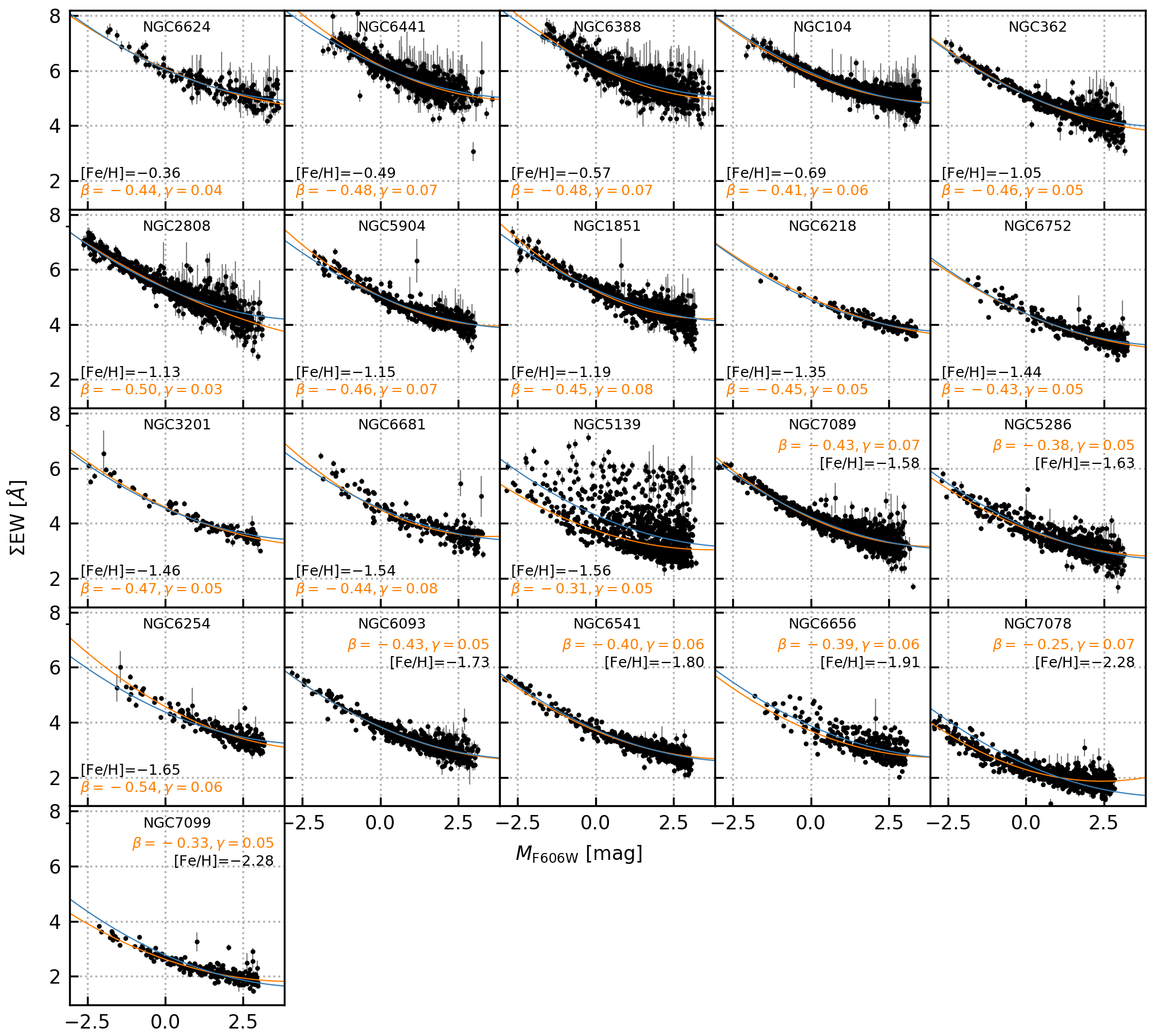}
 \caption{Similar to Figs.~\ref{fig:calib_rew_above_hb} and \ref{fig:calib_rew_all}, but the sum of the equivalent widths $\Sigma \mathrm{EW}$ of the two strongest Ca lines is plotted over the absolute magnitude in F606W $M_\mathrm{F606W}$ for all RGB stars. Quadratic fits to each individual cluster are shown in orange, while a global fit, where the same values for $\beta$ and $\gamma$ are used (giving $\beta=-0.426 \pm 0.002$ and $\gamma=0.053 \pm 0.001$), is plotted for each cluster in blue.}
 \label{fig:calib_rew_M}
\end{figure*}

\begin{figure*}
 \includegraphics{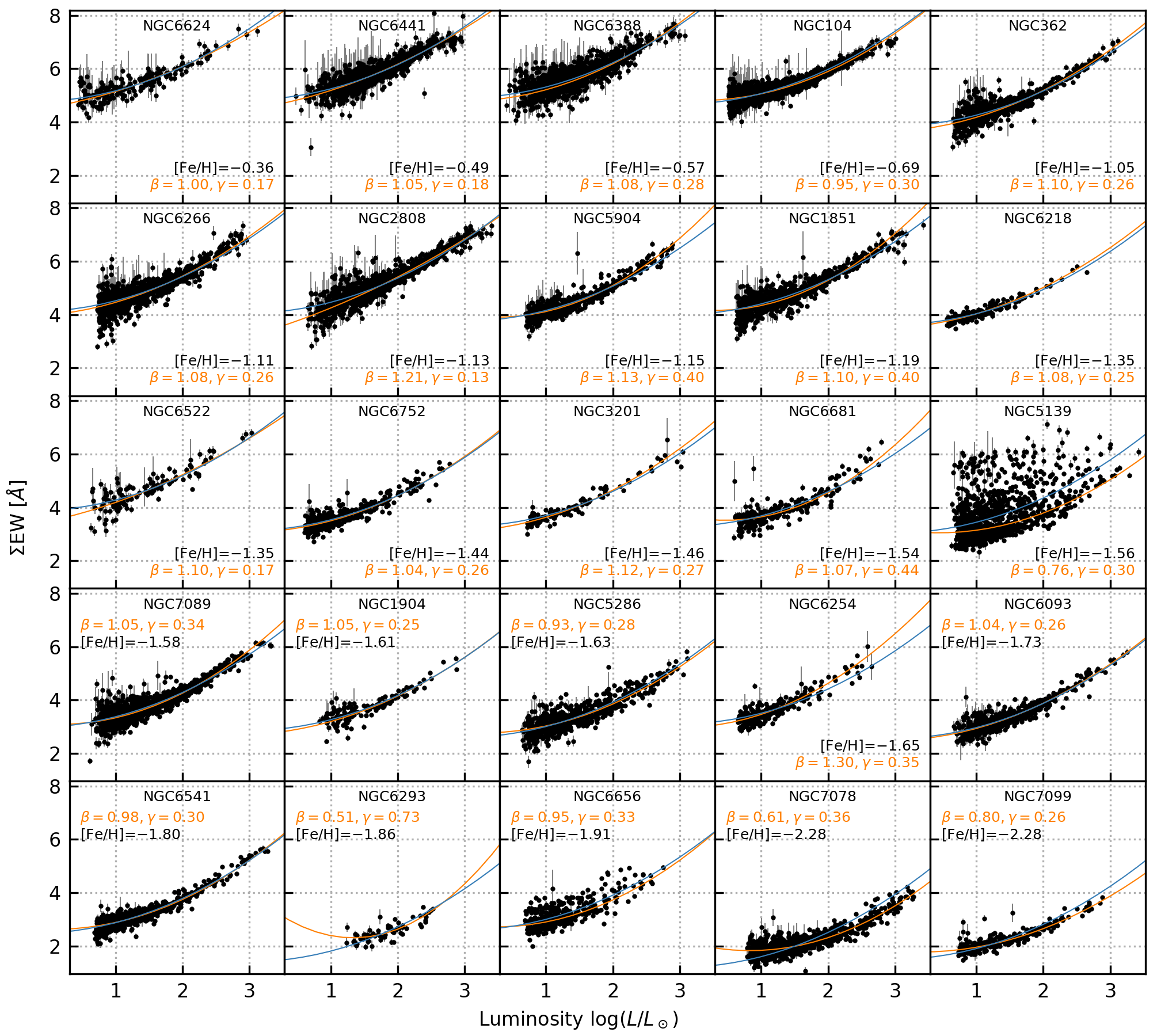}
 \caption{Similar to Figs.~\ref{fig:calib_rew_above_hb} and \ref{fig:calib_rew_all}, but the sum of the equivalent widths $\Sigma \mathrm{EW}$ of the two strongest Ca lines is plotted over $\log L/L_\odot$ for all RGB stars. Quadratic fits to each individual cluster are shown in orange, while a global fit, where the same values for $\beta$ and $\gamma$ are used (giving $\beta=1.006 \pm 0.005$ and $\gamma=0.260 \pm 0.007$), is plotted for each cluster in blue.}
 \label{fig:calib_rew_lum}
\end{figure*}

\end{appendix}



\end{document}